\documentclass{aa}
\usepackage{natbib}
\bibpunct{(}{)}{;}{a}{}{,}
\usepackage{graphicx,color}
\usepackage{txfonts,multirow}

\newcommand{\msunyr}{\ensuremath{\mathit{M}_{\odot}{\rm yr}^{-1}}}   % msun/yr
\newcommand{\kms}{\ensuremath{{\rm km\,s^{-1}}}}                   % $\kms$ec
\newcommand{\msun}{\ensuremath{\mathit{M}_{\odot}}}   % msun
\newcommand{\mini}{\ensuremath{M_{\rm ini}}}                         %M ini
\newcommand{\lsun}{\ensuremath{\mathit{L}_{\odot}}}                  % solar luminosity
\newcommand{\rsun}{\ensuremath{\mathit{R}_{\odot}}}                  % solar radius
\newcommand{\lstar}{\ensuremath{\mathit{L}_{\star}}}                 % stellar luminosity
\newcommand{\mdot}{\ensuremath{\dot{M}}}                             % mass loss rate
\newcommand{\mstar}{\ensuremath{\mathit{M}_{\star}}}                 % stellar mass
\newcommand{\rstar}{\ensuremath{\mathit{R}_{\star}}}                 % stellar radius
\newcommand{\teff}{\ensuremath{\mathit{T}_{\rm eff}}}                % effectieve temperatuur
\newcommand{\reff}{\ensuremath{\mathit{R}_{\rm phot}}}                % effectieve temperatuur
\newcommand{\vinf}{\ensuremath{\upsilon_{\infty}}}                          % maximale uistroomsnelheid
\newcommand{\vesc}{\ensuremath{\upsilon_{\rm esc}}}                         % escape velocity
\newcommand{\K}{\ensuremath{\mathrm{K}}}                 % stellar T
\newcommand{\tauross}{\ensuremath{\tau_{\mathrm{Ross}}}}                 % optical depth of the Lyman continuum
\newcommand{\teffg}{\ensuremath{T_{\mathrm{eff,GVA}}}}                 % optical depth of the Lyman continuum
\newcommand{\teffgcorr}{\ensuremath{T_{\mathrm{eff,GVAcorr}}}}                 % optical depth of the Lyman continuum
\newcommand{\teffc}{\ensuremath{T_{\mathrm{eff,CMF}}}}                 % optical depth of the Lyman continuum

\newcommand{\xsurf}{\ensuremath{X_{\mathrm{sur}}}}                 % optical depth of the Lyman continuum
\newcommand{\ysurf}{\ensuremath{Y_{\mathrm{sur}}}}                 % optical depth of the Lyman continuum
\newcommand{\csurf}{\ensuremath{C_{\mathrm{sur}}}}                 % optical depth of the Lyman continuum
\newcommand{\nsurf}{\ensuremath{N_{\mathrm{sur}}}}                 % optical depth of the Lyman continuum
\newcommand{\osurf}{\ensuremath{O_{\mathrm{sur}}}}                 % optical depth of the Lyman continuum

\newcommand{\lam}{\ensuremath{\lambda}}                 % optical depth of the Lyman continuum

\newcommand{\mbol}{\ensuremath{\mathit{M}_{\rm bol}}}

\begin{document}

\title{The evolution of massive stars and their spectra}
\subtitle{I. A non-rotating 60~\msun\ star from the zero-age main sequence to the pre-supernova stage}
\author{Jose H. Groh\inst{1},  Georges Meynet\inst{1},  Sylvia Ekstr\"om\inst{1}, and Cyril Georgy\inst{2}}

\institute{
%1
Geneva Observatory, Geneva University, Chemin des Maillettes 51, CH-1290 Sauverny, Switzerland; \email{jose.groh@unige.ch}
%2
\and 
Astrophysics group, EPSAM, Keele University, Lennard-Jones Labs, Keele, ST5 5BG, UK
}
\authorrunning{Groh et al.}
\titlerunning{Evolution of massive stars and their spectra. I. non-rotating 60~\msun}

\date{Received  / Accepted }

\abstract{For the first time, the interior and spectroscopic evolution of a massive star is analyzed from the zero-age main sequence (ZAMS) to the pre-supernova (SN) stage. For this purpose, we combined stellar evolution models using the Geneva code and stellar atmospheric/wind models using CMFGEN. With our approach, we were able to produce observables, such as a synthetic high-resolution spectrum and photometry, thereby aiding the comparison between evolution models and observed data. Here we analyze the evolution of a non-rotating 60~\msun\ star and its spectrum throughout its lifetime. Interestingly, the star has a supergiant appearance (luminosity class I) even at the ZAMS. We find the following evolutionary sequence of spectral types: O3 I (at the ZAMS), O4 I (middle of the H-core burning phase), B supergiant (BSG), B hypergiant (BHG), hot luminous blue variable  (LBV; end of H-core burning), cool LBV (H-shell burning through the beginning of the He-core burning phase), rapid evolution through late WN and early WN, early WC (middle of He-core burning), and WO (end of He-core burning until core collapse). We find the following spectroscopic phase lifetimes:  $3.22\times10^6$ yr for the O-type, $0.34\times 10^5$ yr (BSG),  $0.79\times 10^5$ yr (BHG), $2.35\times10^5$ yr (LBV), $1.05\times10^5$ yr (WN), $2.57\times10^5$ yr (WC), and $3.80\times10^4$ yr (WO). Compared to previous studies, we find a much longer (shorter) duration for the early WN (late WN) phase, as well as a long-lived LBV phase. We show that LBVs arise naturally in single-star evolution models at the end of the MS when the mass-loss rate increases as a consequence of crossing the bistability limit. We discuss the evolution of the spectra, magnitudes, colors, and ionizing flux across the star's lifetime, and the way they are related to the evolution of the interior. We find that the absolute magnitude of the star typically changes by $\sim6$ mag in optical filters across the evolution, with the star becoming significantly fainter in optical filters at the end of the evolution, when it becomes a WO just a few $10^4$ years before the SN explosion. We also discuss the origin of the different spectroscopic phases (i.e., O-type, LBV, WR) and how they are related to evolutionary phases (H-core burning, H-shell burning, He-core burning).
}

\maketitle

\section{\label{intro}Introduction} 

Massive stars are essential constituents of stellar populations and galaxies in the near and far Universe. They are among the most important sources of ionizing photons, energy, and some chemical species, which are ejected into the interstellar medium through powerful stellar winds and during their extraordinary deaths as supernovae (SN) and long gamma-ray bursts (GRB). For these reasons, massive stars are often depicted as cosmic engines, because they are directly or indirectly related to most of the major areas of astrophysical research.

Despite their importance, our current understanding of massive stars is still limited. This inconvenient shortcoming can be explained by many reasons on which we elaborate below. First, the physics of star formation mean that massive stars are rare \citep{salpeter55}. Moreover, their lifetime is short, of a few to tens of millions of years \citep[e. g.,][]{ekstrom12,langer12}. These factors make it challenging to construct evolutionary sequences and relate different classes of massive stars. This is in sharp contrast to what can be done for low-mass stars.

Second, one can also argue that the evolution of massive stars is extremely sensitive to the effects of some physical processes, such as mass loss and rotation \citep{maeder_araa00,heger00}, that have relatively less impact on the evolution of low-mass stars. However, the current implementation of rotation in one-dimensional codes relies on parametrized formulas, and the choice of the diffusion coefficients has a key impact on the evolution \citep{meynet13a}. Likewise, mass-loss recipes arising from first principles are only available for main sequence (MS) objects \citep{vink00,vink01} and a restricted range of Wolf-Rayet (WR) star parameters \citep{grafener08}. Third, binarity seems to affect the evolution of massive stars, given that a large portion of them are in binary systems that will interact during the evolution \citep{sana12}.

Fourth, our understanding of different classes of stars is often built by comparing evolutionary models and observations. However, mass loss may affect the spectra, magnitudes, and colors of massive stars, thus making the comparison between evolutionary models and observations a challenge. In addition to luminosity, effective temperature, and surface gravity, the observables of massive stars can be strongly influenced by a radiatively driven stellar wind that is characteristic of these stars. The effects of mass loss on the observables depend on the initial mass and metallicity, since they are in general more noticeable in MS stars with large initial masses, during the post-MS phase, and at high metallicities. When the wind density is significant,  the mass-loss rate, wind clumping, wind terminal velocity, and velocity law have a strong impact on the spectral morphology. This makes the analysis of a fraction of massive stars a difficult task, and obtaining their fundamental parameters, such as luminosity and effective temperature, is subject to the uncertainties that comes from our limited understanding of mass loss and clumping.  Furthermore, the definition of effective temperature of massive stars with dense winds is problematic and, while referring to an optical depth surface, it does not relate to a hydrostatic surface. This is caused by the atmosphere becoming extended, with the extension being larger the stronger the wind is.  Stellar evolution models are able to predict the stellar parameters only up to the stellar hydrostatic surface, which is not directly reached  by the observations of massive stars when a dense stellar wind is present. Since current evolutionary models do not  thoroughly simulate the physical mechanisms happening at the atmosphere and wind, model predictions of the evolution of massive stars are difficult to be directly compared to observed quantities, such as a spectrum or a photometric measurement.

The main driver of this paper is to improve the comparison between models and observations of massive stars. To properly understand the Physics that govern massive stars, it is urgently necessary to combine stellar evolutionary calculations to radiative transfer models of the stellar atmosphere. Essentially, the atmospheric models allow the physical quantities predicted by the stellar evolution model to be directly compared to observed features. We build on 
earlier studies in this direction that were made by the Geneva group, which  coupled an earlier version of the Geneva stellar evolution code with the ISAWIND atmospheric code  \citep{dekoter93,dekoter97}, creating the Costar models \citep{schaerer96a}. These models focused mainly on the spectroscopic evolution during the Main Sequence (MS; \citealt{schaerer96b,schaerer97a}), and on the effects of mass loss on the evolution and envelope structure of Wolf-Rayet (WR) stars \citep{schaerer96wr}. 

From the Costar models up to now, significant improvements both in the stellar evolution and atmospheric models have been accomplished. From the atmospheric modeling perspective, the main advances have been the inclusion of full line blanketing and line overlap, updated and extended atomic data, and wind clumping.  Solving the radiative transfer across the atmosphere of massive stars is a complex and demanding task, and adequate atmospheric codes became available only in the past decade. Complex radiative transfer models such as CMFGEN \citep{hm98}, PoWR \citep{hamann06}, and FASTWIND \citep{puls06}, that take into account the necessary physics to study the radiation transport across the atmosphere and wind, have been separately employed to analyze observations of O stars (e.g., \citealt{hillier03,martins05,bouret03,bouret05,puls06,marcolino09,najarro11,repolust04,mokiem05,mokiem07b,tramper11}), LBVs (e.g., \citealt{hillier01,ghd06,ghd09,ghd11,ghm12,najarro09}), and WRs (e.g., \citealt{hm99,dessart00,grafener05,sander12}). From the stellar evolution perspective, the main advances have been to include the effects of rotation and magnetic fields, and improve opacities and mass-loss recipes.

As of yet, however, no code has been capable of studying the evolution of the spectra of massive stars throughout their entire evolution, since the modern atmospheric/wind models have never been coupled to stellar evolutionary models. Here we bridge this gap by performing, for the first time, coupled calculations of stellar evolution with the Geneva code and atmospheric and wind modeling with the CMFGEN code. This approach opens up the possibility to investigate stellar evolution based not only on interior properties, but also from a spectroscopic point of view. This allow us to relate interior properties of the star with its appearance to the observer. Our ultimate goal is to provide improved comparison between models and observations of massive stars. 

In this first paper of a series, we analyze the spectroscopic and photometric evolution of a non-rotating 60~\msun\ star at solar metallicity, from the zero-age main sequence (ZAMS) until the pre-SN stage. We choose a non-rotating model to properly disentangle the effects that mass loss and rotation have on the evolution of the spectra of massive stars. The reasons for using a 60~\msun\ star are twofold. First, this initial mass is representative of the qualitative evolution of the most massive stellar models, in the range 50--120~\msun. Second, stars with this initial mass do not evolve through a red supergiant (RSG) phase, which allow us to use a single atmospheric code (CMFGEN) to analyze the whole evolution. 

The same modeling approach described here has been employed in previous papers from our group that analyzed the properties of massive stars just before the SN explosion \citep{gme13,gmg13,gge13}. In \citet{gme13}, we found that rotating stars with initial mass (\mini) in the range 20--25 ~\msun\ end their lives as luminous blue variable (LBV) stars. The fate of single massive stars with $\mini=9-120~\msun$ was investigated in \citet{gmg13}, where we showed that massive stars, depending on their initial mass and rotation, can explode as red supergiants (RSG), yellow hypergiants (YHG), LBVs, and Wolf-Rayet (WR) stars of the WN and WO subtype. We applied these models to investigate the nature of the candidate progenitor of the SN Ib iPTF13bvn, concluding that a single WR star with initial mass $\sim31-35~\msun$ could explain the properties of the progenitor \citep{gge13}. These analyses showed that it is crucial to produce an output spectrum out of evolutionary calculations to properly interpret the observations.

This paper is organized as follows. In Sect.~\ref{model} we describe our modeling approach, while we discuss our definitions of evolutionary and spectroscopic phases in Sect. \ref{evolspec}. We analyze the evolution of a non-rotating 60~\msun\ star in Sect.~\ref{broadevol}. In Sect. \ref{lifetimes} we investigate the lifetimes of different evolutionary  and spectroscopic phases, while in Sect. \ref{origspec} we discuss how evolutionary phases are linked to different spectroscopic phases. Sect. \ref{specevol} analyzes the evolution of the spectra, magnitudes, colors, and ionizing flux across the star's lifetime. Caveats of our analysis are discussed in Sect. \ref{caveat}, and our concluding remarks are presented in Sect. \ref{conc}.

In a series of forthcoming papers, we will present the results for a larger initial mass range and investigate the effects of rotation, metallicity, and magnetic fields on the spectroscopic evolution.

\section{\label{model}Physics of the models}

We compute coupled models using the Geneva stellar evolution and the CMFGEN atmosphere/wind radiative transfer codes. Evolutionary models from the ZAMS to the pre-SN stage comprises tens of thousands of calculations of stellar structures. Given that a typical CMFGEN model takes about half day of CPU time to finish, it is impracticable to compute full atmosphere/wind modeling at each timestep of the evolution and still produce a grid of evolution models. Given the huge computing effort involved, our strategy is to perform post-processing atmospheric/wind radiative transfer on already existing evolutionary calculations. We apply this procedure to 53 stages that are carefully selected to sample the full evolution (see Table \ref{model log}). The advantages are that we can benefit from the Physics included in two of the most up-to-date and advanced codes, produce a grid of evolution models with output spectra, and analyze the grid in a manageable amount of time. The main disadvantage is that our models do not include feedback effects of the wind on the evolution \citep{schaerer96wr} and envelope structure \citep{grafener13}. Below we describe the Geneva stellar evolution code (Sect. \ref{evolcode}), the CMFGEN atmospheric/wind code (Sect. \ref{cmfgen}), how the two codes are combined (Sect.~\ref{combine}), and the criteria used for spectroscopic classification (Sect. \ref{classifyspec}).

\subsection{\label{evolcode} Stellar evolution}

The evolutionary model of a 60~\msun\ star discussed here has been computed by \citet{ekstrom12}  with the Geneva stellar evolution code, as a part of a large grid of models. We refer the interested reader to the aforementioned paper for further details. The main characteristics of the code are summarized below.

The model assumes solar metallicity (Z=0.014) and initial abundances from \citet{asplund09}. For the H-core and He-core burning phases, an overshoot parameter of $d_\mathrm{over}= 0.10 H_P$ is assumed\footnote{This value of $d_\mathrm{over}$ is chosen to reproduce the MS width between 1.35 and 9~\msun, and is smaller than in other evolutionary codes (see comparison in \citealt{martins13}).}, where $H_P$ is the local pressure scale height. Mass loss is a key ingredient in the models, affecting not only the evolution throughout the Hertzprung-Russel (HR) diagram but also the emerging spectrum. Therefore, most conclusions achieved in this paper depend on the mass-loss recipe used in the computations. To select the most suitable mass-loss recipe, criteria based on chemical abundances and the effective temperature estimated by the Geneva code (\teffg; see Sect. \ref{temp}) are used. The relevant criteria and respective mass-loss recipes relevant for the 60~\msun\ model discussed here are: \\
-- surface H abundance (\xsurf)  $> 0.3$ and $\log(\teffg/\K)>3.9)$: \citet{vink01};\\
-- $\xsurf < 0.3$ and $3.900 < \log(\teffg/\K) \leq 4.000$): \citet{vink01};\\
-- $\xsurf < 0.3$ and $4.000 < \log(\teffg/\K) \leq 4.477$): \citet{nl00};\\
-- $\xsurf < 0.3$ and $4.477 < \log(\teffg/\K) \leq 4.845$): \citet{grafener08} or \citet{vink01}, whichever gives the highest value of $\mdot$;\\
-- $\xsurf < 0.3$ and $4.845 \geq \log(\teffg/\K)$: \citet{nl00};\\
-- $\log(\teffg/\K) \leq 3.9)$: \citet{dejager88}.

The 60~\msun\ star evolution model used here has been computed up to the Si burning phase. Here we discuss the results up to the end of C burning, since no appreciable changes in the surface properties are seen beyond this phase up to core collapse.

\subsection{\label{cmfgen} Atmospheric and wind modeling}

To compute the output spectra\footnote{The spectra computed here (in vacuum wavelengths) and evolutionary models are public available through the webpage  \url{http://obswww.unige.ch/Recherche/evol/-Database-}} we use the atmospheric radiative transfer code CMFGEN \citep{hm98}. CMFGEN is a  spherically-symmetric, fully line blanketed code that computes line and continuum formation in non-local thermodynamical equilibrium. Since the evolutionary model discussed here has no rotation, the use of spherical symmetry seems justified\footnote{Note, however, that wind inhomogeneities could break the spherical symmetry if they have a large scale length.}. CMFGEN computes a self-consistent radiative transfer including the stellar hydrostatic surface and the wind. CMFGEN is suitable for stars with $\teffg > 7500~\K$,  a condition that is satisfied during almost all the lifetime of a non-rotating 60~\msun\ star. For the $10^4$ years when $\teffg < 7500~\K$, no spectra is computed, and the synthetic photometry (Sect. \ref{specevol}) is linearly interpolated in age between stages 27 and 28 (see Sect. \ref{broadevol}). 

A CMFGEN model needs as input the luminosity (\lstar), effective temperature at a reference optical depth, mass (\mstar), and surface chemical abundances. For consistency, we adopt in CMFGEN the same $\mdot$ recipe as that used by the Geneva evolution code.The momentum equation of the wind is not solved and a velocity structure $v(r)$ needs to be  adopted. For the wind part, we assume a standard $\beta$-type law, with $\beta=1$ if $\teffg > 21000~\K$ and $\beta=2.5$ otherwise. A hydrostatic solution is computed iteratively for the subsonic portion and is applied up to 0.75 of the sonic speed, where the hydrostatic and wind solutions are merged. This scheme has been computed for all models except those with $\teffg < 9000~\K$. These models are cool LBVs that have dense winds and an optical depth larger than 100 at the sonic point. In these cases, the computation of the hydrostatic structure failed and we employed an approximate solution following a scale-height approach. The wind terminal velocity ($\vinf$) is computed using the parametrization from \citet{kudritzki00} for OB stars and LBVs, and from \citet{nl00} for WR stars of the WN and WC type. For WO stars, an iterative scheme is adopted. We initially compute a spectrum with the value of $\vinf$ as given by the \citet{nl00} recipe, which is typically at most $\sim2800~\kms$. If a WO-type spectrum arises, we recompute a spectrum with $\vinf=5000$~\kms\, which is more representative of the observed Galactic WO stars \citep{drew04,sander12}.

Optically-thin wind clumping is included via a volume filling factor ($f$) approach, which assumes dense clumps and a void interclump medium. The wind is also assumed to be unclumped close to the stellar surface and to acquire full clumpiness at large radii. The variation of $f$ as a function of distance from the center of the star ($r$) is given by 
\begin{equation}
\label{clump}
f(r)=f_\infty+(1-f_\infty)\exp[-v(r)/v_c]\,\,, 
\end{equation} 
where $f_\infty$ is the filling factor and $v_c$  is the velocity at which clumps start to form. For O stars we assume $f_\infty=0.2$ and $v_c=30~\kms$. The assumed $f$ is what is typically needed to bring in agreement the observed wind momentum of O stars with the \citealt{vink01} \citep{repolust04,mokiem07b}. For subsequent evolutionary phases we assume $f_\infty=0.1$, which is characteristic of WR stars. We use $v_c=20~\kms$  for B stars and LBVs, and $v_c=200~\kms$ for WR stars. We discuss how wind clumping affects our results in Sect. \ref{mdotclump}.

X-rays are included only for OB-type stars and, for simplicity, we assume fixed x-ray temperatures and filling factors for all OB star models. A two-component plasma is assumed (see \citealt{hm98,pauldrach94,najarro11}), one with temperature $T=2\times10^6\K$ and filling factor of 6.0$\times10^{-2}$, and the other with $T=6\times10^6\K$ and filling factor 8.0$\times10^{-3}$. For both components, we assume a velocity of 500~\kms\ for the shocks to become important.

\begin{figure}
\center
\resizebox{0.99\hsize}{!}{\includegraphics{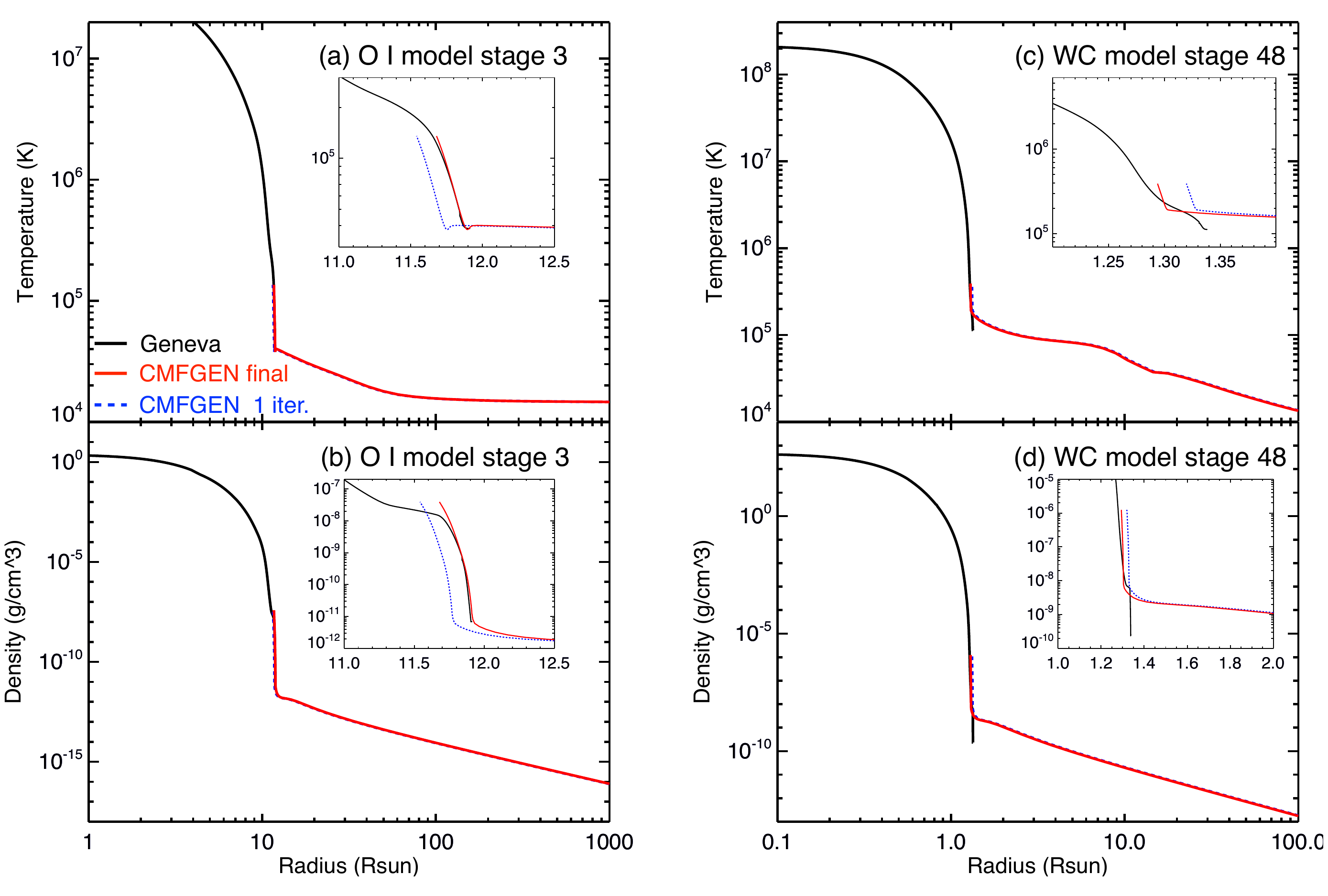}}\\
\caption{\label{tstruct} Temperature (panel $a$) and density ($b$) as a function of radius for a non-rotating star with initial mass of 60~\msun\ at metallicity $Z=0.014$ for a O I model at stage 3 (as defined in Sect. \ref{broadevol}). The solution obtained by the Geneva code is shown in black, while the final CMFGEN solution is displayed in solid red. A one-iteration CMFGEN solution used for initial guessing of the T structure is shown by the dashed blue line. A zoom-in on the connecting region of the Geneva and CMFGEN solutions (at $\tauross=10$) is shown in the inset. The temperature and density structures of a model at stage 48 (WC) is displayed in panels $c$ and $d$, respectively.}
\end{figure}

\subsection{\label{combine} Combining the Geneva code and CMFGEN solutions}

The stellar structure calculations with the Geneva code produce as output \teffg,  \lstar, \mstar, and surface abundances, among others. They are used as input for the CMFGEN model and  we use the temperature and density structures of the stellar envelope to merge the CMFGEN and stellar structure solutions. They are merged at a Rosseland optical depth (\tauross) of 10, to ensure that the envelope solution computed by the Geneva code is used in the inner part of the atmosphere ($\tauross \geq 10$), and that the CMFGEN solution is employed in the outer atmosphere/wind ($\tauross < 10$).

In practice, the merging is achieved by running a CMFGEN model for one iteration, assuming for this initial iteration only that \teffg\ corresponds to the effective temperature at \tauross=10. The output T structure from the one-iteration CMFGEN model is then compared to the T structure of the Geneva code. An inconsistent and non-continuous solution will generally be obtained, which means that the radius of the CMFGEN model needs to be adjusted in order that the two T solutions match at $\tauross=10$. The procedure is illustrated in Fig. \ref{tstruct} for an O star model at the MS and a WR model of the WC subtype at the He-core burning phase. As one can see, our approach guarantees that the temperature and density structure are continuous from the center of the star through the envelope up to the stellar wind, at several hundred \rstar.

The procedure above allows us to compute the values of the effective temperature out of evolutionary models in a novel way. We use the CMFGEN results to determine $\tauross (r)$ and, with that, we find the photospheric radius as $\reff=r(\tauross=2/3)$. The effective temperature is calculated as
\begin{equation}
\label{teffeq}
\teffc=\left({\frac{\lstar}{4 \pi \sigma \reff^2}}\right)^{1/4},
\end{equation}
where $\sigma$ is the Steffan-Boltzmann constant.

 \subsubsection{\label{temp} Comparison with previous determinations of the effective temperature by the Geneva code}

Previous papers using the Geneva code present evolutionary tracks using $\teffg$  \citep[e.g.,][for the more recent ones]{ekstrom12,georgy12a}. In this subsection, we investigate how it compares with our revised $\teffc$.  In particular, the optical depth ($\tau$) scale computed by CMFGEN may differ from that obtained using the simplified approach from the Geneva code. Let us recall that \teffg\ is one of the boundary conditions needed to solve the usual set of stellar structure equations and is determined using an iterative, numerical scheme. A plane parallel, gray-atmosphere is assumed by the Geneva code. We refer the reader to Chapter 24 of \citet{maeder09} and Chapters 11 and 12 of \citet{kip13} for thorough reviews.

When the optical depth of the wind is appreciable, for instance when the star is a WR, the $\tau=2/3$ surface moves outwards in comparison with the optically-thin case \citep{deloore82}. To correct for this effect, the Geneva code uses a simple scheme that takes into account both electron-scattering and line opacities. The correction is based on the relationship from \citet{langer89}\footnote{Note that the original results from \citet{langer89} assume electron-scattering opacity only.} to compute the radius where $\tau=2/3$ as:
\begin{equation}
R_\mathrm{phot,GVA}=R_{\star,GVA}+3 \bar{\kappa} \mdot/(8 \pi \vinf),
\end{equation}
where $R_{\star,GVA}$ is the radius associated with \teffg\ via the Steffan-Boltzmann relation. This equation is valid for $\beta=2$, and $\vinf=2000~\kms$ is assumed in the Geneva code. Here, $\bar{\kappa}$ is a flux-weighted mean opacity, which is computed using the modified radiation-driven wind theory \citep{kudritzki89} as $\bar{\kappa}=\sigma_{\mathrm e} (1+M)$, where $\sigma_{\mathrm e}$ is the electron-scattering opacity and $M$ is the force-multiplier parameter \citep{cak}. The value of $M$ depends on the line-force parameters $k$, $\alpha$, and $\delta$, which are assumed constant in the Geneva code with values $k=0.124$, $\alpha=0.64$, and $\delta=0.07$\footnote{These values correspond to those of an O star with $\teff=50000~\K$ \citep{pauldrach86}.}. The effective temperature corrected for the wind optical depth is then estimated as 
\begin{equation}
\label{testeq}
\teffgcorr=\left({\frac{\lstar}{4 \pi \sigma R_\mathrm{phot,GVA}^2}}\right)^{1/4}.
\end{equation}
In practice, \teffgcorr\ is computed by the Geneva code only at phases when the H surface abundance is less than 0.3 by mass, corresponding roughly to the WR phase \citep{ekstrom12}. We refer the reader also to Sect. 2.7 of \citet{schaller92} for further details.

\begin{figure}
\center
\resizebox{0.980\hsize}{!}{\includegraphics{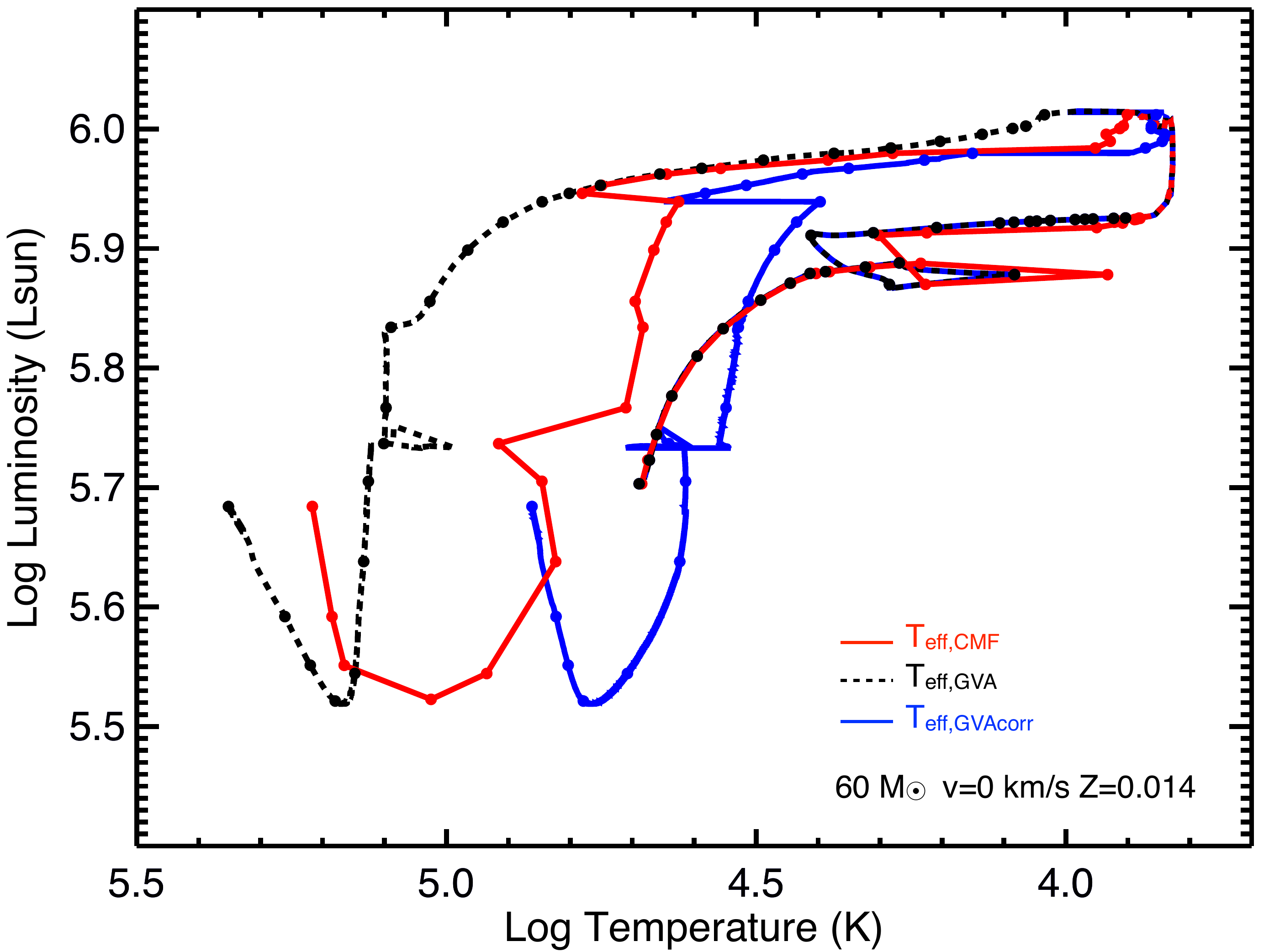}}\\
\caption{\label{teffcomp} Evolutionary tracks of a non-rotating 60~\msun\ star at $Z=0.014$. The track with effective temperatures computed using our revised procedure (\teffc) is shown in red, while previous estimates using the Geneva code are shown in dashed black (\teffg; not corrected for the wind optical depth) and blue (\teffgcorr; corrected for the wind optical depth). The filled circles correspond to timesteps for which a CMFGEN model was computed.}
\end{figure}

Figure \ref{teffcomp} presents the 60~\msun\ evolutionary track in the HR diagram computed with the three effective temperature definitions described above. Important differences can be readily noted. First, in the regime where the winds are optically thin in the continuum, the values of $\teffc$ are similar to those of $\teffg$. This occurs from the beginning of the evolution up to stage 9. Second, throughout the rest of the evolution, $\teffc$ is lower than $\teffg$, indicating that the optical depth of the wind is non-negligible. Third, the wind optical depth becomes important even when $\xsurf>0.3$. This means that \teffgcorr\ is not applied at all phases where a dense wind is present, such as when the star is an LBV. Fourth, in the regime of optically-thick winds, \teffgcorr\  is systematically lower than \teffc. The difference amounts from thousands to several tens of thousands of K, depending on the parameter space. The reasons for the difference are the simple assumptions behind the computation of \teffgcorr. Among them, we highlight that the opacity is computed using constant values of $k$, $\alpha$, and $\delta$, which are more appropriate to O stars \citep{pauldrach86} and may not apply to WR stars \citep{schmutz97,grafener05}. In addition, a high value of $\beta=2$ is assumed when calculating \teffgcorr, while the computation of $\teffc$ uses $\beta=1$. We note that detailed spectroscopic analyses  \citep[e.g.,][]{hm99,sander12} and hydrodynamical models of WR winds \citep{grafener05} indicate that $\beta=1$ is preferred. A higher value of $\beta$ produces a shallow density structure at the base of the wind, shifting the photosphere outwards compared to the $\beta=1$ case. 

{\it In summary, \teffgcorr\ does not seem to provide a good estimate of the effects of a dense wind on the effective temperature.} We also reinforce that $\teffg$ does not  correspond to T at a fixed $\tau$, since the wind contribution to the $\tau$ scale is variable during the evolution. Therefore, \teffg\ should not be compared to $T$ derived from atmospheric modeling at high optical depth (e.g., \tauross=10). Instead, whenever available, the values of $\teffc$ should be preferred over \teffgcorr\ or \teffg\ when comparing evolutionary models to observations of massive stars. Since CMFGEN models could not be computed if $\teffg<7500~\K$, we assume throughout the rest of this paper
\[
\teff = 
\left\{
\begin{array}{ll}
      \teffg & \mathrm{if}~ \teffg \leq 7500 \K \\
      \teffc & \mathrm{if }~\teffg > 7500 \K \\
\end{array} 
\right.
\]

\subsection{\label{classifyspec} Spectroscopic classification}

To classify the synthetic spectra, we use well-established quantitative criteria for O and WR  stars. For O stars, we use the calibrations from \citet{mathys88}, which relates the \ion{He}{i} $\lambda4473$/ \ion{He}{ii} $\lambda4543$ line ratio to spectral types\footnote{We quote vacuum wavelengths in this paper.}. The luminosity classes I, III, and V are related to the ratio of \ion{Si}{iv} $\lambda4090$ to \ion{He}{i} $\lambda4714$ for spectral types later than O7, and to the strength of \ion{He}{ii} $\lambda4687$ for spectral types earlier than O7 \citep{contialschuler71,contileep74,mathys88}. We supplement the quantitative spectral types with the morphological nomenclature devised by \citet{walborn71,walborn73}  and \citet{walborn02}, and recently updated by \citet{sota11}. For WR stars, we employ the criteria from \citet{crowther98} for WO and WC stars, and those from \citet{ssm96} and \citet{crowther95a} for WN stars. 

For early B stars, we employ the morphological classification from \citet{walborn71} and \citet{walborn90}. These are based on the relative strength between \ion{Si}{iv} $\lambda4117$ and \ion{He}{i} $4122$,  \ion{Si}{iii} $\lambda4554$ to \ion{He}{i} $\lambda4389$, and  \ion{Si}{iii} $\lambda4554$ to \ion{Si}{iv} $\lambda4090$. As is common in the literature, we associate the Ia$^+$ luminosity class to B-type hypergiants (BHG; \citealt{vg82b}). BHGs are differentiated from BSGs based on the strength of the H Balmer line emission.  BHG have these lines in emission, usually with a P-Cygni profile, while normal BSGs have Balmer lines in absorption, with the possible exception of H$\alpha$ \citep{lennon92}.

However, we argue here that the BHG classification do not include all stars with $\teff < 25000~\K$ and dense winds. For increasing wind densities, stars in the $\teff$ range of what would be called a `B star'  develop strong (P Cygni) emission line spectrum, which are only seen in LBVs. It would be misleading to classify these stars as BHGs, since LBVs have spectral and wind properties markedly distinct from BHGs \citep{clark12a}. Despite having similar $\teff$ as BHGs, LBVs typically have lower $\vinf$ and higher $\mdot$ than BHGs \citep{clark12a}. However, there seems to be a smooth transition (and even some overlap) between BHG and LBV properties, which makes it challenging to draw a firm line dividing the two classes. It seems clear that there is a progression in wind density from B supergiants to B hypergiants to LBVs \citep{clark12a}. Here we arbitrarily choose the strength of H$\alpha$ as a proxy for the wind density, and classify models that have the intensity of the H$\alpha$ peak above 4 times the continuum as LBVs. 

Therefore, models that have spectrum similar to observed bona-fide LBVs, such as AG Car, P Cygni, and HR Car, have their spectral type listed as LBVs. While we recognize that formally there is no ``LBV'' spectral classification, we opted to use this classification since there is no objective spectral classification criteria of stars with winds denser than those of BHGs and that have $8000 \K \lesssim \teff \lesssim 25000 \K$. The spectra of these stars have  been commonly referred to in the literature as ``P Cygni-type'', ``iron'', and ``slash'' stars (see, e.g., \citealt{wf2000,clark12a}). In addition, we subdivide the LBVs in `hot LBV' and `cool LBV' according to the spectral morphology, following in general lines the classification scheme from \citet{massey07} and \citet{clark12a}. Hot LBVs have strong or moderate \ion{He}{i} lines showing P Cygni profiles. Cool LBVs have weak or no  \ion{He}{i} absorption lines and strong emission from low-ionization species such as  \ion{Fe}{ii},  \ion{N}{i}, and  \ion{Ti}{ii}.

\section{\label{evolspec} Evolutionary and spectroscopic phases}

Unfortunately, often the terms ``evolutionary phase" and ``spectroscopic phase" have been indiscriminately used in the literature, causing confusion. Here we would like to ask the reader to appreciate the difference between the two terms, which sometimes have had different meanings for stellar evolution theorists, stellar atmosphere theorists, and observers working on massive stars. In particular, we note that spectroscopic phases such as RSG, WR, or LBV, should {\it not} be used as a synonym of evolutionary phase.

{\it Evolutionary phases} are defined here in terms of nuclear burning stages and the stellar structure, which are properties inherent to the stellar interior. As such, these are not directly accessible by observations and, to determine the evolutionary phases of stars, one necessarily has to rely on stellar evolution models. The evolutionary phase names are self-explanatory and are H-core burning (or Main Sequence), H-shell burning, He-core plus H-shell burning, He-core burning, He-shell burning, C-core plus He-shell burning, C-core plus He-shell burning plus H-shell burning, and so on. From stellar evolution theory, evolutionary phases can be unambiguously determined and chronologically ordered.

\begin{figure*}
\center
\resizebox{0.980\hsize}{!}{\includegraphics{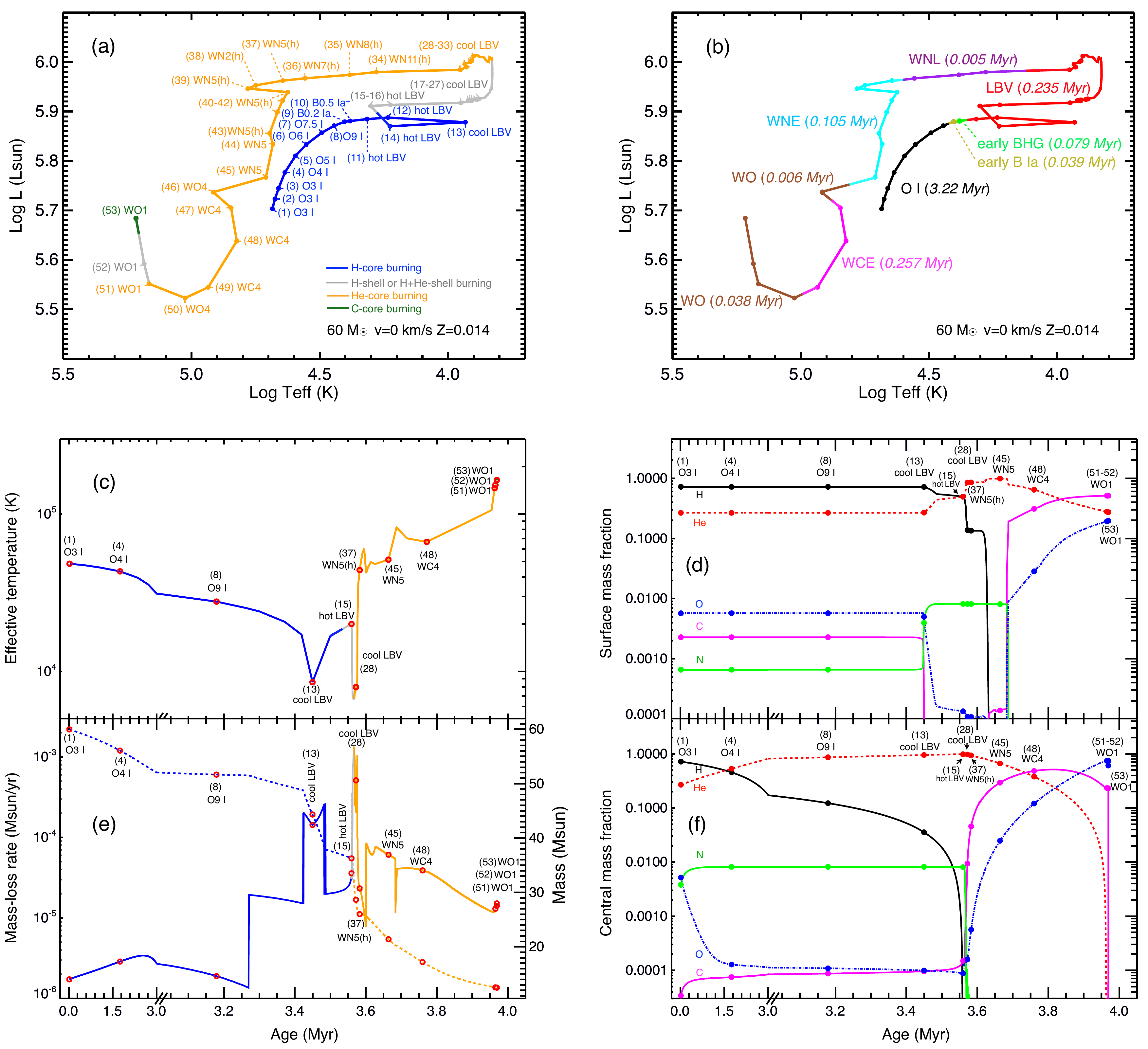}}\\
\caption{\label{hrd1} {\it (a):} HR diagram showing the evolutionary track of a non-rotating star with initial mass of 60~\msun\ at  metallicity $Z=0.014$,  using our revised values of \teff. The color code corresponds to the evolutionary phases of a massive star, with H-core burning in blue, He-core burning  in orange, C-core burning in green, and H and/or He-shell burning in gray.  {\it (b):} Similar to $a$, but color coded according to the spectroscopic phases. Lifetimes of each phase are indicated in parenthesis. {\it (c):} Evolution of $\teff$ as a function of age. The color code is the same as in {\it a}. {\it (d):} Surface abundances of H (black), He (red), C (magenta), N (green), and O (blue) as a function of age. {\it (e):} Mass-loss rate (left axis) and mass (right axis) as a function of age. Color code is the same as in $a$. {\it (f):} Central abundances as a function of age, with the same color-coding as in $d$. Spectral types at selected timesteps are indicated. }
\end{figure*}

{\it Spectroscopic phases} are related to the duration when a certain spectral type appears during the evolution of a star, i.e.,  they are defined here in terms of the spectral appearance of stars.  Spectroscopic phases are in general determined by the surface properties, i.\, e., the physical conditions in the atmosphere and wind. Atmospheric properties are comprised by quantities such as $\teff$, surface gravity, and surface abundances, and wind properties by $\mdot$, $\vinf$, velocity law, and  wind clumping. Spectroscopic phases are labelled here using the spectral type and luminosity class when necessary, corresponding to:
\begin{itemize}
\item O-type spectroscopic phase, B-type, etc (following the usual MK spectral type determination), which can be subdivided according to the luminosity class (I, Ia, Ia$^+$,Ib, Iab, II, III, IV, V). More generally, they can be grouped in O supergiants (all O stars with luminosity class I, Ia, Ib, Iab), or B-type supergiants (BSG), or yellow supergiants (all AFG supergiants), or red supergiants (all KM supergiants); or B-type hypergiants (BHG; B stars with luminosity class Ia$^+$), or yellow hypergiants (YHG, luminosity class Ia$^+$); 
\item Wolf-Rayet (WR) phase, which is subdivided in WN (with N lines), WC (with C lines), or WO (with strong O and C lines). WN and WC stars are further subdivided in WNE (early WN), corresponding to spectral types WN1 to WN5; WNL (late WN, spectral types WN6 to WN 11); WCE (early WC, subtypes WC1 to WC5), and WCL (late WC, subtypes WC6 to WC9), following \citet{crowther07};
\item Luminous Blue variable (LBV) phase, when the spectrum has similar characteristics to those of well-known LBVs such as AG Car, P Cyg, Eta Car, and S Dor. Depending on the spectral lines that are present, they are subdivided here in hot or cool LBVs, as described in Sect. \ref{classifyspec}.
\end{itemize}

Therefore, spectroscopic phases reflect the surface conditions of the star, and may be linked to different evolutionary phases depending on the initial mass, metallicity, rotation, and magnetic fields. To relate evolutionary and spectroscopic phases, a calibration is needed by computing output spectra from stellar evolution models, which is one of the goals of this paper.

\section{\label{broadevol} The evolution of a non-rotating 60~\msun\ star and its various spectral types }

\begin{figure*}
\center
%\resizebox{0.99\hsize}{!}{\includegraphics{spec_selected_montage.pdf}}\\
\resizebox{0.95\hsize}{!}{\includegraphics{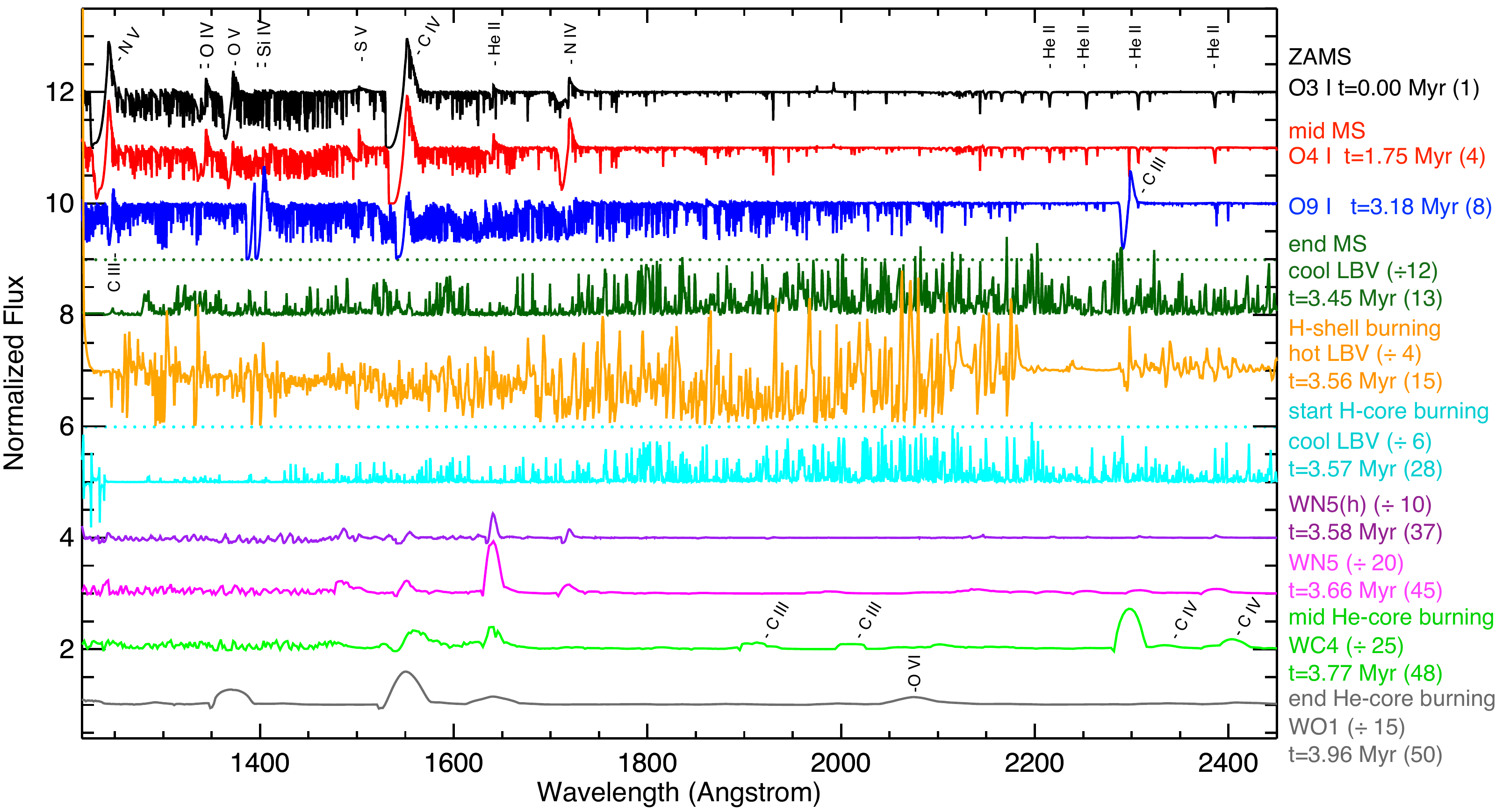}}\\
\resizebox{0.95\hsize}{!}{\includegraphics{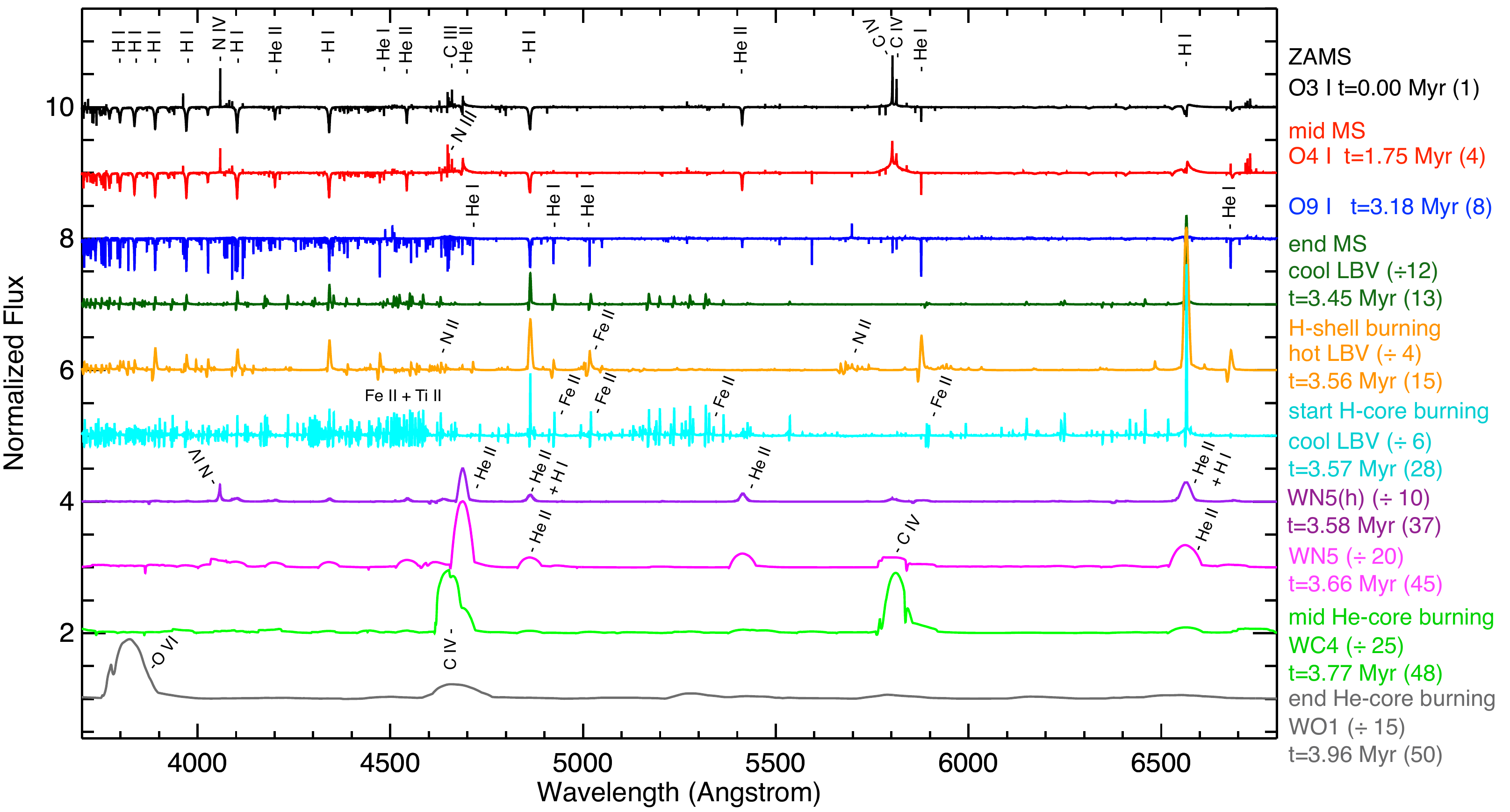}}\\
\caption{\label{spec1} Evolution of the ultraviolet (a) and optical spectra (b) of a non-rotating 60~\msun\ star. The evolution proceeds from top to bottom, with labels indicating the evolutionary phase, spectral type, scale factor when appropriate, age, and model stage according to Table \ref{model log}. Note that certain spectra have been scaled for the sake of displaying the full range of UV and optical emission lines.}
\end{figure*}

In this Section we investigate how a star with \mini=60~\msun\ evolves from a spectroscopic point of view and how the surface properties are linked with the evolution of the stellar interior. There is a long record in the literature about the interior evolution of non-rotating massive stars with $\mini\sim60~\msun$, and the interior evolution described here is qualitatively similar to that described in, e.g., \citet{chiosi86}. A number of recent papers discuss this subject, including the latest advances in modeling of the stellar interior, and we refer the reader to \citet{maeder11}, \citet{brott11}, \citet{ekstrom12}, \citet{georgy12a}, \citet{langer12}, and \citet{chieffi13}.

Figure \ref{hrd1}a shows the evolutionary track of a 60~\msun\ star in the Hertzsprung-Russell (HR) diagram. Spectra were computed at 53 stages, which are indicated in Fig. \ref{hrd1}a together with the corresponding spectral type. Table \ref{model log} presents the fundamental parameters of the models at each stage. Figure \ref{hrd1}b illustrates the effects of the stellar wind on the determination of \teff. Because of the wind optical depth, $\teff <\teffg$, and the difference between the two depends on $\mdot$, \vinf, \lstar, and \rstar. These values of \teff\ can be directly compared to those determined from observations of massive stars, as long as the observed value of \teff\ refers to the surface where \tauross=2/3.  

Figure \ref{hrd1} also shows the evolution of the surface temperature as a function of age (panel $c$), the surface abundances  in mass fraction\footnote{Abundances are quoted in mass fraction in this paper.} ($d$), mass-loss rate and mass ($e$), and central abundances in mass fraction ($f$). Selected stages and corresponding spectral types are indicated, focusing on stages of either relative long duration or related to significant stages of the interior evolution, such as the different nuclear burning phases. 
 
Figure \ref{spec1} displays sections of the ultraviolet and optical spectrum at selected stages, illustrating the broad evolution. Regarding the evolution of the optical spectrum, we find that the H and He absorption lines that dominate the O-type spectra (black, red, and blue tracing in Fig. \ref{spec1}b) progressively turn into emission as the star evolves and the stellar wind becomes denser (green, orange, and cyan). The H emission lines disappear at the WNE phase because of the H exhaustion at the surface (purple), leaving \ion{He}{ii} and \ion{N}{iv} lines as the dominant features in the spectrum. When the star evolves to the WC phase (green), N lines vanish and C lines appear as a result of the He burning products (C and O) being exposed at the surface. At the end of its life, the star shows a WO spectral type (dark gray), with strong \ion{O}{vi} lines dominating the spectrum as a result of the extremely high surface temperatures due to the surface contraction.

We find the following evolutionary sequence for a non-rotating 60~\msun\ star, where we quote the main spectroscopic phases only and their respective evolutionary phase in parenthesis:\\
\\
{\it O3 I (ZAMS) $\rightarrow$ O4 I (mid H-core burning) $\rightarrow$ hot LBV (end H-core burning) $\rightarrow$ cool LBV (start He-core burning) $\rightarrow$ WNL $\rightarrow$ WNE $\rightarrow$ WC (mid He-core burning) $\rightarrow$ WO (end He-core burning until core collapse). }
 \\
 
In the next subsections, we analyze in detail the different evolutionary stages and the spectroscopic phases associated to them.

\subsection{\label{msdetail} H-core burning: the Main Sequence evolution (stages 1--15)}

The MS is characterized by H-core burning, which in the case of the non-rotating 60~\msun\ star takes 3.53 Myr to finish \citep{ekstrom12}. As we elaborate below, we find that when this model is in the MS evolutionary phase, the following spectroscopic phases appear: O I, BSG, BHG, and LBV (Fig. \ref{msfig}g,h).

\subsubsection{\label{oi} An O3 I during 40\% of the MS phase (stages 1--3)}
Our models indicate that the star appears at the Zero Age Main Sequence (ZAMS) as an O3 If$^*$c star ($\teff=48390~\K$, $\lstar=504582~\lsun$, stage 1, Fig. \ref{spec1}), and not as a luminosity class V star. This is because the luminosity criterium for early O stars is based on the equivalent width of the \ion{He}{ii} $\lambda$4687 line \citep{contialschuler71,walborn71}, which in our model is filled by wind emission even at the ZAMS for a 60~\msun\ star. Therefore, the star has a supergiant appearance even at the ZAMS. It is also worth noting that the star has strong \ion{N}{iv} \lam4058 and \ion{C}{iii} \lam\lam\lam4658,4651,4653 emissions, which justifies the O3 If$^*$c classification. The supergiant appearance is caused by the high value of \lstar\ since, at this regime, $\mdot \propto \lstar^2$ \citep[][see also e.\,g. \citealt{dejager88,vink00}]{abbott82}. The strong wind affects the spectrum, with lines turning progressively from absorption into emission. Here we show that this effect starts to occur for a 60~\msun\ star when using the \citet{vink01} mass-loss rate prescription and $f_\infty=0.2$. This result is also in agreement with the observational findings from \citet{crowther10} that very massive stars (above 100~\msun) have an O I or even WNh spectral type at the ZAMS at the LMC metallicity (see also \citealt{dekoter97,martins08}). In our models, \ion{He}{ii} $\lambda$4687 would be partially filled by wind emission even assuming an unclumped wind, and one would infer a luminosity class III at the ZAMS (see discussion in Sect. \ref{mdotclump}.)

\begin{figure*}
\center
\resizebox{0.32\hsize}{!}{\includegraphics{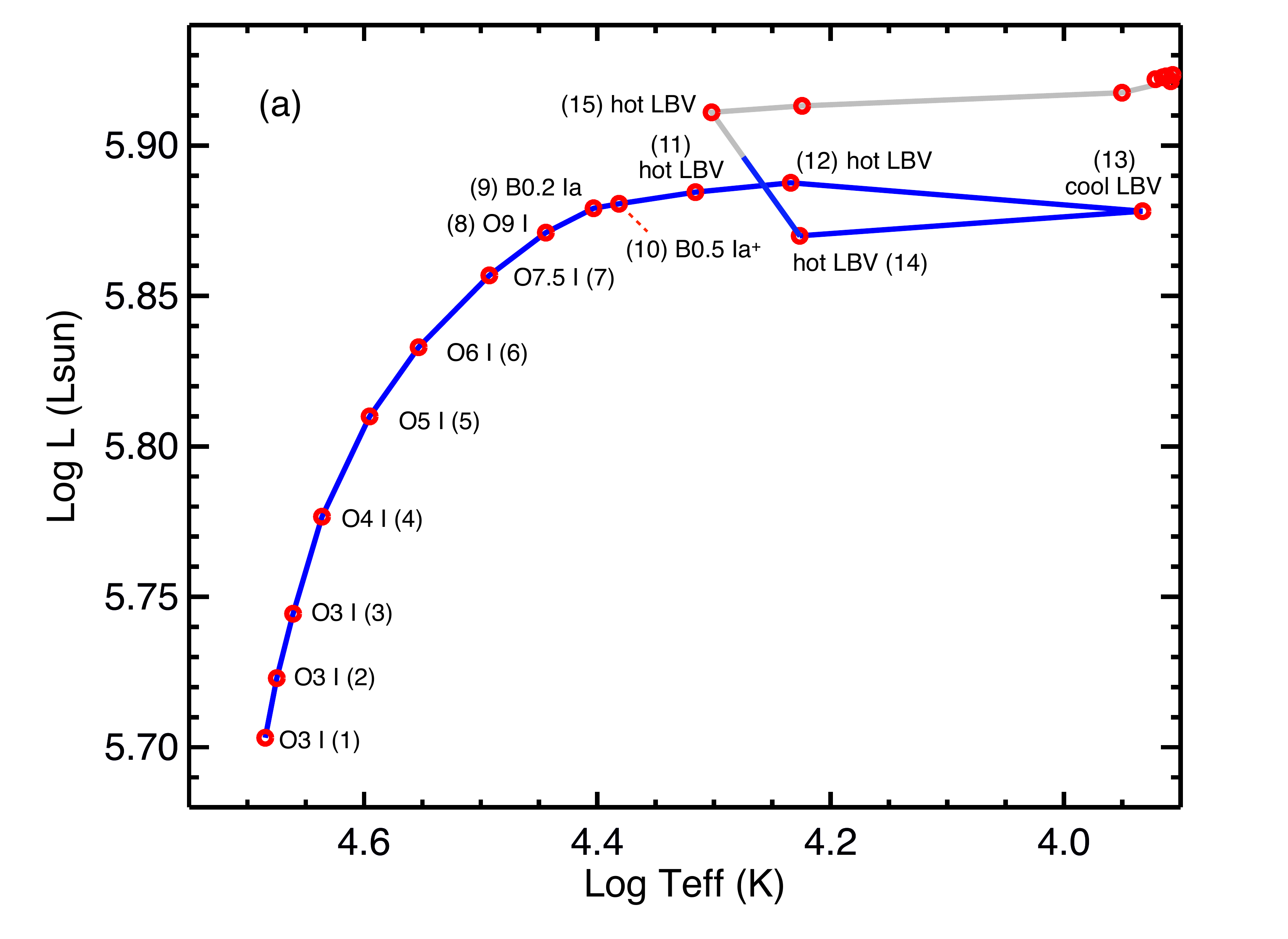}} 
\resizebox{0.32\hsize}{!}{\includegraphics{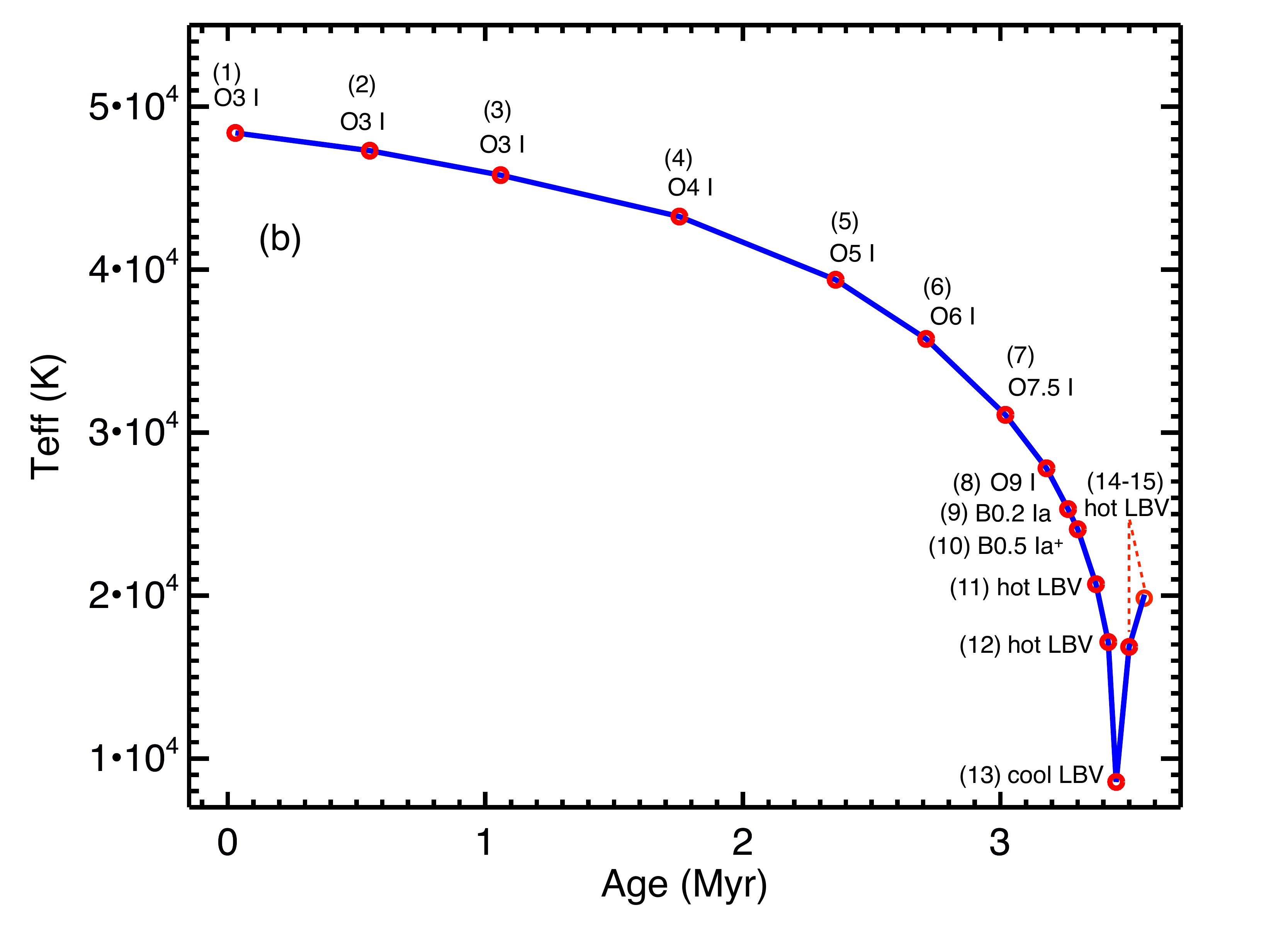}}
\resizebox{0.32\hsize}{!}{\includegraphics{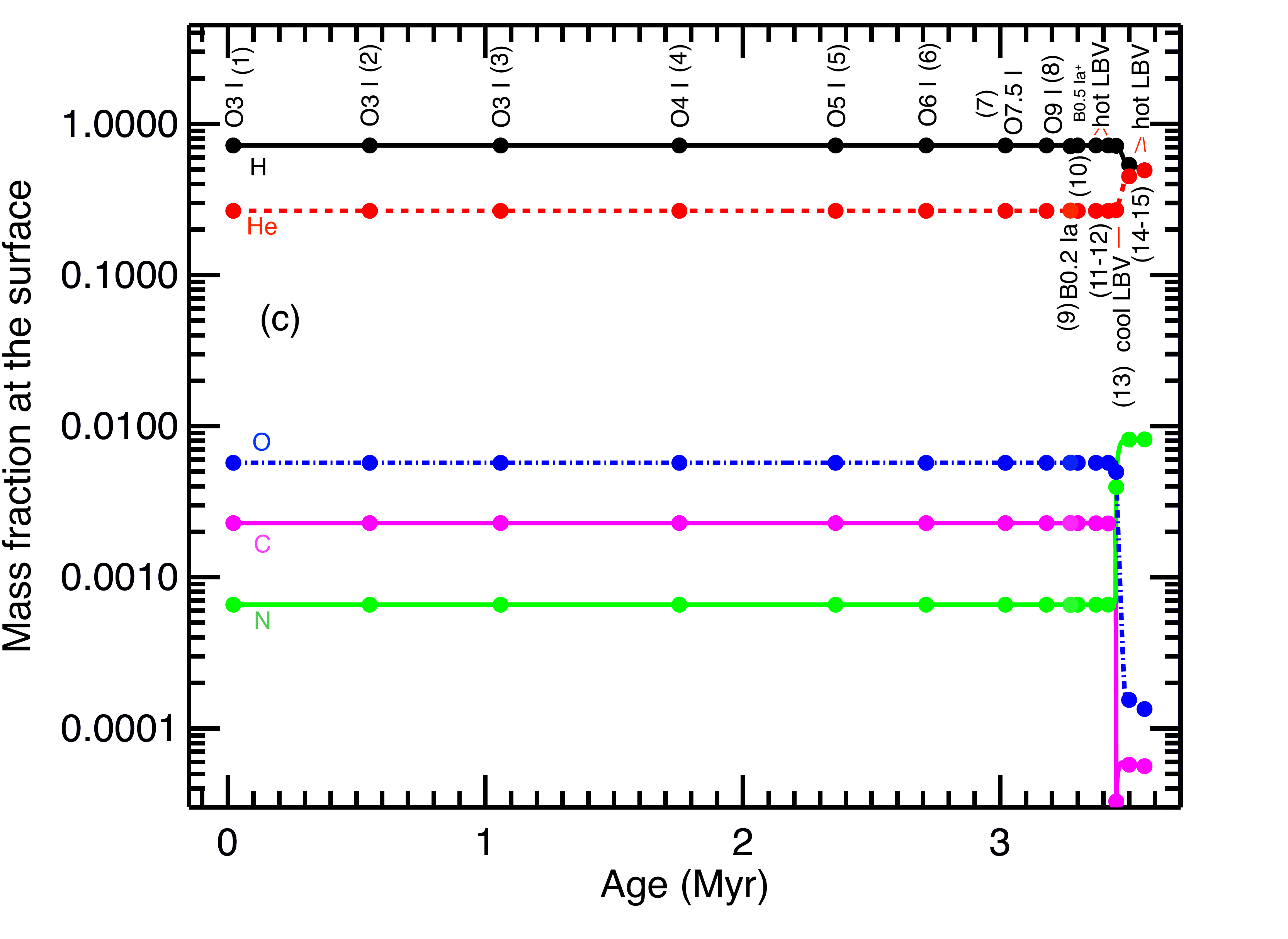}}\\
\resizebox{0.32\hsize}{!}{\includegraphics{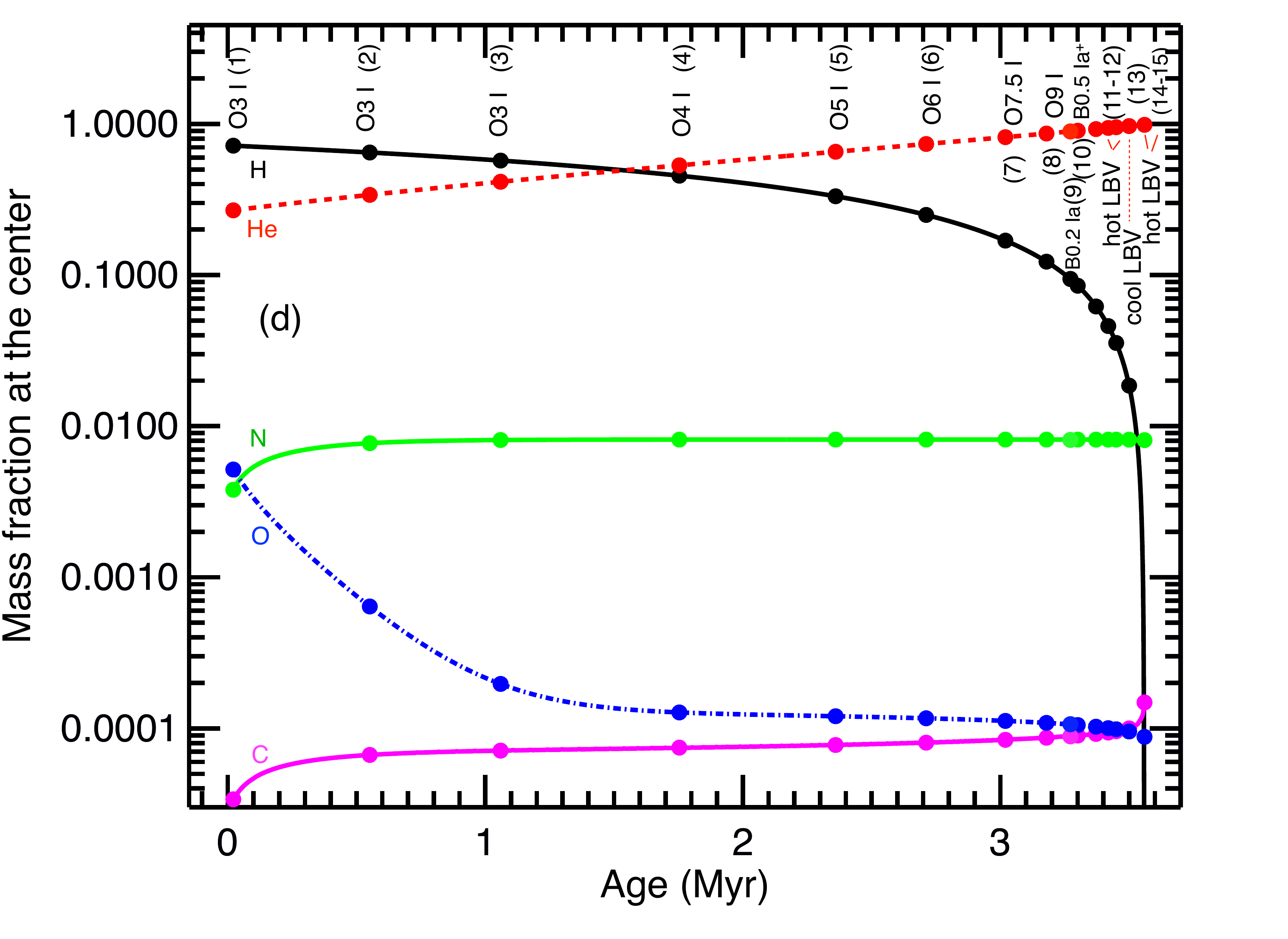}}
\resizebox{0.32\hsize}{!}{\includegraphics{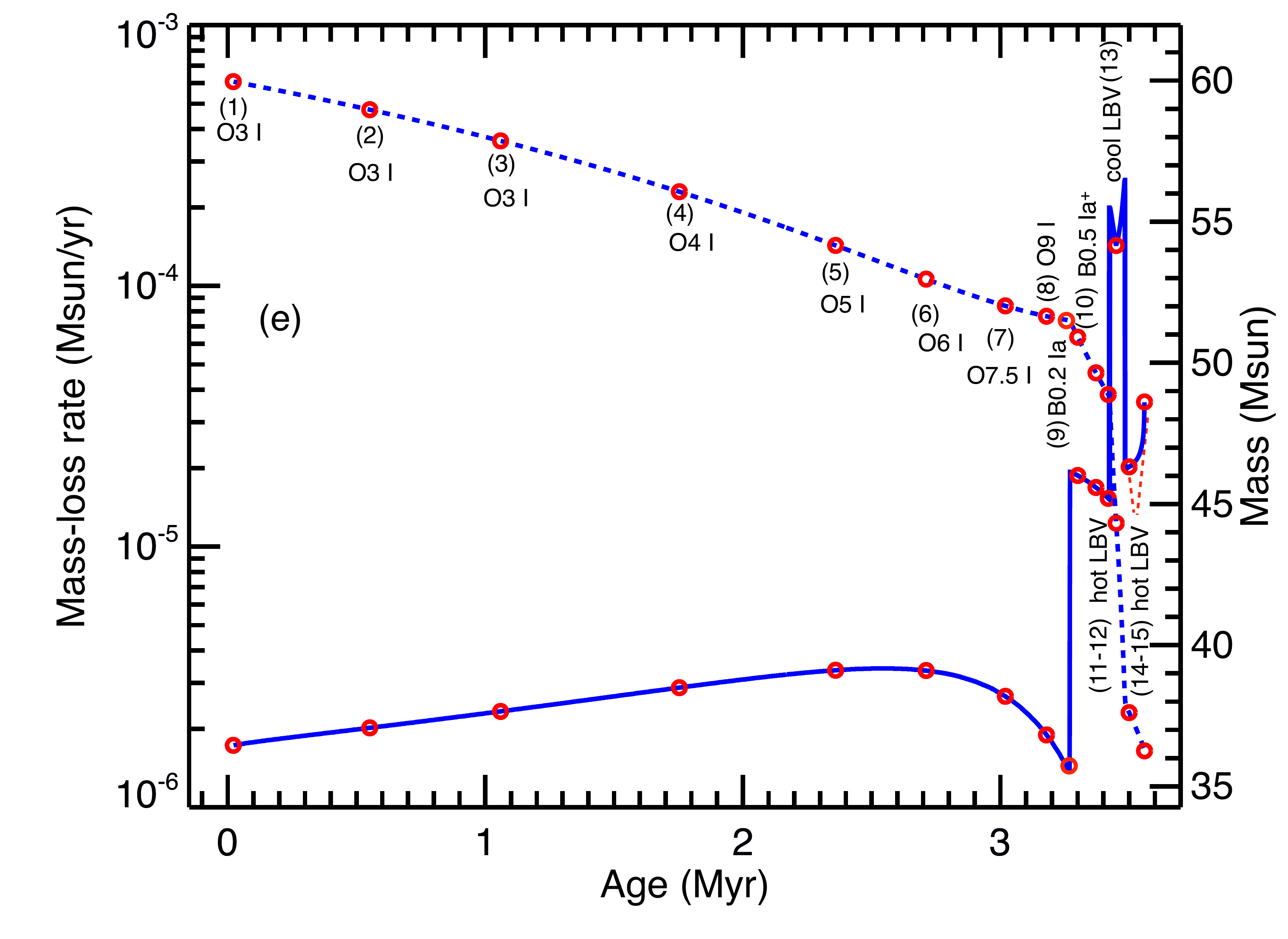}}
\resizebox{0.32\hsize}{!}{\includegraphics{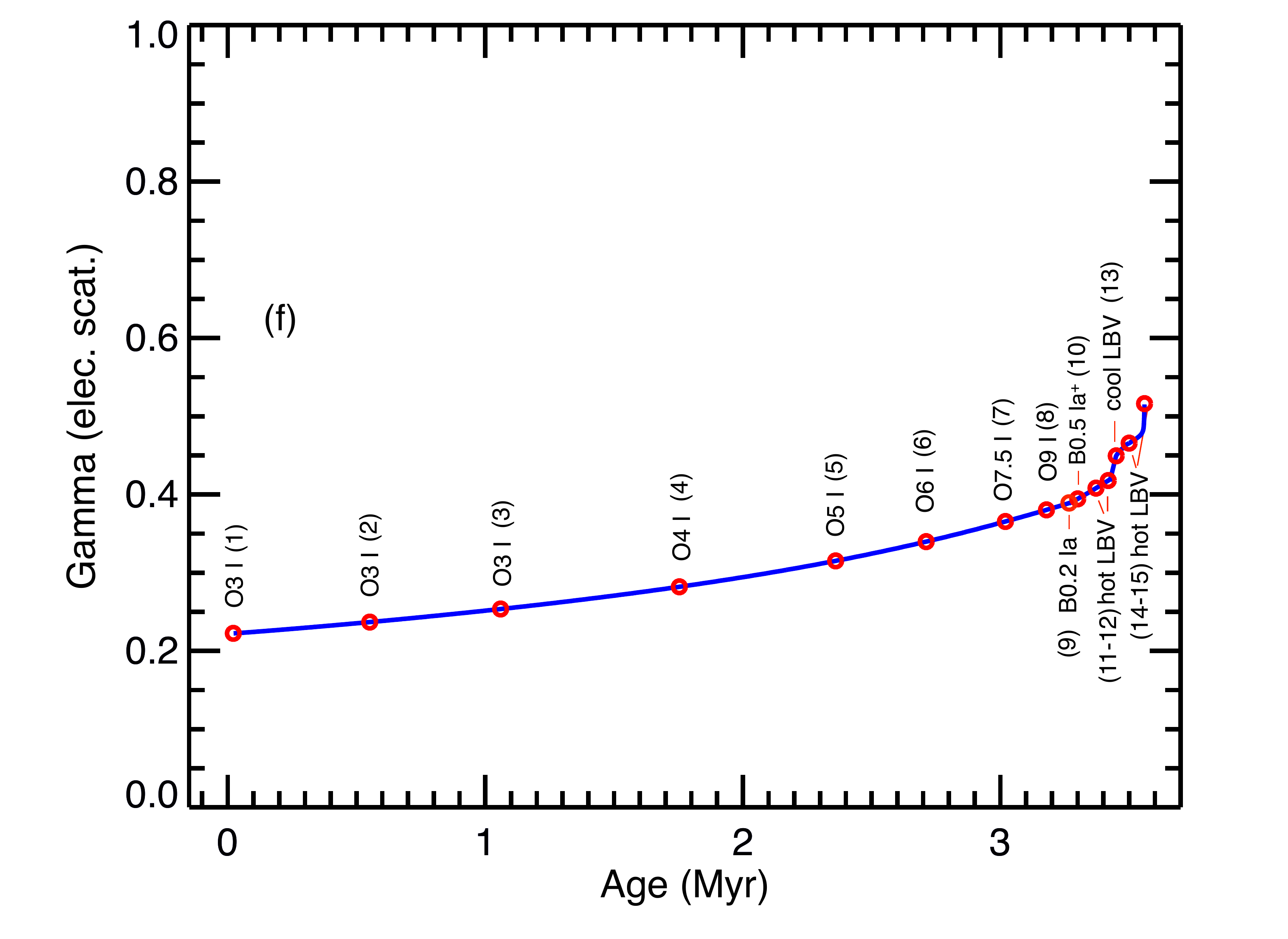}}\\
\resizebox{0.90\hsize}{!}{\includegraphics{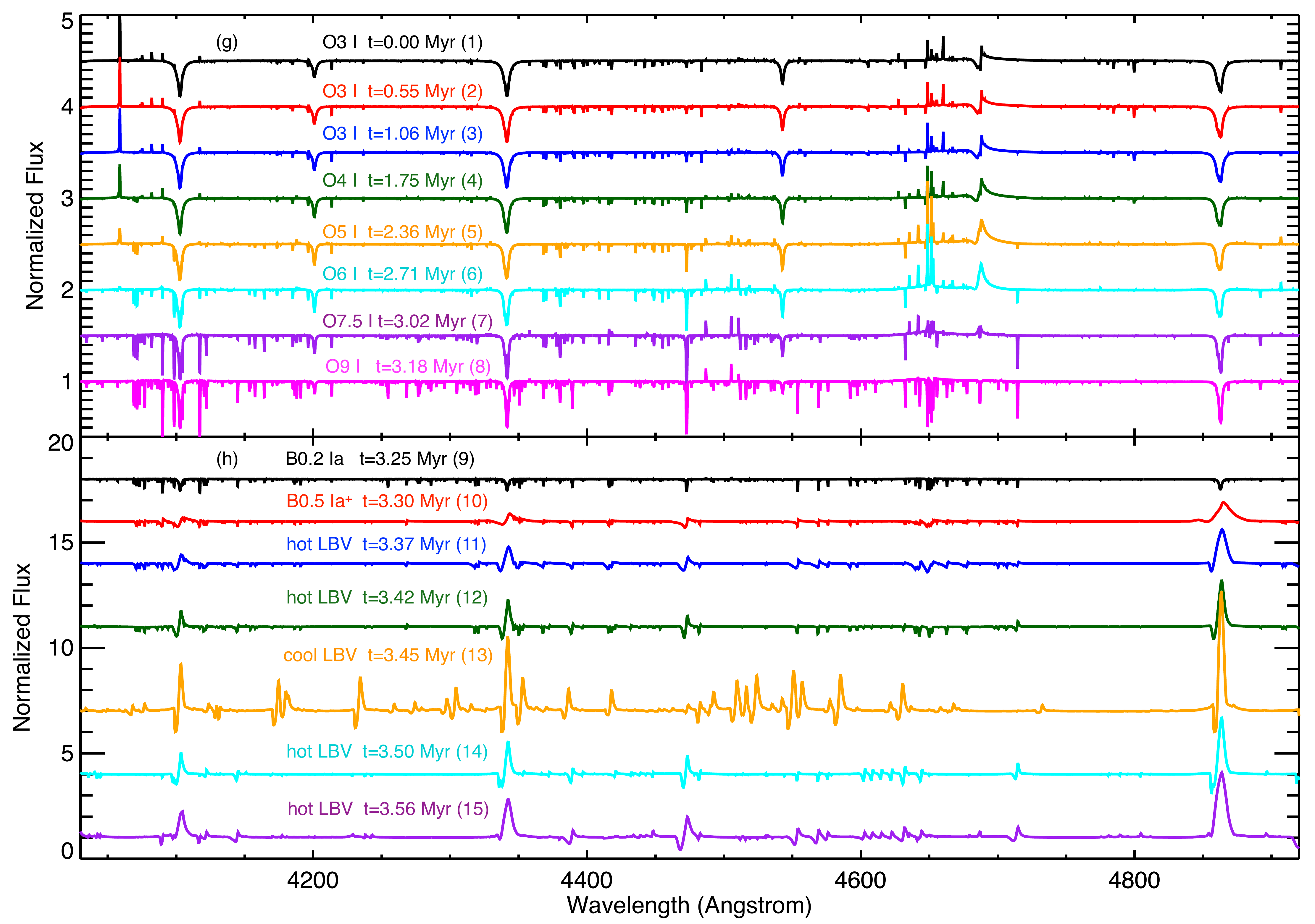}}
\caption{\label{msfig} {\it (a)} through {\it (e)}: Similar to Fig. \ref{hrd1}, but focusing on the MS evolution. {\it (f)}: Evolution of the Eddington parameter at the surface, computed for electron-scattering opacity. {\it (g)} and {\it (h)}: Similar to Fig. \ref{spec1}, but focusing on the MS evolution.}
\end{figure*}

During the MS (stages 1 through 15 in Fig. \ref{msfig}a), the convective core contracts, the stellar envelope expands, and \rstar\ increases. There is also an increase in $\lstar$ (Fig. \ref{msfig}a), which is caused by the increase in mean molecular weight as H burns into He. As a combination of these two effects, \teff\ decreases (Fig. \ref{msfig}b) and the spectral type shifts to later types (Fig. \ref{msfig}g). Because of the relatively high \lstar\ and \mdot\, the \ion{He}{ii} $\lambda$4687 line remains contaminated by wind emission, meaning that the star appears as a supergiant with luminosity class I. During the MS, the \mdot\ behavior is regulated by its dependence on $\lstar$ and $\teff$ (see \citealt{vink00}). Up to stage 6, $\mdot$ increases because of the increase in $\lstar$ and the shallow dependence of $\mdot$ on $\teff$ in the range 35000--50000~\K. 
Between stages 6 and 9 and until the star reaches the bistability limit (Fig. \ref{msfig}e), $\mdot$ decreases because $\lstar$ is roughly constant and $\teff$ decreases. In addition, since $\vinf$ is thought to depend on $\vesc$ \citep{cak}, $\vinf$ decreases as a result of the decrease in $\vesc$ as \rstar\ increases. Mass loss and the increase in \lstar\ during the MS causes the Eddington parameter to increase \citep[Fig. \ref{msfig}f;][]{vink99,vink11}, which also makes $\vinf$ smaller.

\begin{figure*}
\center
\resizebox{0.32\hsize}{!}{\includegraphics{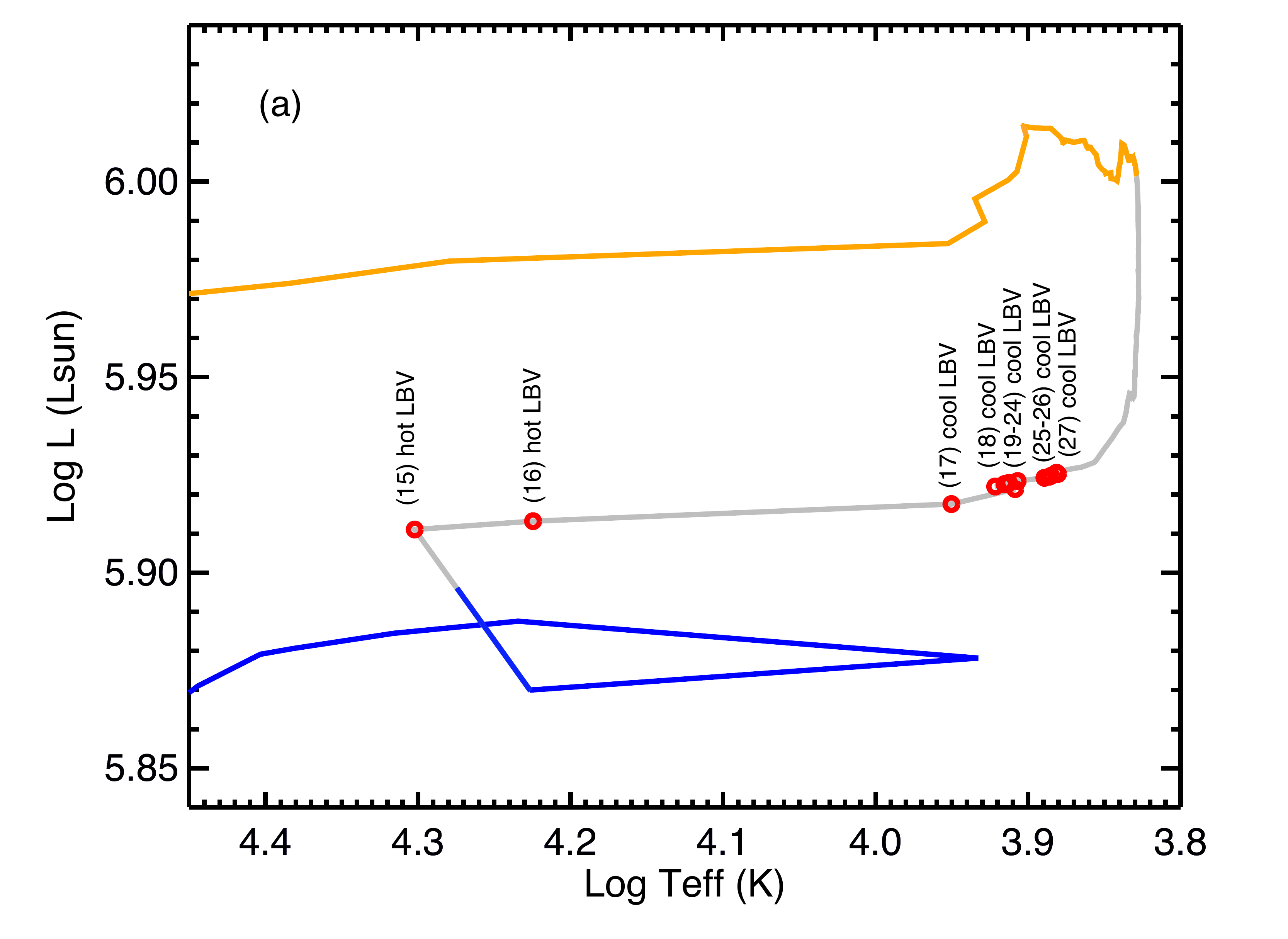}}
\resizebox{0.32\hsize}{!}{\includegraphics{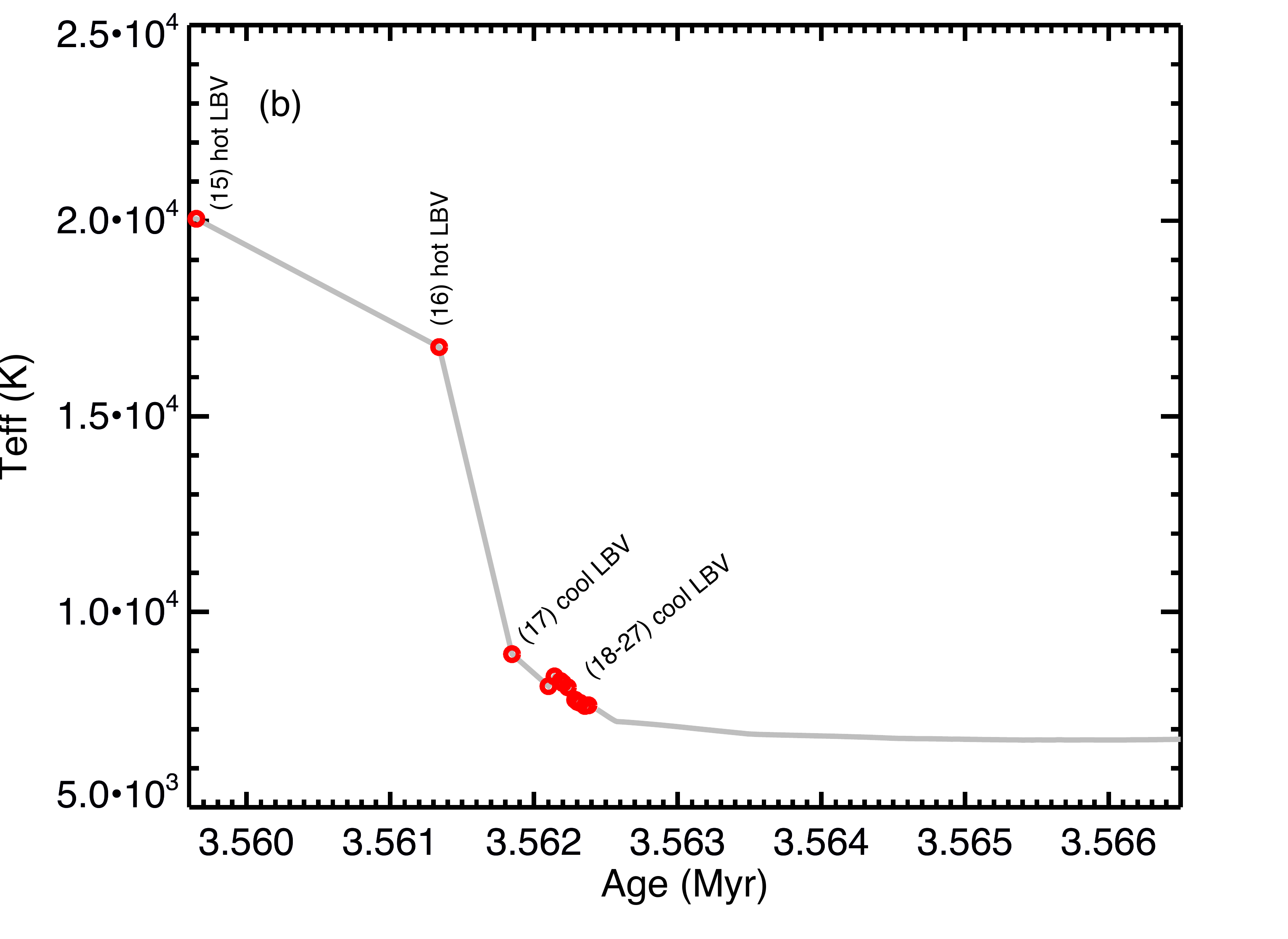}}
\resizebox{0.32\hsize}{!}{\includegraphics{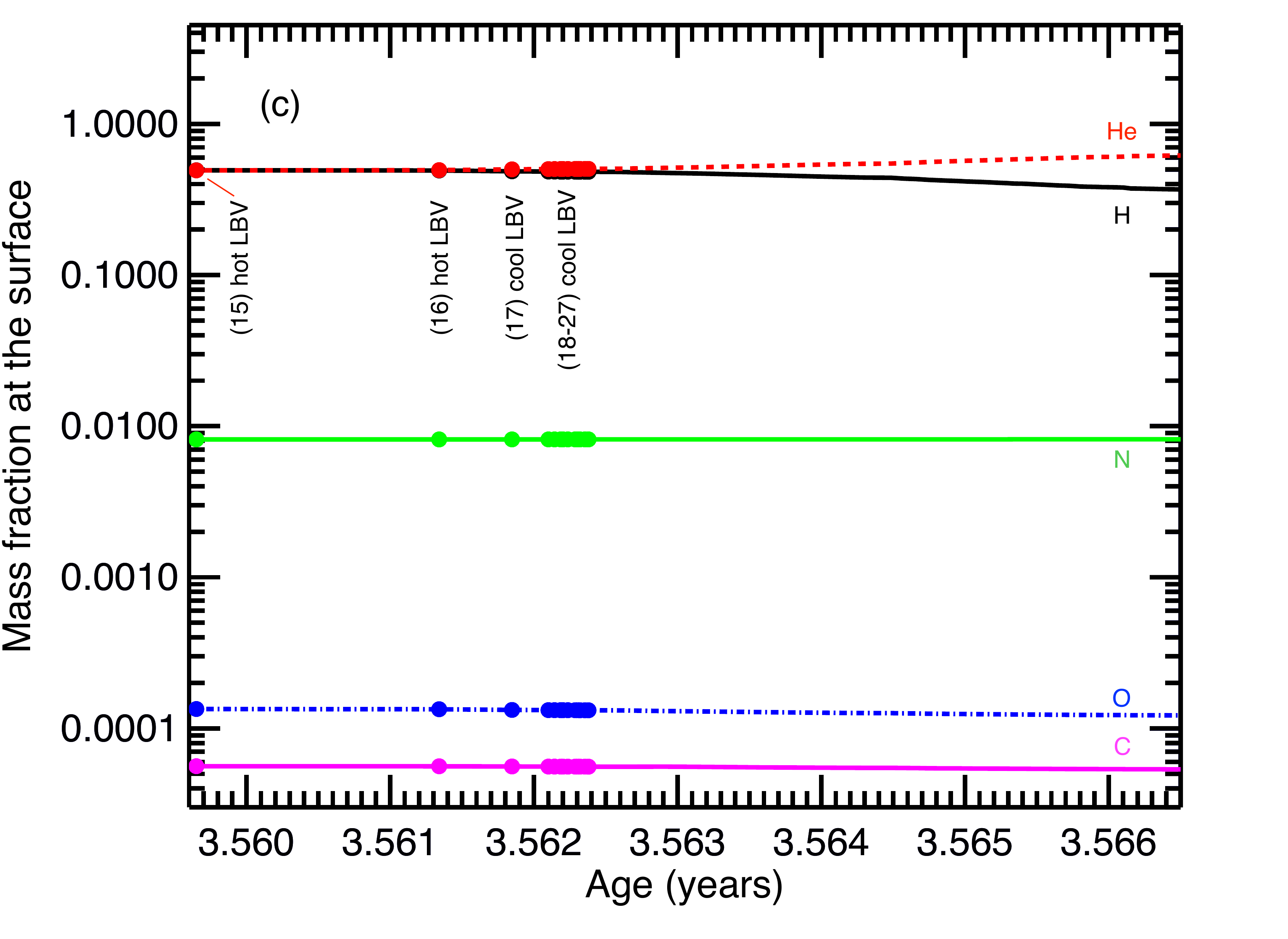}}\\
\resizebox{0.32\hsize}{!}{\includegraphics{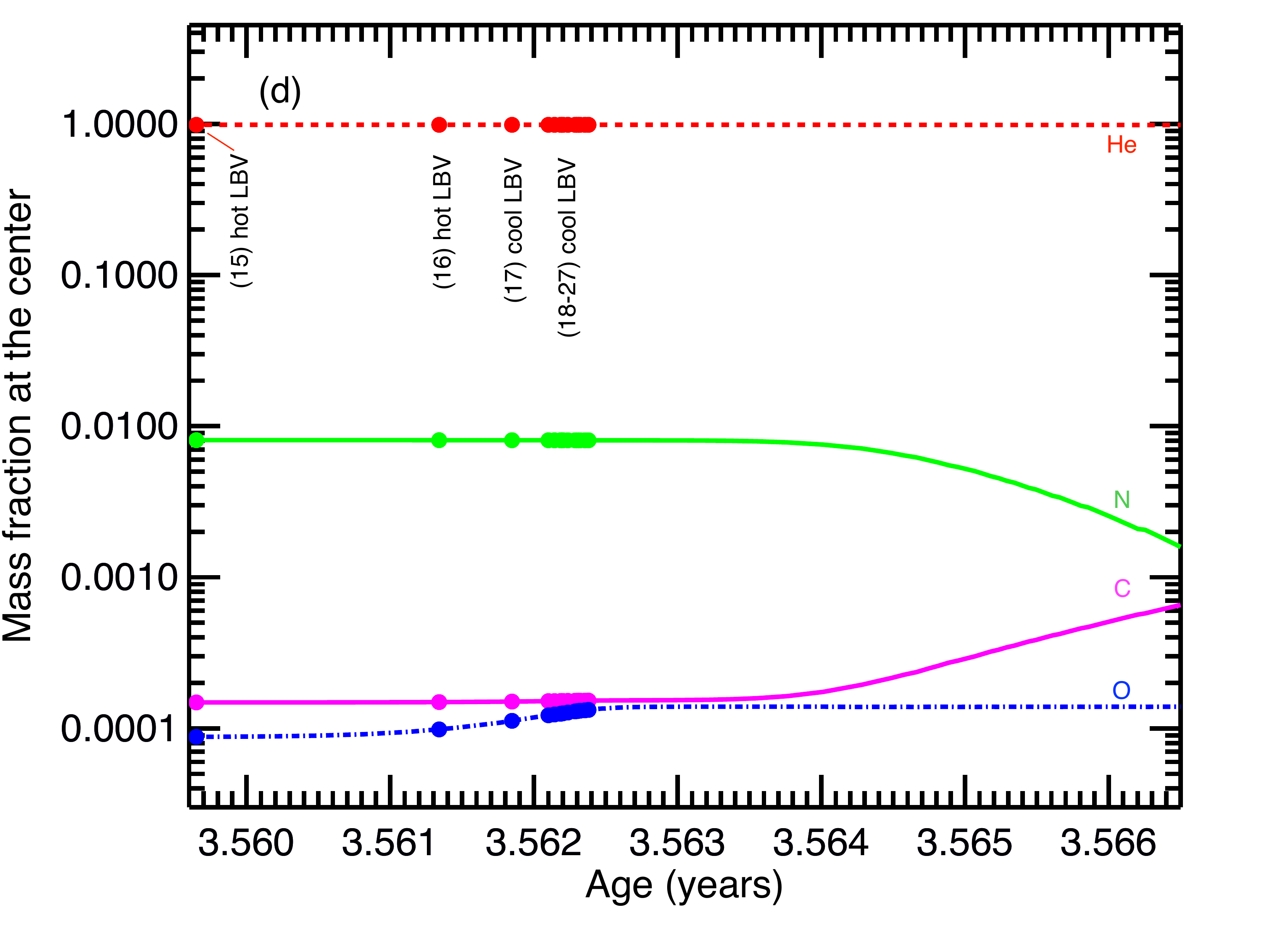}}
\resizebox{0.32\hsize}{!}{\includegraphics{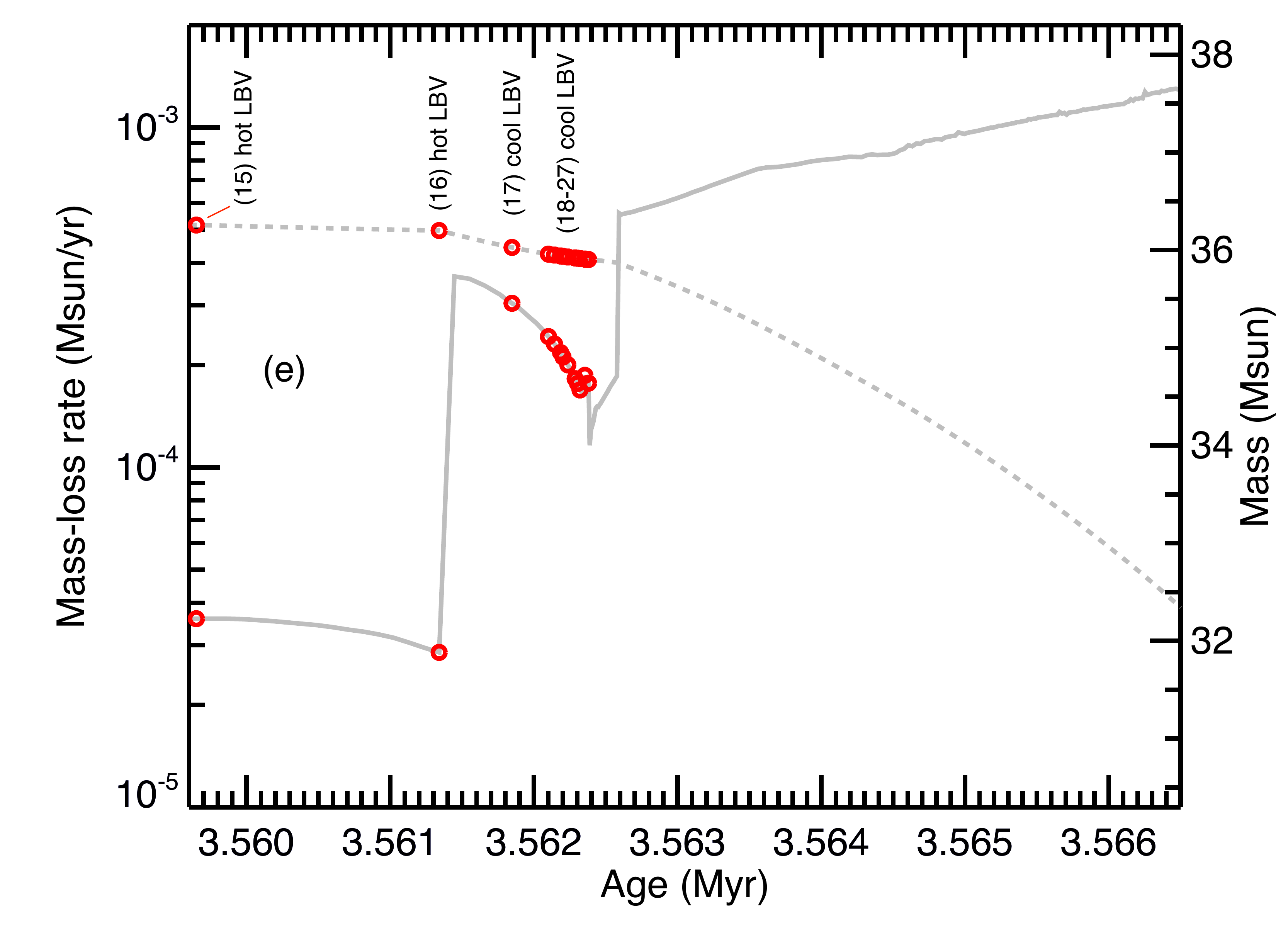}}
\resizebox{0.32\hsize}{!}{\includegraphics{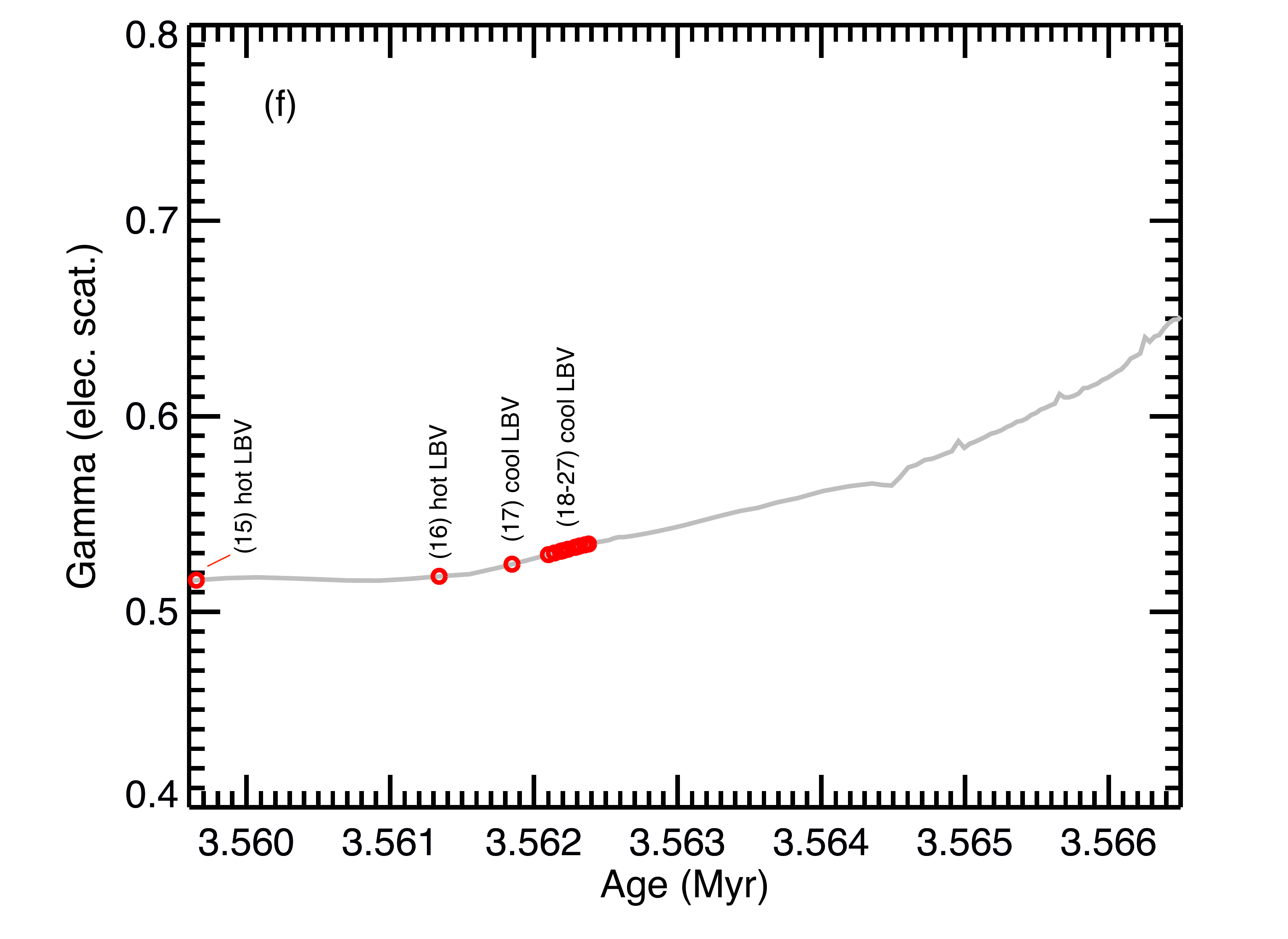}}\\
\resizebox{0.99\hsize}{!}{\includegraphics{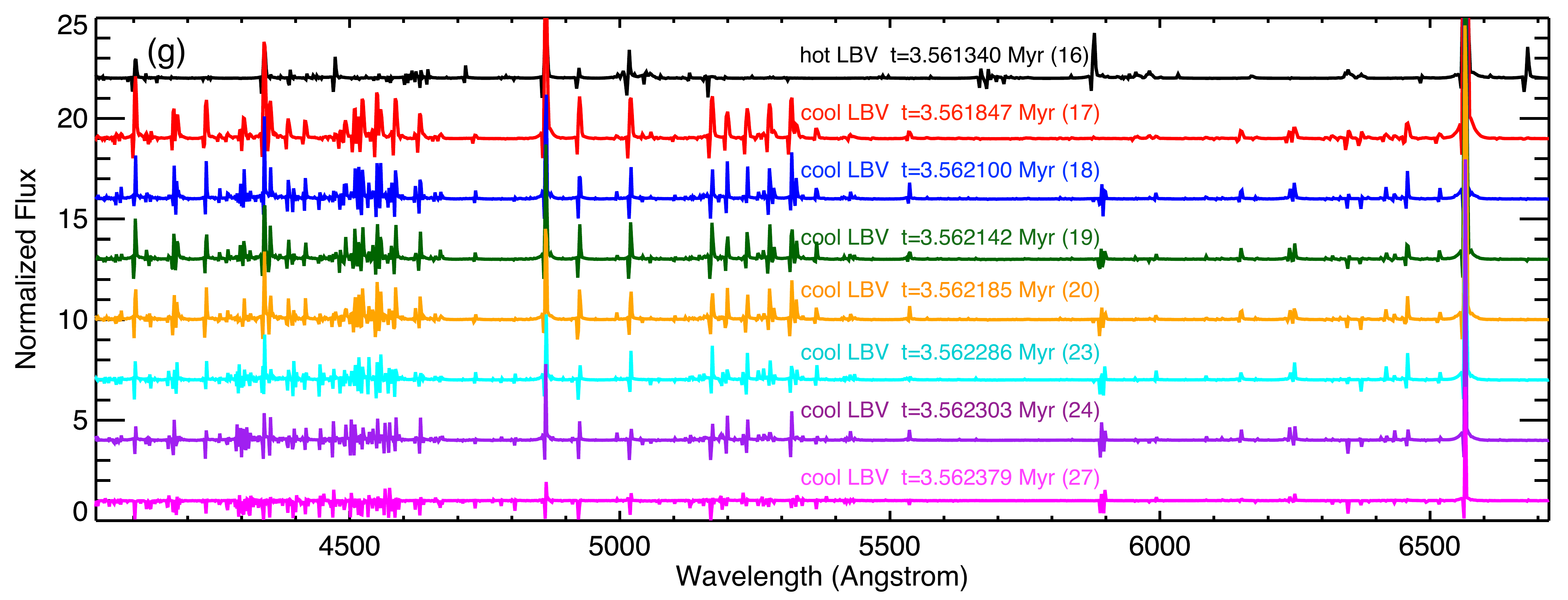}}
\caption{\label{hshellfig} Similar to Fig. \ref{msfig}, but focusing on the evolution during H shell burning. }
\end{figure*}

We obtain that the increase in $\lstar$ and decrease in \teff\ is modest for the first 1.44 Myr (stages 1 through 3), and the small changes are not enough to change its spectral type. The star remains with an O3 If$^*$c spectral type for about $40\%$ of the MS lifetime (36\% of the total lifetime). However, important changes occur in the stellar interior, with an increase in He abundance, at the expense of the decrease of the H content, as H burns (Fig. \ref{msfig}d).  

\subsubsection{From O4 I to B0.2 I during the next 50\% of the MS (stages 4--7)}
Up to the middle of the MS (age=1.8 Myr, stage 4),  the spectroscopic appearance of a 60~\msun\ star does not change significantly. We determine an O4 If$^*$c spectral type, which lasts  for 0.66 Myr (Fig. \ref{msfig}g).  It is interesting to notice that while the surface abundances and temperature have changed little (Fig. \ref{msfig}b,c) , the interior properties have been greatly modified. The H abundance at the center (X$_c$) has decreased from 0.72 to 0.45, while the He abundance at the center (Y$_c$) has increased from 0.26 to 0.53 (Fig. \ref{msfig}d).
 
As the star evolves on the MS, it becomes even more luminous and cooler, showing an O5 Ifc (for 0.47 Myr) and O6 Iafc (for 0.32 Myr) spectral types (Fig. \ref{msfig}g). The total duration of the phase when the star shows an early O I spectral type (O3 I to O6 I) is thus 2.89 Myr (stages 1 through 6), when the star loses 7.6~\msun\ due to its stellar wind (Fig. \ref{msfig}e).

Eventually, the increased $\lstar$ and $\mdot$ produces an O7.5 Iafc spectral type (stage 7), with \ion{He}{ii} $\lambda$4687 still significantly filled by wind emission (Fig. \ref{msfig}g). Note that the $fc$ morphological classification lasts for a significant amount of time during the O I phase, since the surface C abundance still remains close to the solar values. The star continues to increase in $\lstar$ (Fig. \ref{msfig}a), with $\mdot$ slightly decreasing as $\teff$ decreases (Fig. \ref{msfig}e), shifting to later spectral types (Fig. \ref{msfig}g). For instance, it shows an O9~Iab spectral type when its age is t=3.18 Myr (stage 8). Therefore, the non-rotating 60~\msun\ star is a late O supergiant star (luminosity class I) for 0.36 Myr, with the star losing 0.9~\msun\ during this period (stages 6--8; Fig. \ref{msfig}e). At the end of the late O I spectroscopic phase (stage 8), $\lstar$ and \mdot\ have increased by 50\% and 10\% relatively to the initial values, respectively (Fig. \ref{msfig}a,e), while $\vinf$ has decreased to 45\% of the ZAMS value. The surface composition remains essentially unchanged compared to the initial state (Fig. \ref{msfig}c), while at the center, H continues to decrease (X$_c$=0.12) and He to increase (Y$_c$=0.86) (Fig. \ref{msfig}d). Later, the decrease in \teff\ produces a B0.2 Ia spectral type (stage 9). We estimate that the B supergiant phase is short, lasting for about 0.039 Myr and during which the star loses 0.09~\msun. 

\subsubsection{The encounter of the bistability limit: producing BHG and LBV spectra in the last 10\% of the MS phase (stages 10--15)}
\label{bistability}

Due to the core contraction and corresponding envelope expansion, at some point the surface of the star is cool enough to encounter the bistability limit of line-driven winds.  Note that reaching the bistability limit during the MS is dependent on the choice of the overshooting parameter, mass-loss recipe, and initial rotation, since all these factors affect the width of the MS band (see e\,g. the reviews by \citealt{chiosi86} and \citealt{maeder_araa00}).

This is a crucial point in the evolution of a massive star because \mdot\ increases significantly according to the \citet{vink00} prescription. The first bistability jump occurs when $\teff\sim21000-25000~\K$ \citep{pauldrach90,vink99,vink00,vink02,ghd09,ghd11}, when Fe recombines from Fe$^{3+}$ to Fe$^{2+}$ in the inner wind, resulting in increased $\mdot$. According to the \citet{vink99,vink00} parametrization incorporated in our stellar evolution models, $\mdot$ increases by a factor of $\sim10$ as the star crosses from the hot (stage 9) to the cool side (stage 10) of the bistability. The factor of $\sim10$ arises because the Geneva models follow the \citet{vink00} recommendation to change the ratio $\vinf/\vesc$ from 2.6 (hot side) to 1.3 (cool side) of the bistability limit \citep{vink99,lamers95}. For the non-rotating 60~\msun\ star, this occurs at t=3.27 Myr, when the star becomes a BHG with a B0.2--0.5 Ia$^+$ spectral type (Fig. \ref{msfig}g). We estimate that the star remains as a BHG for 0.079 Myr and loses 1.3~\msun\ during this period.

For the non-rotating 60~\msun\ model, we find that when the star crosses the bistability limit, the wind optical depth increases, the photosphere becomes extended and is formed in an expanding layer. Most of the spectral lines are affected by the presence of the wind, with some of them developing a classical P-Cygni profile. Because of the high \mdot\ and low $\vinf$, coupled with the relatively low $\teff$, the spectrum markedly resembles that of a hot LBV (stages 11 and 12). This suggests that some stars with spectra similar to LBVs may still be in the end stages of H-core burning, MS objects. When the star first shows an LBV spectrum, the H content at the center is X=0.06 (Fig. \ref{msfig}d). 

Further on, the star crosses the second bistability jump,  when Fe$^{2+}$ recombines to Fe$^+$ in the inner wind and another increase by factor of $~\sim10$ occurs (\citealt{vink01}; stage 12, Fig. \ref{msfig}e). At this point, the star has reached its coolest point in the MS at t=3.41 Myr (stage 13), showing a cool LBV spectral type (Fig. \ref{msfig}g) and having $\mdot=1.4\times10^{-4}~\msunyr$ and $\vinf=303~\kms$.

Up to this point, the surface abundances are essentially unchanged compared to the initial values on the ZAMS (Fig. \ref{msfig}c), with $X_\mathrm{sur}=0.72$, $Y_\mathrm{sur}=0.27$, $C_\mathrm{sur}=2.3\times10^{-3}$, $N_\mathrm{sur}=6.6\times10^{-4}$, and $O_\mathrm{sur}=5.7\times10^{-3}$ (Fig. \ref{msfig}c). However, because of the increase in $\mdot$ that occurs when the star crosses to the cool side of the bistability limit, \mstar\ rapidly decreases and the surface abundances begin to change from this point in the evolution onwards.  First, N is rapidly enriched at the surface while C is depleted, so the star will appear as an LBV with solar N and C abundance only for a very short amount of time. For instance, at stage 13, when the star has crossed the second bistability jump, severe modifications have already occurred, with  $C_\mathrm{sur}=3.3\times10^{-5}$, $N_\mathrm{sur}=4.0\times10^{-3}$, and $O_\mathrm{sur}=5.0\times10^{-3}$. The H and He surface abundances take a longer timescale to change, and we find H and He surface abundances similar to the initial values, with $X_\mathrm{sur}=0.72$ and  $Y_\mathrm{sur}=0.27$ (Fig. \ref{msfig}c). 

When the star returns to the blue side of the HR diagram (stage 14) after reaching its coolest point on the MS (stage 13), the surface shows the products of the CNO cycle, with an enrichment of He and N and depleted with H, C and O (Fig. \ref{msfig}c).  Because of the increase in $\mdot$ (Fig. \ref{msfig}e), the Eddington parameter  at the hydrostatic radius becomes noticeably higher than during the preceding stages of the MS, reaching $\Gamma=0.40-0.52$ (Fig. \ref{msfig}f). The star shows a hot LBV spectrum for the remaining of its MS lifetime, up to t=5.53 Myr (Fig. \ref{msfig}g), but with a progressively increase in He and decrease of H contents at the surface (Fig. \ref{msfig}c). At the end of the MS (between stages 14 and 15), the star still has a hot LBV spectrum (Fig. \ref{msfig}g) and the surface abundances shows the products of the CNO cycle, with $\xsurf=0.49$, $\ysurf=0.49$, $\csurf=5.6\times10^{-5}$, $\nsurf=8.2\times10^{-3}$, and $\osurf=1.3\times10^{-4}$ (Fig. \ref{msfig}c). The total duration of the LBV phase during the MS in our model is 0.15 Myr (see discussion in Sect. \ref{lifetimes}). During the LBV phase that occurs during the MS, the star loses 14.0~\msun\ of material though steady stellar winds (Fig. \ref{msfig}e).

Thus, the appearance of an LBV spectrum during the end of the MS is a consequence of the increase in $\mdot$ that occurs when the star crosses the bistability limit. We reinforce that the LBV phase is thus a {\it prediction} of our models and not an assumption put by hand in the modeling.

\subsection{H-shell burning (stages 15--27)}

When H burning at the center halts (stage 15), the core contracts and the envelope expands (stages 15 through 27). Since this occurs on a Kelvin-Helmholtz timescale (0.007 Myr for the non-rotating 60~\msun\ star), $\lstar$ remains roughly constant (Fig. \ref{hshellfig}a). As the core contracts, H begins to burn in a shell around the core. As a result of the envelope expansion, the star becomes cooler at the surface (Fig. \ref{hshellfig}b). The star crosses the second bistability limit, which causes further increase in $\mdot$ and decrease in $\vinf$ (Fig. \ref{hshellfig}e). As a consequence, the star remains as an LBV during H-shell burning, with the spectrum shifting from hot LBV to cool LBV as the envelope expands (Fig. \ref{hshellfig}g). The total mass lost during the H-shell burning phase is 4.0~\msun\ (Fig. \ref{hshellfig}e).

During the H-shell burning, because of the extreme \mdot, the surface abundances vary from $\xsurf=0.49$ and $\ysurf=0.49$ at the beginning to $\xsurf=0.37$ and $\ysurf=0.62$ at the end of this phase (Fig. \ref{hshellfig}c). No noticeable changes in the central abundances occur, since H is burnt in a shell (Fig. \ref{hshellfig}d).

At the end of the He-core contraction (between stages 27 and 28), immediately before the He-core burning begins, the star shows a cool LBV spectrum (Fig. \ref{spec1}). The CNO abundances have already reached the equilibrium value of the CNO cycle and remain the same as those of the end of the MS ($\csurf=5.6\times10^{-5}$, $\nsurf=8.2\times10^{-3}$, and $\osurf=1.3\times10^{-4}$; Fig. \ref{hrd1}d).

\subsection{\label{hecore}He-core burning (stages 28--51)}

As the He-core contracts, the central T increases. When the central T is high enough, He-core burning begins. At this point (stage 28), a sizable amount of H is still present on top of the core (Fig. \ref{hecorefig}c), and part of that burns in a shell around the He core. Thus, the star initiates He-core burning with a layer of H that burns in a shell, and below we discuss its consequences in the spectrum.

\subsubsection{From cool LBV to WN: losing the H envelope (stages 28--45)}

\begin{figure*}
\resizebox{0.32\hsize}{!}{\includegraphics{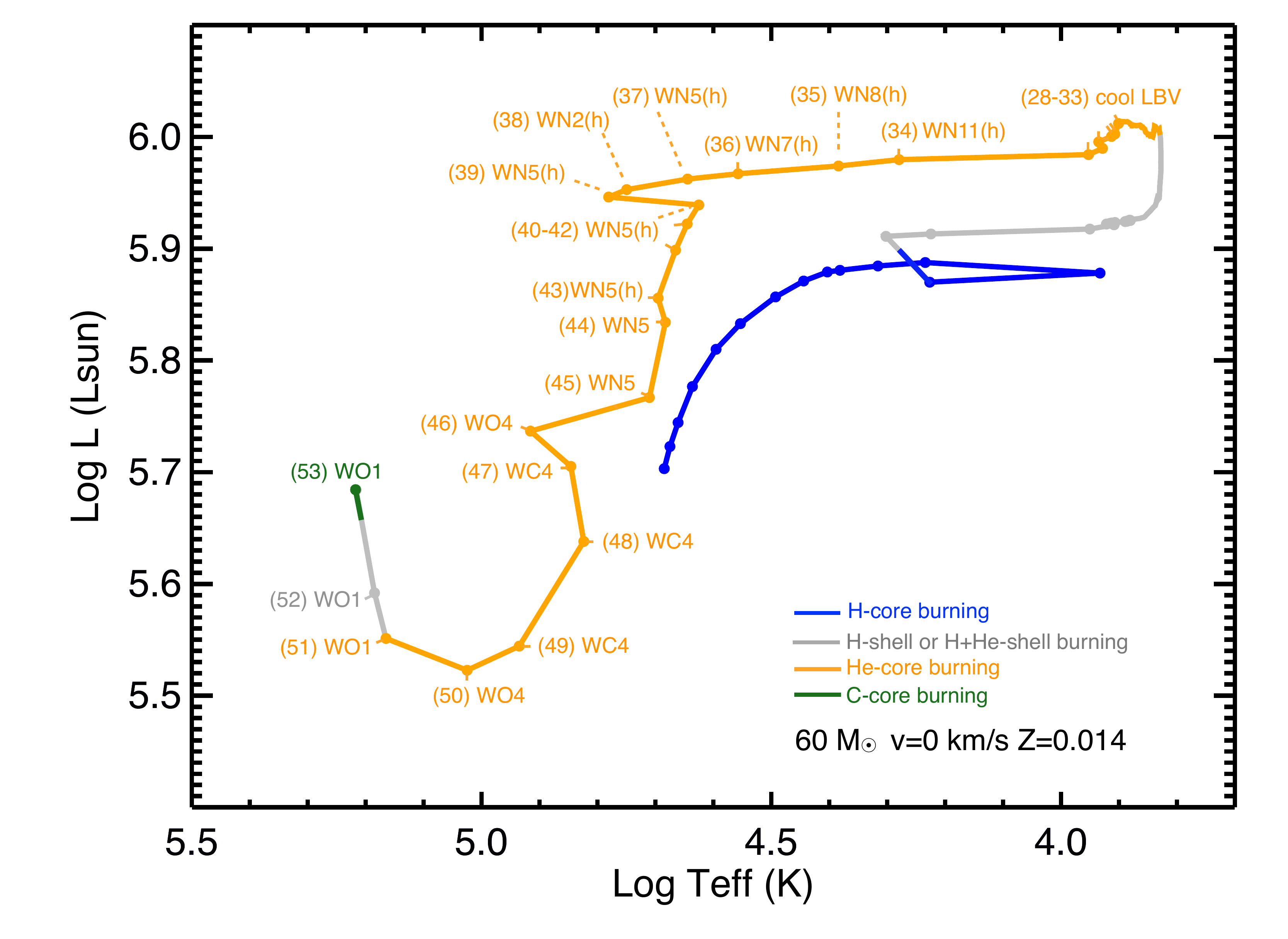}}
\resizebox{0.32\hsize}{!}{\includegraphics{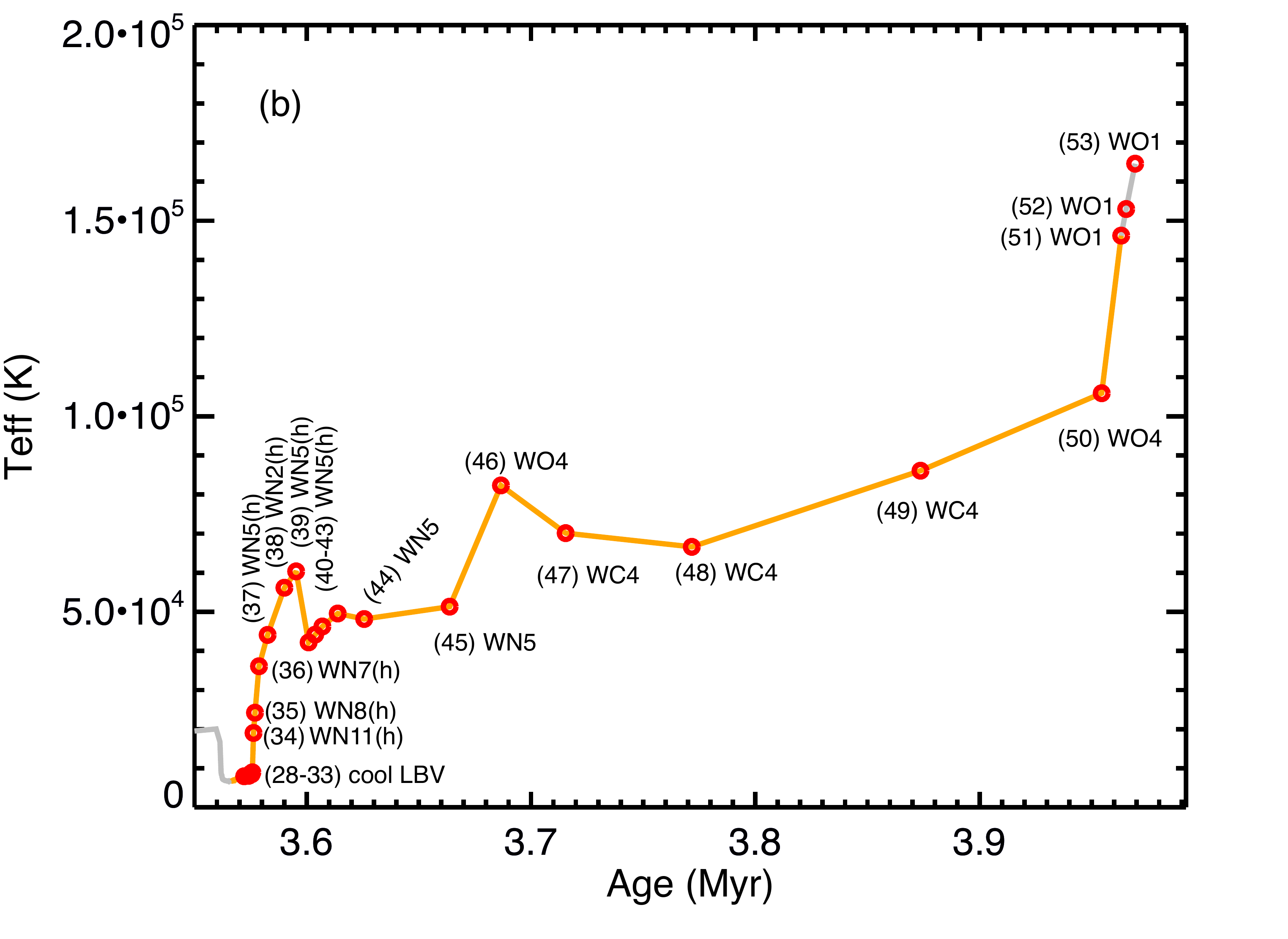}}
\resizebox{0.32\hsize}{!}{\includegraphics{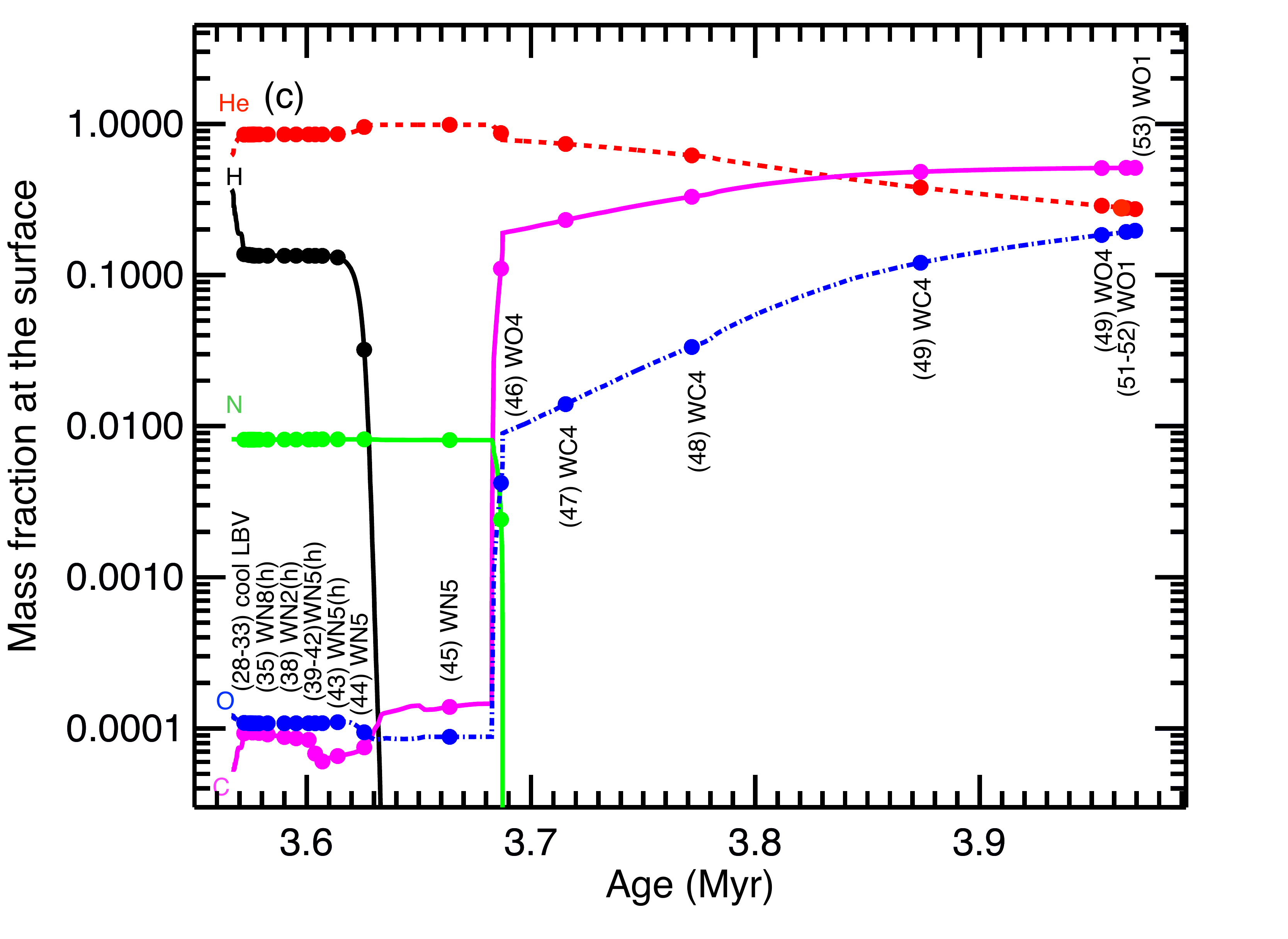}}\\
\resizebox{0.32\hsize}{!}{\includegraphics{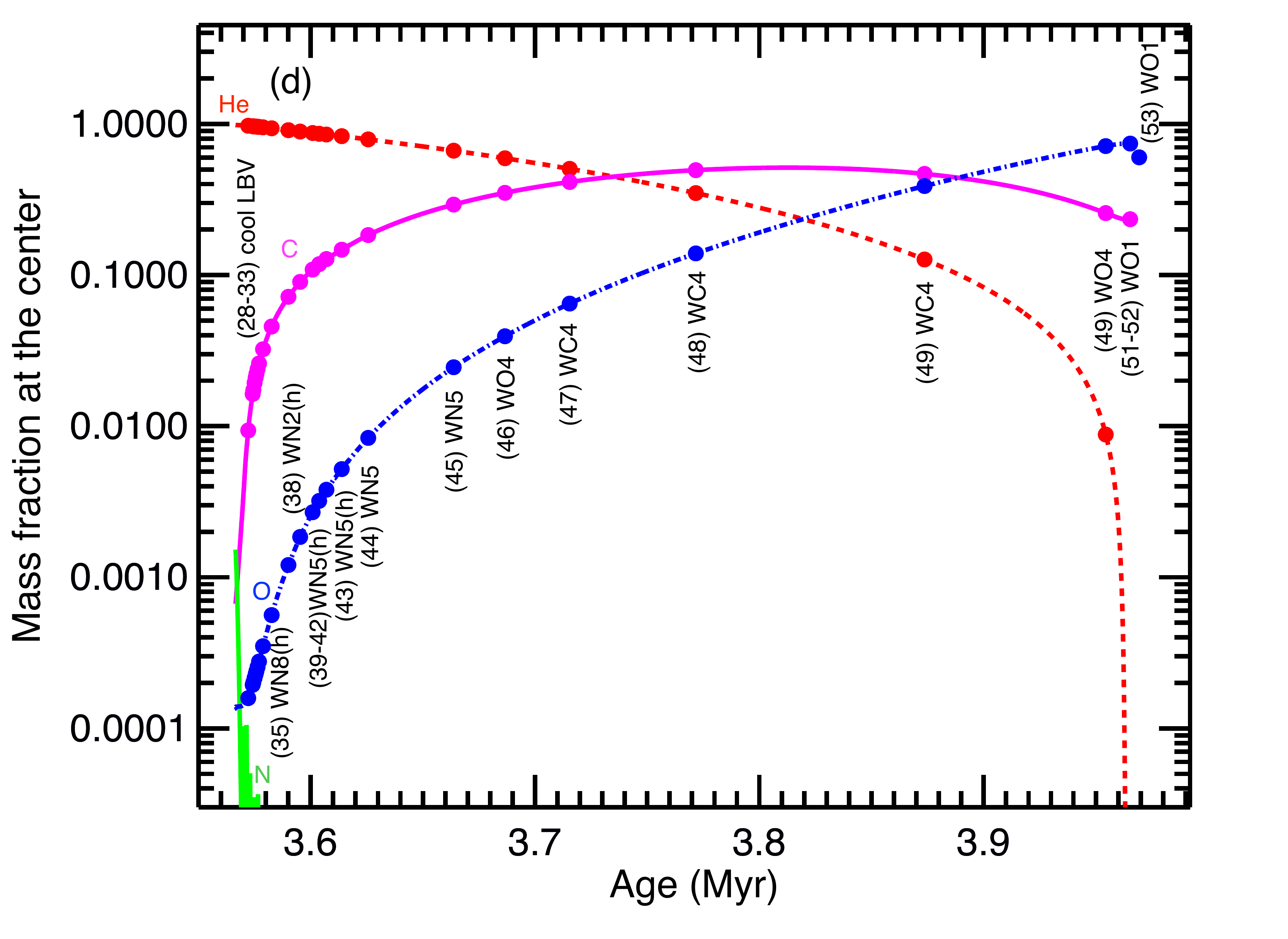}}
\resizebox{0.32\hsize}{!}{\includegraphics{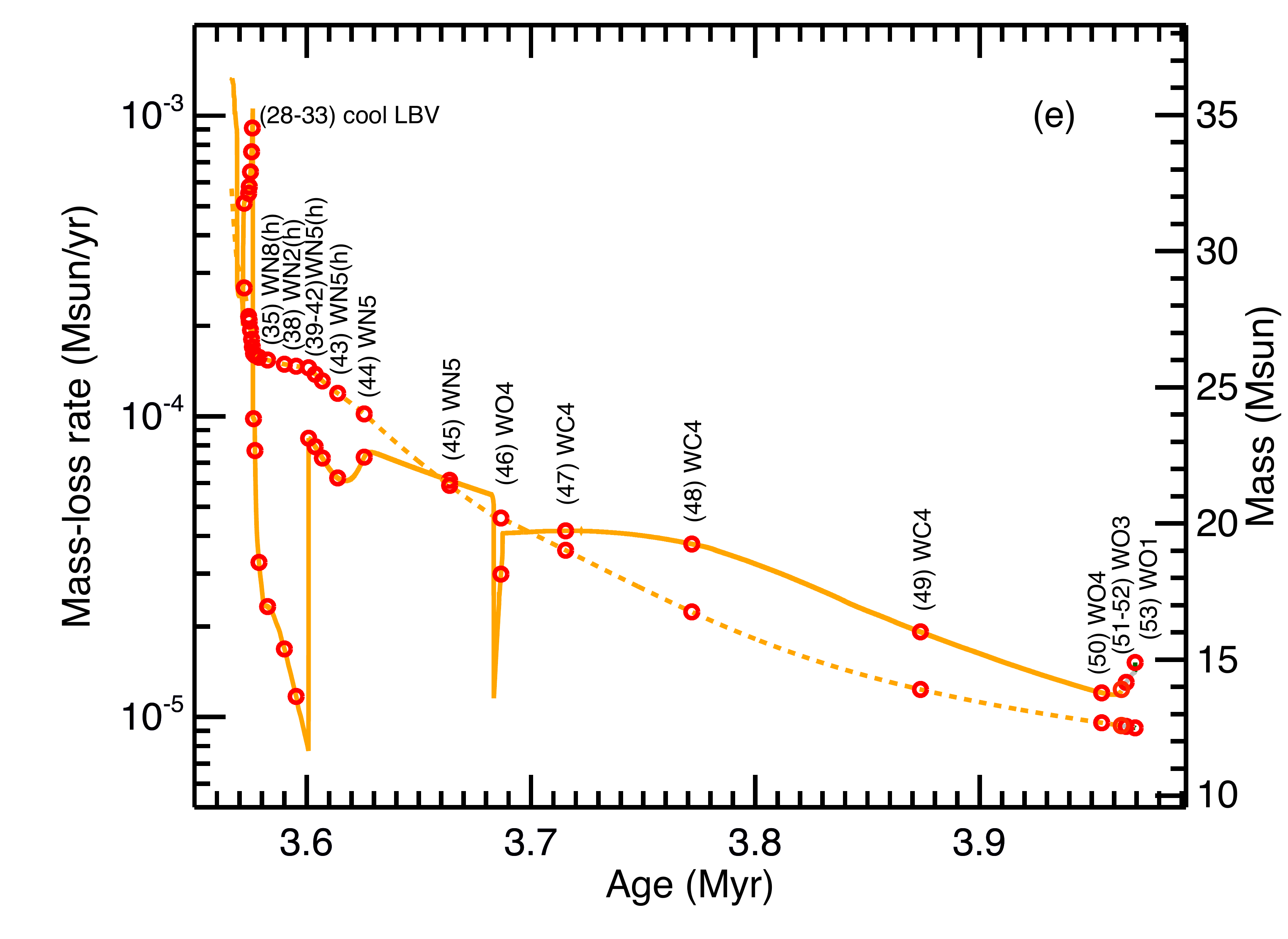}}
\resizebox{0.32\hsize}{!}{\includegraphics{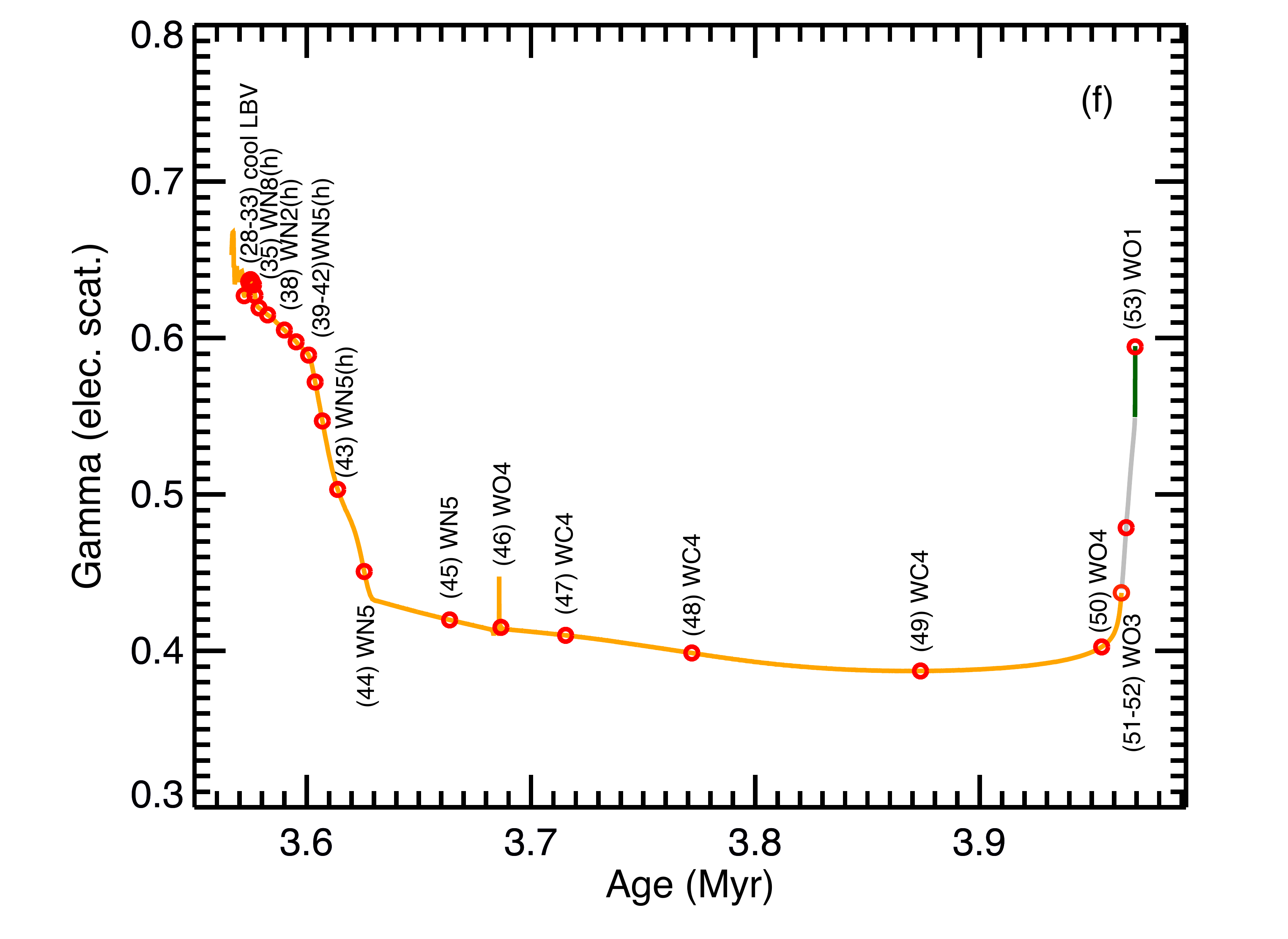}}\\
\caption{\label{hecorefig} Similar to Fig. \ref{hrd1}, but focusing on the evolution during He-core burning and advanced stages.
 }
\end{figure*}

\begin{figure*}
\center
\resizebox{0.89\hsize}{!}{\includegraphics{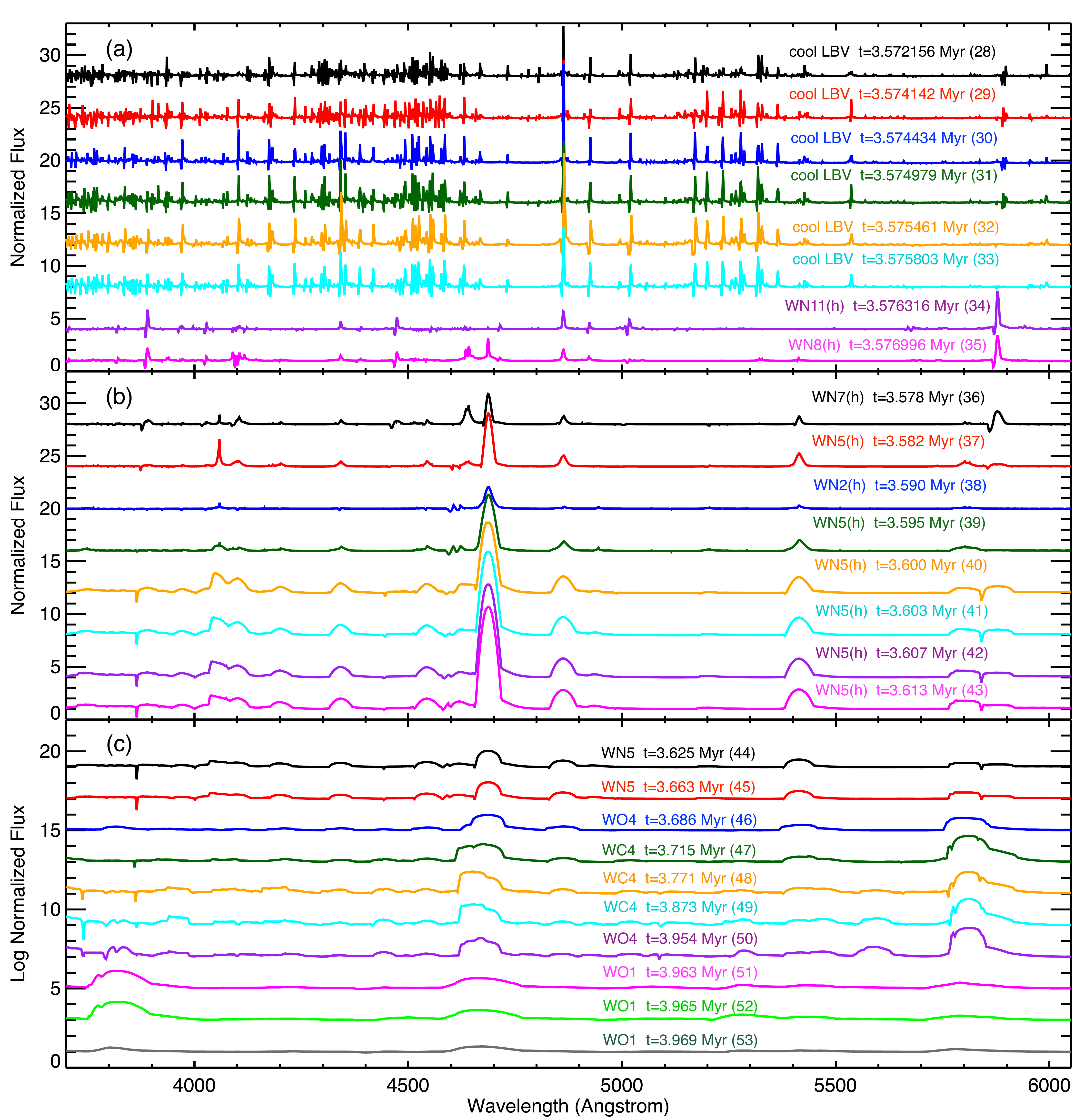}}\\
\caption{\label{hespec} Evolution of the optical spectra of a non-rotating 60~\msun\ star during the He-core burning and advanced stages. The evolution proceeds from top to bottom, with labels indicating the spectral type, age, and model ID according to Table \ref{model log}. In panel $(c)$, the fluxes are shown in logarithm scale for the purpose of displaying the full range of optical emission lines.}
\end{figure*}

The star enters this phase showing an extremely cool LBV spectrum (Fig. \ref{hespec}). As soon as He-core burning starts, the surface temperature increases (Fig. \ref{hecorefig}b) and the star rapidly evolves to the blue (Fig. \ref{hecorefig}a). This is caused by the decrease of the opacity as the chemical composition changes, and by the reduction of the inflation of the envelope as the mass of the H-shell burning layer decreases. The star shows a cool LBV spectrum for a brief period of time (0.010 Myr, stages 28 through 33), but substantial mass loss occurs in this period, with the star losing 5.94~\msun\ (Fig. \ref{hecorefig}e). The star proceeds its evolution presenting quickly-evolving WNL spectral types that, for the non-rotating 60~\msun\ star, occurs with small amounts of H still present on the surface ($\xsurf=0.13$; Fig. \ref{hecorefig}c). This warrants a WN(h) classification \citep{ssm96}. The spectral type evolves from WN11(h) (stage 34) to WN7(h) (stage 36), as shown in Fig. \ref{hespec}. This is a brief phase that lasts for 0.005 Myr, during which 0.31~\msun\ of material is lost (Fig. \ref{hecorefig}e).

The evolution to the blue causes a rapid display of spectral types and, on a short timescale ($\sim0.030$ Myr, stages 33 to 43), the star presents spectra of a cool LBV, hot LBV, WNL, and WNE with small amounts of H (Fig. \ref{hespec}). The surface abundances present little evolution (Fig. \ref{hecorefig}c), and the main changes in spectral type arise from the progressive increase in the surface temperature.

The size of the convective core increases during He burning, with the envelope contracting and the surface temperature increasing (Fig. \ref{hecorefig}b). This causes the star to eventually display early WN spectral types (Fig. \ref{hespec}). Interestingly, the model shows a rather constant WN5(h) spectral type throughout this phase\footnote{Except for a short amount of time when a WN2(h) spectrum appears, which is caused by a momentary diminishing of $\mdot$.} (Fig. \ref{hespec}). The first 40\% of the WNE phase (0.043 Myr) develops with H still present on the surface ($\xsurf=0.13$; Fig. \ref{hecorefig}c), with the star losing 1.94~\msun\ (Fig. \ref{hecorefig}e). The strong mass loss reduces the mass of the H-burning shell, reducing \lstar. The remainder 60\% (0.060 Myr) occurs without H at the surface (Fig. \ref{hecorefig}c) and 3.81~\msun\ is lost (Fig. \ref{hecorefig}e). We note that there is a decrease in $\lstar$ during the WNE phase (Fig. \ref{hecorefig}a) because of the significant decrease in $\mstar$ due to strong mass loss (Fig. \ref{hecorefig}e). 

\subsubsection{A long-lived WC and short WO phases (stages 46--51)}

Because of mass loss, eventually the layers rich in He, C, and O appear at the surface, while the N abundance dramatically drops at the surface (stage 46, Fig. \ref{hecorefig}c). These changes in the surface abundances mark the end of the WN phase. At this point, the star is in a short transition phase that lasts for 0.006 Myr, and a WO4 spectral type appears (Fig. \ref{hespec}). This is caused by a momentary reduction of $\mdot$. Once \mdot\ increases again, It quickly evolves to an early WC (WC4) spectral type that lasts for a significant amount of time (0.244 Myr; stages 47 through 49). During this period, the surface temperature steadily increases  (Fig. \ref{hecorefig}b), while $\lstar$ decreases (Fig. \ref{hecorefig}a) since the star loses 7.02~\msun\ due to the action of strong stellar winds (Fig. \ref{hecorefig}e). The main change in the surface abundances is the increase in O (Fig. \ref{hecorefig}c).

Close to end of He-core burning (stage 50), several changes in the fundamental properties of the star occurs. There is an increase in $\teff$ (Fig. \ref{hecorefig}b) and O abundance at the surface (Fig. \ref{hecorefig}c), and decrease of $\mdot$ (Fig. \ref{hecorefig}e). These changes cause the star to show a WO4 and subsequently a WO1 spectral type (Fig. \ref{hecorefig}b), which is a brief phase that lasts 0.030 Myr, when the star loses 0.57~\msun\ (Fig. \ref{hecorefig}e).

The sequence WO -- WC -- WO seen at the end of the evolution indicates that the difference between WC and WO stars results mainly from ionization effects, which are regulated by $\teff$ and $\mdot$, rather than by a large increase of the O abundance at the surface as the star evolves from WC to WO. This result is in line with suggestions from observational studies of WO stars \citep{crowther99b,crowther00,tramper13}.

\subsection{He-shell burning,  C-core burning, and advanced stages}

At the end of He-core burning (stage 51), the core and envelope contract, with He being ignited in a shell. This is an extremely short evolutionary phase (0.006 Myr; stage 51), where the contraction causes an increase in the surface temperature (Fig. \ref{hecorefig}b). In combination with the high O surface abundance, this causes the star shows a WO1 spectral type (Fig. \ref{hespec}). When the central temperature is high enough, C-core burning begins. It lasts for 0.002 Myr, the star loses 0.02~\msun, and the spectral type continuing to reflect early WO subtypes (WO1; Fig. \ref{hespec}). Beyond C-core burning (stage 53), our models indicate only modest changes of 0.01 dex in $\log \teff$ and $\log \lstar$, showing that the star will remain as a WO1 until core collapse \citep{gmg13}.

\begin{sidewaystable*}
\begin{minipage}{\textwidth}
\caption{Fundamental properties of the non-rotating 60 \msun\ star at Z=0.014 at selected phases of its evolution. From left to right, the columns correspond to the stage ID, evolutionary phase,  spectral type, age, mass (\mstar), bolometric luminosity (\lstar), effective temperature estimated by the Geneva code (\teffg), effective temperature computed by CMFGEN at $\tauross=2/3$ (\teff), radius of the hydrostatic surface (\rstar), mass-loss rate (\mdot), wind terminal velocity (\vinf), steepness of the velocity law ($\beta$),  and H, He, C, N, and O abundances at the surface and center, in mass fraction. The code for the evolutionary phases are: H-c=H-core burning, H-sh=H-shell burning, He-c=He-core burning, and C-c=C-core burning. }

\label{model log}
\scriptsize
\centering
\vspace{0.1cm}
\begin{tabular}{l c c c c c c c c c c c c c c c c c c c c c c}
\hline\hline
ID &  Evol. & Spec. & Age & \mstar & \lstar & \teffg & \teff & \rstar & \mdot & \vinf & $\beta$ & X$_{sur}$ & Y$_{sur}$ & C$_{sur}$ & N$_{sur}$ & O$_{sur}$ & X$_{c}$ & Y$_{c}$ & C$_{c}$ & N$_{c}$ & O$_{c}$ \\ 
& Phase & Type  & (yr) & (\msun) & (\lsun) & (\K) & (\K) &(\rsun) & (\msunyr) & (\kms) &  & & & & & & & & & &    \\
\hline 
      1 & H-c & O3If$^*$c &        29532 & 59.95 &  504821 &  48853 &  48390 &    9.95 &   1.73E-06 &    3543 &   1.0 &  0.72 &  0.27 &   2.28E-03 &   6.59E-04 &   5.72E-03 &  0.72 &  0.27 &   3.55E-05 &   4.02E-03 &   4.90E-03 \\
      2 & H-c & O3If$^*$c &       551447 & 58.97 &  528380 &  47031 &  47300 &   10.98 &   2.02E-06 &    3314 &   1.0 &  0.72 &  0.27 &   2.28E-03 &   6.59E-04 &   5.72E-03 &  0.65 &  0.34 &   6.67E-05 &   7.71E-03 &   6.41E-04 \\
      3 & H-c & O3If$^*$c &      1059410 & 57.86 &  555083 &  45796 &  45800 &   11.87 &   2.33E-06 &    3122 &   1.0 &  0.72 &  0.27 &   2.28E-03 &   6.59E-04 &   5.72E-03 &  0.57 &  0.41 &   7.12E-05 &   8.09E-03 &   1.97E-04 \\
      4 & H-c & O4If$^*$c &      1753470 & 56.06 &  597948 &  43261 &  43260 &   13.80 &   2.88E-06 &    2795 &   1.0 &  0.72 &  0.27 &   2.28E-03 &   6.59E-04 &   5.72E-03 &  0.45 &  0.53 &   7.46E-05 &   8.15E-03 &   1.27E-04 \\
      5 & H-c & O5Ifc &      2360274 & 54.16 &  645563 &  39375 &  39380 &   17.31 &   3.36E-06 &    2396 &   1.0 &  0.72 &  0.27 &   2.28E-03 &   6.59E-04 &   5.72E-03 &  0.33 &  0.65 &   7.76E-05 &   8.15E-03 &   1.20E-04 \\
      6 & H-c & O6Iafc &      2711788 & 52.97 &  680682 &  35749 &  35750 &   21.57 &   3.35E-06 &    2084 &   1.0 &  0.72 &  0.27 &   2.28E-03 &   6.59E-04 &   5.72E-03 &  0.25 &  0.74 &   8.04E-05 &   8.15E-03 &   1.17E-04 \\
      7 & H-c & O7.5Iafc &      3019444 & 52.02 &  719178 &  31088 &  31090 &   29.31 &   2.67E-06 &    1737 &   1.0 &  0.72 &  0.27 &   2.28E-03 &   6.59E-04 &   5.72E-03 &  0.17 &  0.82 &   8.40E-05 &   8.15E-03 &   1.12E-04 \\
      8 & H-c & O9Iab &      3178984 & 51.65 &  743097 &  27864 &  27810 &   37.09 &   1.90E-06 &    1520 &   1.0 &  0.72 &  0.27 &   2.28E-03 &   6.59E-04 &   5.72E-03 &  0.12 &  0.86 &   8.67E-05 &   8.15E-03 &   1.09E-04 \\
      9 & H-c & B0.2 Ia &      3262672 & 51.51 &  757148 &  25884 &  25310 &   43.39 &   1.40E-06 &    1394 &   1.0 &  0.72 &  0.27 &   2.28E-03 &   6.59E-04 &   5.72E-03 &  0.10 &  0.89 &   8.88E-05 &   8.15E-03 &   1.06E-04 \\
     10 & H-c & B0.5 Ia$^+$ &      3300425 & 50.91 &  759715 &  24443 &  24070 &   48.74 &   1.87E-05 &    1301 &   1.0 &  0.72 &  0.27 &   2.28E-03 &   6.59E-04 &   5.72E-03 &  0.08 &  0.90 &   8.97E-05 &   8.15E-03 &   1.05E-04 \\
     11 & H-c & hot LBV &      3371211 & 49.65 &  766547 &  21072 &  20700 &   65.87 &   1.69E-05 &    1093 &   1.0 &  0.72 &  0.27 &   2.28E-03 &   6.59E-04 &   5.72E-03 &  0.06 &  0.92 &   9.18E-05 &   8.15E-03 &   1.03E-04 \\
     12 & H-c & hot LBV &      3418874 & 48.88 &  772591 &  18573 &  17150 &   85.12 &   1.53E-05 &     500 &   1.0 &  0.72 &  0.27 &   2.27E-03 &   6.60E-04 &   5.72E-03 &  0.05 &  0.94 &   9.39E-05 &   8.15E-03 &   1.00E-04 \\
     13 & H-c & cool LBV &      3449657 & 44.32 &  755429 &  12105 &   8564 &  198.14 &   1.43E-04 &     303 &   1.0 &  0.72 &  0.27 &   3.28E-05 &   3.96E-03 &   4.98E-03 &  0.04 &  0.95 &   9.57E-05 &   8.14E-03 &   9.87E-05 \\
     14 & H-c & hot LBV &      3499862 & 37.61 &  741291 &  19271 &  16850 &   77.45 &   2.02E-05 &     440 &   2.5 &  0.54 &  0.45 &   5.75E-05 &   8.15E-03 &   1.54E-04 &  0.02 &  0.97 &   9.99E-05 &   8.14E-03 &   9.54E-05 \\
     15 & H-c & hot LBV &      3559650 & 36.26 &  814797 &  25785 &  20050 &   45.35 &   3.59E-05 &    1018 &   1.0 &  0.49 &  0.49 &   5.62E-05 &   8.17E-03 &   1.34E-04 &  0.00 &  0.99 &   1.48E-04 &   8.08E-03 &   8.79E-05 \\
     16 & H-sh & hot LBV &      3561340 & 36.20 &  818839 &  20457 &  16770 &   72.24 &   2.86E-05 &     424 &   2.5 &  0.49 &  0.49 &   5.61E-05 &   8.17E-03 &   1.34E-04 &  0.00 &  0.99 &   1.49E-04 &   8.07E-03 &   9.84E-05 \\
     17 & H-sh & cool LBV &      3561847 & 36.03 &  827076 &  16164 &   8918 &  116.29 &   3.04E-04 &     331 &   2.5 &  0.49 &  0.50 &   5.60E-05 &   8.17E-03 &   1.32E-04 &  0.00 &  0.99 &   1.50E-04 &   8.07E-03 &   1.12E-04 \\
     18 & H-sh & cool LBV &      3562100 & 35.96 &  834422 &  12785 &   8099 &  186.68 &   2.43E-04 &     260 &   2.5 &  0.48 &  0.50 &   5.59E-05 &   8.17E-03 &   1.32E-04 &  0.00 &  0.99 &   1.51E-04 &   8.06E-03 &   1.22E-04 \\
     19 & H-sh & cool LBV &      3562142 & 35.95 &  835687 &  12130 &   8348 &  207.57 &   2.31E-04 &     246 &   2.5 &  0.48 &  0.50 &   5.59E-05 &   8.17E-03 &   1.32E-04 &  0.00 &  0.99 &   1.52E-04 &   8.06E-03 &   1.24E-04 \\
     20 & H-sh & cool LBV &      3562185 & 35.94 &  836924 &  11444 &   8231 &  233.36 &   2.18E-04 &     232 &   2.5 &  0.48 &  0.50 &   5.58E-05 &   8.17E-03 &   1.32E-04 &  0.00 &  0.99 &   1.52E-04 &   8.06E-03 &   1.25E-04 \\
     21 & H-sh & cool LBV &      3562201 & 35.94 &  837454 &  11141 &   8176 &  246.28 &   2.11E-04 &     226 &   2.5 &  0.48 &  0.50 &   5.58E-05 &   8.17E-03 &   1.32E-04 &  0.00 &  0.99 &   1.52E-04 &   8.06E-03 &   1.26E-04 \\
     22 & H-sh & cool LBV &      3562235 & 35.93 &  838434 &  10594 &   8069 &  272.52 &   2.00E-04 &     214 &   2.5 &  0.48 &  0.50 &   5.58E-05 &   8.17E-03 &   1.32E-04 &  0.00 &  0.99 &   1.52E-04 &   8.06E-03 &   1.27E-04 \\
     23 & H-sh & cool LBV &      3562286 & 35.92 &  839972 &   9664 &   7750 &  327.83 &   1.82E-04 &     139 &   2.5 &  0.48 &  0.50 &   5.58E-05 &   8.17E-03 &   1.32E-04 &  0.00 &  0.99 &   1.52E-04 &   8.06E-03 &   1.30E-04 \\
     24 & H-sh & cool LBV &      3562303 & 35.92 &  840531 &   9317 &   7696 &  352.77 &   1.77E-04 &     134 &   2.5 &  0.48 &  0.50 &   5.58E-05 &   8.17E-03 &   1.32E-04 &  0.00 &  0.99 &   1.52E-04 &   8.06E-03 &   1.30E-04 \\
     25 & H-sh & cool LBV &      3562320 & 35.91 &  840990 &   9038 &   7673 &  375.03 &   1.69E-04 &     130 &   2.5 &  0.48 &  0.50 &   5.58E-05 &   8.17E-03 &   1.32E-04 &  0.00 &  0.99 &   1.52E-04 &   8.06E-03 &   1.31E-04 \\
     26 & H-sh & cool LBV &      3562354 & 35.91 &  841934 &   8373 &   7593 &  437.23 &   1.87E-04 &     120 &   2.5 &  0.48 &  0.50 &   5.58E-05 &   8.17E-03 &   1.32E-04 &  0.00 &  0.99 &   1.53E-04 &   8.06E-03 &   1.32E-04 \\
     27 & H-sh & cool LBV &      3562379 & 35.90 &  842525 &   8011 &   7610 &  477.81 &   1.77E-04 &     115 &   2.5 &  0.48 &  0.50 &   5.58E-05 &   8.17E-03 &   1.32E-04 &  0.00 &  0.99 &   1.53E-04 &   8.06E-03 &   1.32E-04 \\
     28 & He-c+H-sh & cool LBV &      3572156 & 28.65 & 1027971 &  10827 &   7958 &  288.92 &   5.12E-04 &     118 &   2.5 &  0.14 &  0.85 &   9.28E-05 &   8.14E-03 &   1.09E-04 &  0.00 &  0.97 &   9.37E-03 &   2.18E-06 &   1.59E-04 \\
     29 & He-c+H-sh & cool LBV &      3574142 & 27.60 & 1005669 &  11606 &   8077 &  248.72 &   5.52E-04 &     124 &   2.5 &  0.14 &  0.85 &   9.78E-05 &   8.13E-03 &   1.08E-04 &  0.00 &  0.97 &   1.63E-02 &   1.04E-06 &   1.94E-04 \\
     30 & He-c+H-sh & cool LBV &      3574434 & 27.43 & 1001146 &  12182 &   8184 &  225.23 &   5.81E-04 &     129 &   2.5 &  0.14 &  0.85 &   9.81E-05 &   8.13E-03 &   1.08E-04 &  0.00 &  0.96 &   1.73E-02 &   1.07E-05 &   2.01E-04 \\
     31 & He-c+H-sh & cool LBV &      3574979 & 27.10 &  989874 &  13648 &   8607 &  178.44 &   6.50E-04 &      20 &   2.5 &  0.14 &  0.85 &   9.68E-05 &   8.13E-03 &   1.08E-04 &  0.00 &  0.96 &   1.93E-02 &   2.62E-06 &   2.15E-04 \\
     32 & He-c+H-sh & cool LBV &      3575461 & 26.76 &  976661 &  15961 &   8479 &  129.60 &   7.59E-04 &      20 &   2.5 &  0.14 &  0.85 &   9.54E-05 &   8.14E-03 &   1.08E-04 &  0.00 &  0.96 &   2.09E-02 &   5.91E-06 &   2.28E-04 \\
     33 & He-c+H-sh & cool LBV &      3575803 & 26.48 &  964242 &  19166 &   8964 &   89.30 &   9.10E-04 &     203 &   2.5 &  0.13 &  0.85 &   9.39E-05 &   8.14E-03 &   1.08E-04 &  0.00 &  0.96 &   2.21E-02 &   2.00E-06 &   2.39E-04 \\
     34 & He-c+H-sh & WN11(h) &      3576316 & 26.24 &  954291 &  23678 &  19030 &   58.21 &   9.82E-05 &     351 &   1.0 &  0.13 &  0.85 &   9.48E-05 &   8.14E-03 &   1.08E-04 &  0.00 &  0.96 &   2.38E-02 &   2.00E-05 &   2.55E-04 \\
     35 & He-c, H-sh & WN8(h) &      3576966 & 26.18 &  941859 &  30783 &  24210 &   34.21 &   7.69E-05 &     461 &   1.0 &  0.13 &  0.85 &   9.41E-05 &   8.14E-03 &   1.08E-04 &  0.00 &  0.96 &   2.60E-02 &   2.85E-06 &   2.77E-04 \\
     36 & He-c, H-sh & WN7(h) &      3578756 & 26.10 &  927045 &  38708 &  36090 &   21.47 &   3.27E-05 &    1113 &   1.0 &  0.13 &  0.85 &   9.32E-05 &   8.14E-03 &   1.08E-04 &  0.00 &  0.95 &   3.23E-02 &   4.55E-07 &   3.50E-04 \\
     37 & He-c, H-sh & WN5(h) &      3582601 & 26.00 &  916946 &  45243 &  44100 &   15.63 &   2.33E-05 &    1310 &   1.0 &  0.13 &  0.85 &   9.12E-05 &   8.14E-03 &   1.08E-04 &  0.00 &  0.94 &   4.56E-02 &   6.15E-08 &   5.62E-04 \\
     38 & He-c, H-sh & WN2(h) &      3590081 & 25.85 &  897212 &  56419 &  56170 &    9.94 &   1.68E-05 &    1658 &   1.0 &  0.13 &  0.85 &   8.76E-05 &   8.15E-03 &   1.08E-04 &  0.00 &  0.91 &   7.18E-02 &   1.72E-06 &   1.21E-03 \\
     39 & He-c, H-sh & WN5(h) &      3595309 & 25.77 &  883519 &  63351 &  60350 &    7.82 &   1.17E-05 &    1884 &   1.0 &  0.13 &  0.85 &   8.60E-05 &   8.15E-03 &   1.08E-04 &  0.00 &  0.89 &   9.01E-02 &   7.07E-10 &   1.85E-03 \\
     40 & He-c, H-sh & WN5(h) &      3600871 & 25.72 &  869168 &  70075 &  42180 &    6.34 &   8.45E-05 &    2000 &   1.0 &  0.13 &  0.85 &   8.40E-05 &   8.15E-03 &   1.08E-04 &  0.00 &  0.87 &   1.09E-01 &   3.70E-07 &   2.69E-03 \\
     41 & He-c, H-sh & WN5(h) &      3603786 & 25.48 &  835922 &  81012 &  44170 &    4.65 &   7.93E-05 &    2000 &   1.0 &  0.13 &  0.85 &   6.81E-05 &   8.18E-03 &   1.08E-04 &  0.00 &  0.86 &   1.18E-01 &   6.16E-07 &   3.20E-03 \\
     42 & He-c, H-sh & WN5(h) &      3607017 & 25.23 &  791919 &  92368 &  46260 &    3.48 &   7.26E-05 &    2000 &   1.0 &  0.13 &  0.85 &   6.04E-05 &   8.19E-03 &   1.08E-04 &  0.00 &  0.85 &   1.28E-01 &   2.90E-06 &   3.79E-03 \\
     43 & He-c, H-sh & WN5(h) &      3613846 & 24.78 &  717440 & 106349 &  49580 &    2.50 &   6.24E-05 &    2000 &   1.0 &  0.13 &  0.86 &   6.56E-05 &   8.18E-03 &   1.10E-04 &  0.00 &  0.83 &   1.47E-01 &   1.93E-07 &   5.20E-03 \\
     44 & He-c, H-sh & WN5 &      3625652 & 24.02 &  682455 & 122874 &  48170 &    1.83 &   7.31E-05 &    2000 &   1.0 &  0.03 &  0.95 &   7.49E-05 &   8.18E-03 &   9.42E-05 &  0.00 &  0.79 &   1.84E-01 &   2.00E-11 &   8.36E-03 \\
     45 & He-c & WN5 &      3663742 & 21.40 &  584484 & 125165 &  51330 &    1.63 &   6.12E-05 &    2000 &   1.0 &  0.00 &  0.99 &   1.39E-04 &   8.09E-03 &   8.79E-05 &  0.00 &  0.66 &   2.92E-01 &   1.21E-13 &   2.45E-02 \\
     46 & He-c & WO4 &      3686605 & 20.20 &  545448 & 126228 &  82330 &    1.55 &   2.99E-05 &    2475 &   1.0 &  0.00 &  0.87 &   1.10E-01 &   2.41E-03 &   4.21E-03 &  0.00 &  0.59 &   3.50E-01 &   6.83E-15 &   3.93E-02 \\
     47 & He-c & WC4 &      3715449 & 19.02 &  507199 & 133655 &  70130 &    1.33 &   4.16E-05 &    2394 &   1.0 &  0.00 &  0.74 &   2.31E-01 &   1.32E-10 &   1.40E-02 &  0.00 &  0.50 &   4.13E-01 &   1.88E-16 &   6.47E-02 \\
     48 & He-c & WC4 &      3771686 & 16.75 &  434538 & 136007 &  66620 &    1.19 &   3.76E-05 &    2185 &   1.0 &  0.00 &  0.62 &   3.30E-01 &   6.61E-14 &   3.34E-02 &  0.00 &  0.35 &   4.94E-01 &   2.07E-19 &   1.39E-01 \\
     49 & He-c & WC4 &      3873640 & 13.90 &  350243 & 140542 &  86090 &    1.00 &   1.92E-05 &    1928 &   1.0 &  0.00 &  0.38 &   4.82E-01 &   1.75E-17 &   1.21E-01 &  0.00 &  0.13 &   4.67E-01 &   7.87E-22 &   3.88E-01 \\
     50 & He-c & WO4 &      3954404 & 12.68 &  332128 & 151199 & 105900 &    0.84 &   1.20E-05 &    1918 &   1.0 &  0.00 &  0.29 &   5.10E-01 &   3.81E-19 &   1.84E-01 &  0.00 &  0.01 &   2.57E-01 &   2.78E-29 &   7.13E-01 \\
     51 & He-c & WO1 &      3963096 & 12.58 &  355836 & 165881 & 146200 &    0.72 &   1.22E-05 &    5000 &   1.0 &  0.00 &  0.28 &   5.11E-01 &   2.53E-19 &   1.91E-01 &  0.00 &  0.00 &   2.34E-01 &   1.23E-29 &   7.41E-01 \\
     52 & He-sh & WO1 &      3965312 & 12.55 &  390822 & 182276 & 153000 &    0.63 &   1.30E-05 &    5000 &   1.0 &  0.00 &  0.28 &   5.12E-01 &   2.27E-19 &   1.93E-01 &  0.00 &  0.00 &   2.34E-01 &   3.26E-29 &   7.41E-01 \\
     53 & end C-c & WO1 &      3969319 & 12.50 &  483177 & 224855 & 164600 &    0.46 &   1.52E-05 &    5000 &   1.0 &  0.00 &  0.27 &   5.12E-01 &   1.77E-19 &   1.97E-01 &  0.00 &  0.00 &   1.76E-05 &   3.26E-29 &   6.01E-01 \\
\hline
\end{tabular}
\end{minipage}
\end{sidewaystable*}

\section{\label{lifetimes} Lifetimes of different evolutionary phases: spectra vs. surface chemical abundance and $\teff$ criteria}

Up to now, in the absence of a spectrum, the duration of different spectroscopic phases has been estimated based on chemical abundance and $\teff$ criteria \citep{smithmaeder91,meynet03}. Let us analyze the lifetime of the spectroscopic phases that we found based on the computation of the spectrum, and how those compares to the previous estimates. Here we compare with the results from \citet{georgy12a}, which used the same 60~\msun\ evolutionary model employed here but employed chemical abundance and $\teff$ criteria. Our results for the lifetimes are shown in Table \ref{duration}, while Table \ref{compareduration} compares the duration of the spectroscopic phases of our models with previous studies.

We obtained that the O-type spectroscopic phase lasts for $3.22\times10^6$ yr, which is similar to the value estimated based on the $\teff$ and abundance criteria ($3.00\times10^6$ yr). This small difference arises because \citet{georgy12a} assumes that O stars have $\log (\teff/K) > 4.5$, while we find that the latest O supergiant models have $\log (\teff/K)  \simeq 4.45$ (in agreement with \citealt{martins05}). This makes the O-type phase slightly longer.

\begin{table*}
\center
\caption{Spectral types, duration, and mass lost at different evolutionary stages of a non-rotating 60~\msun\ star. }
\footnotesize
\label{duration}
\begin{tabular}{llcccr}
\hline
\hline         
Evolutionary stage 		& Spectral type & Duration & Duration    & Mass lost & Mass lost\\
                           		          &                         &  (Myr)        & (\% of total)& (\msun)     & (\% of total loss)\\                         
\hline
MS (core H burning)  & early O I               &   2.894 &    72.9 &    7.62 &    16.1  \\
"                    & late O I                &   0.329 &     8.3 &    0.81 &     1.7  \\
"                    & BSG                     &   0.039 &     1.0 &    0.06 &     0.1  \\
"	   			     & BHG                     &   0.079 &     2.0 &    1.34 &     2.8  \\
"        			 & LBV                     &   0.217 &     5.5 &   13.85 &    29.3  \\
H shell burning      & LBV                     &   0.009 &     0.2 &    4.60 &     9.7  \\
core He burning      & LBV                     &   0.009 &     0.2 &    5.37 &    11.4  \\
"				     &  WNL with H             &   0.005 &     0.1 &    0.31 &     0.7  \\
"			       	 &  WNE with H             &   0.045 &     1.1 &    2.02 &     4.3  \\
"				     &  WNE no H               &   0.055 &     1.4 &    3.62 &     7.7  \\
"			    	 &  WO          	       &   0.006 &     0.2 &    0.23 &     0.5  \\
"			     	 &  WCE 		           &   0.257 &     6.5 &    7.38 &    15.6  \\
" 			         &  WO                     &   0.030 &     0.7 &    0.57 &     0.9 \\
He shell burning	 &  WO          		&         0.006  &0.1                  &    0.07 & 0.1       \\
core C burning  and beyond        &  WO     &        0.002   & <0.1		     &   0.02   & <0.1  \\
\hline
\hline
\end{tabular}
\end{table*}

According to our new models, the LBV phase of the non-rotating 60~\msun\ star lasts for $2.35\times10^5$~yr. This is $~\sim7\%$ of the lifetime of the O-type phase and  $~\sim59\%$ of the WR lifetime for that particular initial mass and rotation. The lifetime of the LBV phase in our models is $\sim5 $ times longer than the  $\sim4\times10^4$ yr that is usually assumed based on the number counting of known LBVs compared to WR stars in the LMC \citep{bohannan97}. However, we caution that an analysis of models comprising the full range of initial masses and rotation is needed to draw firm conclusions about the LBV lifetime. For instance, we foresee that our most massive models ($120~\msun$) will have much shorter LBV phases, and so will our rotating models. Thus, all we can safely say at the moment is that the duration of the LBV phase depends on the initial mass and rotation. Also, with the criteria employed here to classify a star as  an LBV (Sect. \ref{classifyspec}), the lifetime of the LBV phase also depends on the mass-loss recipe and clumping factor used. As discussed in Sect. \ref{bistability}, the presence of a bistability jump in \mdot\ is what causes a BHG/LBV spectrum to appear. If the jump in \mdot\  is weaker than theoretically predicted, as suggested by \citep{crowther06}, this would favor the presence of BSGs, BHGs, and A-type supergiants and hypergiants, decreasing the duration of the  LBV phase.

We found that the duration of the WR phase is also slightly modified when a spectroscopic criterium is used.  We find a slightly shorter lifetime of $3.95\times10^5$ yr compared to \citet{georgy12a}, because part of the track that was classified as WNL by \citet{georgy12a} at the beginning of He-core burning are actually LBVs in our models. The difference in the total lifetime of the WR phase is small because the evolution of the surface properties is fast at the beginning of He-core burning (Sect. \ref{hecore}). The duration of the WN phase is also similar to what has been obtained before using an abundance criterium.

However, the duration of the WNE and WNL phases are strongly modified using the spectroscopic criterium. The WNL lifetime decreases by a factor of 15.25, while the WNE duration increases by a factor of 2. These huge changes occur because the H abundance at the surface, which has been employed in previous stellar evolution studies as a diagnostic \citep[e.\,g.][]{georgy12a}, is not a good tracer of the WNE or WNL status. The WNL or WNE appearance is regulated by the ionization structure of the wind, which is determined by $\teff$ and \lstar, and ultimately sets whether a WNE or WNL spectrum arises \citep[e.g.,][]{hillier87a}. If the decrease of the WNL and increase of the WNE lifetime is confirmed for other initial masses and rotation, this could help to reconcile the theoretical and observed WNL/WNE ratio, which was found to be too high by \citet{georgy12a}.

Our models indicate that the WC lifetime is slightly reduced when spectroscopic criteria are used. The main reason is that a certain surface C abundance is required in order to have a WC spectrum, and that occurs at a higher C abundance than what is assumed by the chemical abundance criterium of \citet{georgy12a}. We also note that no WN/WC transition stars are seen in our models, in agreement with previous observational and theoretical studies \citep{crowther95b,meynet03}.

We found that a short WO phase exists at the end of the evolution of a 60~\msun\ star. This was inexistent when an abundance criterium was used,  but our atmospheric calculations show that because of the extremely high \teff\ and O abundance, a WO spectral type arises at the very end of the evolution of the most massive stars until core collapse \citep{gmg13}.

\begin{table}
\caption{Lifetime of the spectroscopic phases of a non-rotating 60~\msun\ star. Column 2 refers to the previous values found using chemical abundance and $\teff$ criteria (from \citealt{georgy12a}), while  column 3 shows the revised durations obtained in this work through a classification of the synthetic spectra from the evolutionary models. The timescales of the LBV ($\tau_\mathrm{LBV,60,0}$), WR ($\tau_\mathrm{WR,60,0}$), and O-type phases ($\tau_\mathrm{O,60,0}$) for a non-rotating 60~\msun\ star are indicated at the bottom of the table.}
\footnotesize
\begin{center}
\label{compareduration}
\begin{tabular}{lcr}
\hline
\hline         
Spectral type & Previous duration & Revised duration \\
                         &  (chem. abund., yr )        &  (spec. class., yr)   \\                         
\hline
O                   &       $3.00\times10^6$	          & $3.22\times10^6$	                  \\
BHG             &      $-$                                       & $0.79\times 10^5$                  \\
LBV     	   &       $-$                                      & $2.35\times10^5$			\\
WR                &       $3.97\times10^5$           & $3.95\times10^5$	                  \\
WNL              &       $6.13\times10^4$           & $0.49\times10^4$	                  \\
WNE              &       $5.15\times10^4$           & $10.31\times10^4$	         \\
WC                &       $2.85\times10^5$           & $2.57\times10^5$	                  \\
WO                &       $-$                                   & $3.77\times10^4$	                  \\
\hline
$\tau_\mathrm{LBV,60,0}/\tau_\mathrm{O,60,0}=0.073$ &&\\
$\tau_\mathrm{WR,60,0}/\tau_\mathrm{O,60,0}=0.120$ &&\\
$\tau_\mathrm{LBV,60,0}/\tau_\mathrm{WR,60,0}=0.594$ &&\\
\hline
\hline
\end{tabular}
\end{center}
\end{table}

\section{\label{origspec} The origin of the different spectroscopic classes of massive stars}

In this paper we analyze the evolution of the interior and spectra of a star with only one initial mass (60~\msun) and rotation speed (0~\kms). Nevertheless, we are already able to make inferences about what causes different spectroscopic classes to appear during the evolution of a massive star. In particular, we argue that a given spectroscopic phase can be linked to different evolutionary phases depending on initial mass, rotation, and metallicity. We stress that the conclusions presented here are valid at solar metallicity and are dependent on the mass-loss rates across the evolution, since the spectral morphology of massive stars is significantly affected by \mdot. Moreover, the conclusions do not cover all possible single star evolutionary scenarios, and will be augmented as the analysis of models with different initial masses, rotation, and metallicity becomes available. 

The O-star phase seems to be exclusively linked to the H-core burning evolutionary phase, i.e., all O stars are in the MS. This is a direct result of the stellar structure and evolution equations, which determines the value of \teff\ and \lstar\ for a given \mstar. The fact that the star appears as an O-type star during the MS  is also dependent on \mdot.

The BHG phase can be linked either to the end portion of the MS or to post-MS evolutionary phases, depending on the initial mass and rotation. For a 60~\msun\ star without rotation,  the BHG spectroscopic phase is intimately linked to the crossing from the hot side to the cool side of the first bistability limit of line-driven winds around $\teff=21000-25000~\K$. When this occurs, \mdot\ increases by a factor of $\sim10$ according to the \citet{vink01} recipe, and a BHG spectrum appears. For lower initial masses (e.g., 20--25~\msun\ with rotation), we found that the BHG spectroscopic phase occurs only after the RSG phase, during He-core burning \citep{gme13}. This result was obtained using the same technique of combined stellar evolution and atmospheric modeling to obtain spectral types described in this paper. Because BHGs are pre- or post-RSGs, this illustrates that a given spectroscopic phase can be linked to different evolutionary phases, and using spectroscopic phases as a synonym for evolutionary phases will lead to confusion.

Likewise, the LBV spectroscopic phase also seem to be linked to different evolutionary stages, depending on the initial mass and rotation. In the case of a 60~\msun\ star without rotation, our models predict that LBVs are linked to the end of the MS (i.e., H-core burning), H-shell burning, and beginning of He-core burning, before the star evolves and show a WR spectra. We find that the LBV phase is also linked with the star having a high \mdot\ due to the crossing of the bistability limit. However, if the initial mass is lower, LBVs can appear only at the advanced stages of He-core, C-core burning and core collapse ($\sim20-25$~\msun; \citealt{gme13}; see also, e.g., \citealt{kv06}).

Concerning the WN phase with H, the non-rotating 60~\msun\ model indicates that this phase appears at beginning of He-core burning, when there is still a shell of H burning on top of the He core. Since at this stage $X_\mathrm{sur}=0.13$, a spectral type of WN(h) is warranted \footnote{We thank P. Crowther for this remark.}. This is in contrast with the hydrogen-rich WNh spectroscopic phase, which seems to appear during the MS if the initial mass and/or \mdot\ are high enough (likely above 100~\msun; \citealt{dekoter97,martins08,sc08,crowther10}).

\begin{figure*}
\center
\resizebox{0.79\hsize}{!}{\includegraphics{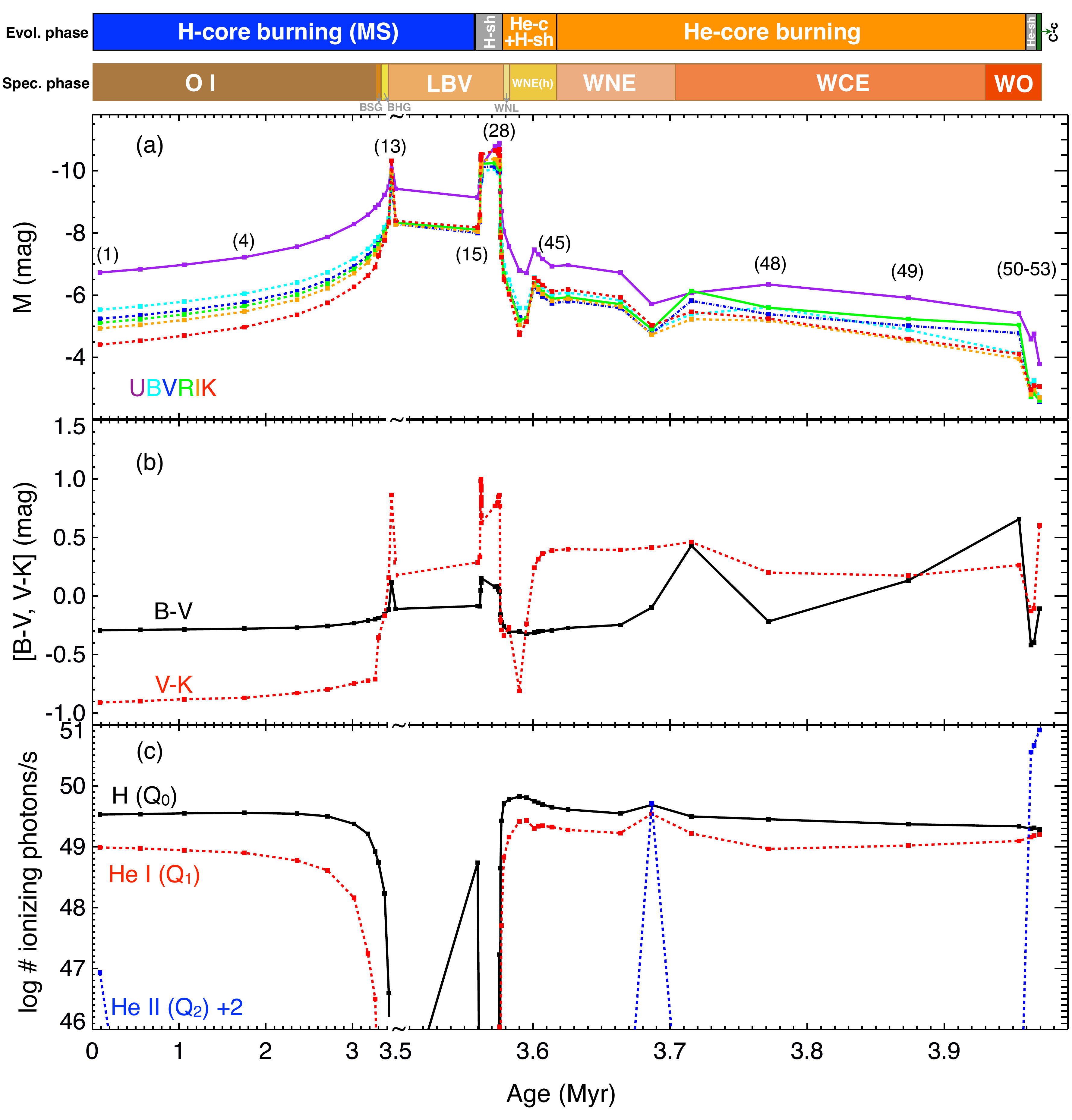}}
\caption{\label{magevol} Evolution of absolute magnitudes (a), $B-V$ and $V-K$ colors (b), and number of \ion{H}{}, \ion{He}{i}, and \ion{He}{ii} ionizing photons emitted per second (c) throughout the lifetime of a non-rotating 60~\msun\ star. At the top we indicate the duration of evolutionary and spectroscopic phases as reference. To make the post-MS evolution clearer, the ordinate scale was changed at 3.5 Myr.}
\end{figure*}

\section{\label{specevol} Evolution of the absolute magnitudes, colors, bolometric correction, and ionizing fluxes }

Our combined stellar evolution and atmospheric modeling allows us to investigate the temporal evolution of the observables of massive stars, such as the spectrum in different wavelength ranges, absolute magnitudes in various filters, colors, bolometric corrections, ionizing fluxes, and \teff. This complements the output from classical stellar evolution models \citep[e.g.,][]{ekstrom12}, which provide the evolution of \teffg\ and \lstar. In the future, the temporal evolution of the observables will be used as input to produce theoretical isochrones and, in more general terms, as input to stellar population synthesis models.

Our models provide as output the high-resolution spectrum from the extreme UV to radio wavelengths. Synthetic photometry was performed using the Chorizos code \citep{chorizos04}, adopting its built-in passband and zero point definitions that were obtained from \citet{cohen03,ma06,holberg06,ma07}.

Following \citet{gmg13}, the absolute magnitudes ($M_P$)  in the modified Vega magnitude system,  for a given filter $P$, are
\begin{equation}
M_{P} =  -2.5\log_{10}\left(\frac{\int P(\lambda)F_{\lambda}(\lambda)\lambda\,d\lambda}
                         {\int P(\lambda)F_{\lambda,\mathrm{Vega}}(\lambda)\lambda\,d\lambda}\right)
                       + {\rm ZP}_{P},
\label{absmageq}
\end{equation}
where  $\lambda$ is the wavelength, $P(\lambda)$ is the sensitivity curve of the system, $F_{\lambda}$ is the model flux at 10 pc, $F_{\lambda,\mathrm{Vega}}$ is the flux of Vega scaled to a distance of 10pc, and ${\rm ZP}_{P}$ is the zero point.

We use the usual relationship to compute the bolometric magnitudes (\mbol), assuming that the solar \mbol\ is 4.74 mag, 
\begin{equation}
M_{bol} =   -2.5\log_{10} (\lstar/\lsun) + 4.74.
\label{mboleq}
\end{equation}
To obtain bolometric corrections in a given filter $P$ (BC$_P$), we use
\begin{equation}
\mathrm{BC}_{P} = \mbol - M_{P}.
\label{bceq}
\end{equation}
Values of BC$_P$ for each of the 53 stages discussed here are shown in Tables~\ref{absmagbc1} and \ref{absmagbc2}.

Massive stars are the main sources of input of ionizing photons into the interstellar medium. Because stars evolve, the number of ionizing photons emitted by a massive star is expected to vary as the surface conditions change during the evolution. Here we quantify this effect by computing the number of photons capable of ionizing \ion{H}{} ($Q_0$), \ion{He}{i} ($Q_1$), and \ion{He}{ii} ($Q_2$) as follows,

\begin{equation}
Q_i=4\pi R_{\star} \int^{\lambda_i}_0 \frac{ \pi \lambda F_\lambda }{hc} \mathrm{d}\lambda \,,
\end{equation}
where $F_\lambda$ is the stellar flux at \rstar\ and $\lambda_\mathrm{0}$, $\lambda_\mathrm{1}$, and $\lambda_\mathrm{2}$ are the wavelength corresponding to the ionization edges of  \ion{H}{} (912~{\AA}), \ion{He}{i}  (574~{\AA}), and \ion{He}{ii}  (228~{\AA}), respectively.

One of the advantages of our models is the flexibility to produce synthetic photometry in different filter systems, avoiding errors that stem from converting between different magnitude systems. Here we quote results in the Johnson-Cousins $UBVRI$ and 2MASS $J$, $H$, and $K_S$ (Table \ref{absmagbc1}), and the  {\it Hubble Space Telescope (HST)}/Wide Field Planetary Camera 2 (WFPC2) $F170W$, $F300W$, $F450W$, $F606W$, $F814W$ filters (Table \ref{absmagbc2}).

The evolution of the non-rotating 60~\msun\ star analyzed in this paper covers a large range of surface temperatures (6700 to 225000~\K; Fig. \ref{hrd1}a), luminosities (0.3 to $1.0\times10^6~\lsun$; Fig. \ref{hrd1}a), mass-loss rates ($10^{-5}$ to $10^{-3}$~\msunyr; Fig. \ref{hrd1}e), and surface abundances (Fig. \ref{hrd1}d). As such, we would expect significant variations of the spectral energy distribution, absolute magnitudes, and colors throughout the evolution.

%Figure \ref{sedevol} shows the spectral energy distribution of the non-rotating 60~\msun\ star at selected timesteps along its evolution. The shape of the spectral energy distribution depends on \teff, \lstar, and wind density via \mdot\ and \vinf.

Figure \ref{magevol}a shows the absolute magnitudes in the WFPC2/$F170W$ and $UBVRIJHK_S$ filters throughout the evolution of a non-rotating 60~\msun\ star. Indeed, one can readily see that absolute magnitudes present strong variations of up to 6 mag during the star's lifetime. The behavior of the absolute magnitudes and bolometric corrections as a function of time is then regulated by how much flux from the star falls within the passband of a given filter. This depends on the stellar (\teff\ and \lstar) and wind properties (\mdot\ and \vinf). Let us analyze the absolute magnitudes and how they broadly vary during the evolution. For this purpose, we refer to the stages  overplotted in Fig. \ref{magevol} (as defined in Fig.~\ref{hrd1}).

During the majority of the MS evolution (stages 1 through 13), there is a brightening in all UV, optical and near-IR filters. This is ultimately caused by the core contraction due to H being burnt into He, which causes an expansion of the envelope and increase of the global mean molecular weight and opacity. As a result, \lstar\ increases and \teff\ decreases, which shifts the spectral energy distribution (SED) towards longer wavelengths and increase the flux in the $UBVRIJHK$ filters. During the OIII and OI spectroscopic phases, the star is much brighter in $UBV$ than $JHK$. When it becomes an LBV (stage 13), the color indexes become closer to 0. At the final portion of the MS (stages 13 through 15), the model becomes fainter in all UV, optical and near-IR filters. This is caused by the increase in $\teff$ as the model returns to the blue. 
 
When the H-shell burning stage begins (stage 15), the star becomes brighter again, reaching its maximum brightness at stage 28.  This is caused by the contraction of the core, which is not burning H anymore. As a result of the rapid core contraction, the envelope expands with constant \lstar\ and decreasing $\teff$, shifting the SED to longer wavelengths and increasing the flux in the optical and near-IR filters.

When He-core burning begins (stage 28), there is another rapid fainting of the model up to when the star becomes a WNEh (stage 37). The ultimate reason is the huge mass loss, which removes the outer layers of the star and decreases the size of the H-shell burning layer. This decreases the inflation of the stellar envelope, increasing $\teff$ and making the star fainter in the optical and near-IR.

 During He-core burning (stages 28 to 51), there is a progressive decrease in the optical an near-IR fluxes as the star becomes hotter and less luminous. This is ultimately caused by mass loss, which diminishes $\mstar$ and, as a consequence, $\lstar$ as well. Mass loss also removes the outer layers of the star, making $\teff$ higher. These factors contribute to shifting the SED to the blue and decreasing the brightness in the optical and near-IR. The decrease in the $U$ filter is much less pronounced than in the other filters. The star becomes even fainter at the end of He-core burning (stages 49 through 50), which is caused by the overall contraction of the star.

From the end of He-core burning (stage 50) until core collapse (stage 53), there is a significant decrease in the optical and near-IR fluxes. This is caused by the contraction of the star as He-core burning ends. There are two competing effects happening. There is an increase in $\lstar$, which has the tendency to make the star brighter in the optical and near-IR filters, but there is an increase in \teff\ as well, which has the tendency to make the star fainter in these filters. We found that the $\teff$ effect dominates and the star becomes much fainter at core collapse than it was at the end stages of He-core burning, This phenomenon is relevant for investigating the detectability of massive stars just before the SN explosion, and explains the overall faintness of SN Ic progenitors \citep[][see also \citealt{yoon12}]{gmg13}.

The variation of the $B-V$ and $V-K$ colors as a function of age is displayed in Fig. \ref{magevol}b. We see that $B-V$ and $V-K$ are roughly constant during the O-type phase, and rapidily change when the star reach the LBV phase. From the beginning of He-core burning onwards, these broadband colors are affected by the presence of emission lines, which dominate the flux in these filters. Therefore, erratic variations are seen depending on the stellar parameters when the star is a WR.

Figure \ref{magevol}c shows the evolution of the number of  \ion{H}{} ($Q_0$), \ion{He}{i} ($Q_1$), and \ion{He}{ii} ($Q_2$) ionizing photons emitted per unit time. We found that $Q_0$ is roughly constant until the end of the MS ($Q_0=10^{49.5}$~photon/s), when the star becomes an LBV with low $\teff$. Our models indicate that $Q_0$ increases again at the beginning of He-core burning, remaining roughly constant until the pre-SN stage.$Q_1$ follows a similar qualitative trend, while $Q_2$ rapidly decreases during the MS and only becomes significant at the transition between the WNE and WCE phase, and before core-collapse, when the star is a WO. However, we stress that the behavior of $Q_2$ and the wind ionization structure are thought to be severely affected by the presence of x-rays \citep{hillier93,pauldrach94,feldmeier97}. Here we include the effects of x-rays only for O-type stars  (see Sect. \ref{cmfgen}), so the values of $Q_2$ should be taken as lower limits during the WR stages.

\begin{sidewaystable*}
%\begin{table*}
\begin{minipage}{\textwidth}
\scriptsize
%\tiny
%\footnotesize
\caption{Evolution of the absolute magnitudes and bolometric corrections of a non-rotating 60~\msun\ star in the Johnson-Cousins $UBVRI$ and 2MASS $JHK_S$ filters.}
\label{absmagbc1}
\centering
\vspace{0.1cm}
\begin{tabular}{l c c c c c c c c c c c c c c c c c c}
%Non-rotating models\\
\hline\hline
Stage & Evol. phase &  Sp. Type & $M_U$& $M_B$ & $M_V$ & $M_R$ & $M_I$ & $M_J$ & $M_H$ & $M_K$ & BC$_U$& BC$_B$ & BC$_V$ & BC$_R$ & BC$_I$ &  $BC_J$ & $BC_H$ & $BC_K$  \\ 
& & & (mag) & (mag)& (mag)& (mag)& (mag)& (mag)& (mag)& (mag)& (mag)& (mag) & (mag)& (mag)& (mag)& (mag)& (mag) & (mag)\\
\\  \hline
 \hline
% Johnson-Cousins + JHK filters
     1 & H-c & O3If$^*$c &  -6.73 &  -5.53 &  -5.24 &  -5.11 &  -4.93 &  -5.29 &  -4.55 &  -4.41 &  -2.79 &  -3.99 &  -4.28 &  -4.41 &  -4.59 &  -4.23 &  -4.97 &  -5.11 \\
      2 & H-c & O3If$^*$c &  -6.83 &  -5.64 &  -5.35 &  -5.23 &  -5.05 &  -5.41 &  -4.68 &  -4.53 &  -2.73 &  -3.92 &  -4.21 &  -4.34 &  -4.52 &  -4.16 &  -4.89 &  -5.03 \\
      3 & H-c & O3If$^*$c &  -6.98 &  -5.79 &  -5.51 &  -5.38 &  -5.20 &  -5.56 &  -4.84 &  -4.70 &  -2.64 &  -3.83 &  -4.11 &  -4.24 &  -4.42 &  -4.06 &  -4.78 &  -4.92 \\
      4 & H-c & O4If$^*$c &  -7.22 &  -6.05 &  -5.77 &  -5.65 &  -5.47 &  -5.82 &  -5.11 &  -4.97 &  -2.48 &  -3.66 &  -3.93 &  -4.05 &  -4.23 &  -3.88 &  -4.59 &  -4.73 \\
      5 & H-c & O5Ifc &  -7.55 &  -6.40 &  -6.13 &  -6.02 &  -5.85 &  -6.18 &  -5.51 &  -5.37 &  -2.23 &  -3.38 &  -3.65 &  -3.77 &  -3.93 &  -3.60 &  -4.28 &  -4.41 \\
      6 & H-c & O6Iafc &  -7.87 &  -6.73 &  -6.48 &  -6.37 &  -6.22 &  -6.53 &  -5.88 &  -5.75 &  -1.97 &  -3.11 &  -3.36 &  -3.47 &  -3.62 &  -3.31 &  -3.96 &  -4.09 \\
      7 & H-c & O7.5Iafc &  -8.28 &  -7.17 &  -6.94 &  -6.84 &  -6.71 &  -6.99 &  -6.39 &  -6.27 &  -1.62 &  -2.73 &  -2.96 &  -3.06 &  -3.20 &  -2.92 &  -3.51 &  -3.64 \\
      8 & H-c & O9Iab &  -8.58 &  -7.49 &  -7.28 &  -7.19 &  -7.06 &  -7.32 &  -6.76 &  -6.63 &  -1.35 &  -2.45 &  -2.66 &  -2.75 &  -2.88 &  -2.62 &  -3.18 &  -3.31 \\  
      9 & H-c & B0.2 Ia &  -8.81 &  -7.73 &  -7.54 &  -7.44 &  -7.31 &  -7.57 &  -7.02 &  -6.90 &  -1.14 &  -2.22 &  -2.42 &  -2.51 &  -2.64 &  -2.38 &  -2.94 &  -3.06 \\
     10 & H-c & B0.5 Ia$^+$ &  -8.91 &  -7.86 &  -7.67 &  -7.64 &  -7.50 &  -7.72 &  -7.32 &  -7.27 &  -1.06 &  -2.10 &  -2.29 &  -2.32 &  -2.46 &  -2.24 &  -2.64 &  -2.69 \\
     11 & H-c & hot LBV &  -9.22 &  -8.19 &  -8.03 &  -8.02 &  -7.91 &  -8.08 &  -7.78 &  -7.77 &  -0.75 &  -1.78 &  -1.94 &  -1.95 &  -2.06 &  -1.89 &  -2.19 &  -2.20 \\
     12 & H-c & hot LBV &  -9.48 &  -8.47 &  -8.35 &  -8.37 &  -8.31 &  -8.39 &  -8.29 &  -8.35 &  -0.50 &  -1.51 &  -1.63 &  -1.61 &  -1.67 &  -1.59 &  -1.69 &  -1.63 \\
     13 & H-c & cool LBV & -10.31 &  -9.53 &  -9.64 &  -9.83 &  -9.96 &  -9.67 & -10.19 & -10.32 &   0.36 &  -0.43 &  -0.31 &  -0.13 &   0.01 &  -0.29 &   0.23 &   0.36 \\
     14 & H-c & hot LBV &  -9.42 &  -8.39 &  -8.28 &  -8.33 &  -8.28 &  -8.32 &  -8.30 &  -8.39 &  -0.51 &  -1.55 &  -1.66 &  -1.61 &  -1.66 &  -1.61 &  -1.63 &  -1.55 \\
     15 & H-c & hot LBV &  -9.14 &  -8.08 &  -8.00 &  -8.10 &  -8.04 &  -8.05 &  -8.11 &  -8.18 &  -0.90 &  -1.96 &  -2.04 &  -1.93 &  -2.00 &  -1.99 &  -1.93 &  -1.86 \\
     16 & H-sh & hot LBV &  -9.49 &  -8.44 &  -8.35 &  -8.44 &  -8.39 &  -8.40 &  -8.47 &  -8.58 &  -0.56 &  -1.60 &  -1.69 &  -1.61 &  -1.65 &  -1.64 &  -1.57 &  -1.46 \\
     17 & H-sh & cool LBV & -10.46 &  -9.62 &  -9.67 &  -9.86 &  -9.98 &  -9.71 & -10.19 & -10.37 &   0.40 &  -0.43 &  -0.39 &  -0.19 &  -0.08 &  -0.35 &   0.14 &   0.31 \\
     18 & H-sh & cool LBV & -10.43 &  -9.71 &  -9.85 & -10.02 & -10.17 &  -9.87 & -10.40 & -10.53 &   0.37 &  -0.35 &  -0.21 &  -0.04 &   0.11 &  -0.19 &   0.34 &   0.47 \\
     19 & H-sh & cool LBV & -10.42 &  -9.70 &  -9.83 &  -9.99 & -10.13 &  -9.85 & -10.35 & -10.47 &   0.36 &  -0.36 &  -0.24 &  -0.07 &   0.06 &  -0.22 &   0.28 &   0.41 \\
     20 & H-sh & cool LBV & -10.42 &  -9.71 &  -9.83 & -10.00 & -10.13 &  -9.86 & -10.35 & -10.48 &   0.35 &  -0.35 &  -0.23 &  -0.07 &   0.07 &  -0.21 &   0.29 &   0.41 \\
     21 & H-sh & cool LBV & -10.42 &  -9.72 &  -9.84 & -10.00 & -10.14 &  -9.86 & -10.35 & -10.48 &   0.35 &  -0.35 &  -0.23 &  -0.07 &   0.07 &  -0.20 &   0.29 &   0.41 \\
     22 & H-sh & cool LBV & -10.42 &  -9.74 &  -9.86 & -10.01 & -10.15 &  -9.88 & -10.36 & -10.48 &   0.35 &  -0.33 &  -0.21 &  -0.06 &   0.08 &  -0.19 &   0.29 &   0.41 \\
     23 & H-sh & cool LBV & -10.40 &  -9.80 &  -9.92 & -10.06 & -10.19 &  -9.94 & -10.38 & -10.50 &   0.33 &  -0.27 &  -0.15 &  -0.01 &   0.12 &  -0.13 &   0.31 &   0.43 \\
     24 & H-sh & cool LBV & -10.38 &  -9.83 &  -9.95 & -10.08 & -10.21 &  -9.97 & -10.39 & -10.50 &   0.31 &  -0.24 &  -0.12 &   0.01 &   0.14 &  -0.10 &   0.32 &   0.43 \\
     25 & H-sh & cool LBV & -10.36 &  -9.86 &  -9.97 & -10.09 & -10.22 &  -9.99 & -10.39 & -10.49 &   0.29 &  -0.22 &  -0.10 &   0.02 &   0.15 &  -0.08 &   0.31 &   0.42 \\
     26 & H-sh & cool LBV & -10.24 &  -9.92 & -10.06 & -10.17 & -10.29 & -10.08 & -10.42 & -10.51 &   0.17 &  -0.15 &  -0.01 &   0.10 &   0.22 &   0.01 &   0.35 &   0.44 \\
     27 & H-sh & cool LBV & -10.16 &  -9.97 & -10.12 & -10.22 & -10.33 & -10.14 & -10.45 & -10.52 &   0.08 &  -0.10 &   0.05 &   0.15 &   0.26 &   0.06 &   0.38 &   0.44 \\
     28 & He-c+H-sh & cool LBV & -10.78 & -10.06 & -10.14 & -10.25 & -10.38 & -10.16 & -10.52 & -10.65 &   0.49 &  -0.23 &  -0.15 &  -0.04 &   0.09 &  -0.13 &   0.23 &   0.36 \\
     29 & He-c+H-sh & cool LBV & -10.77 & -10.02 & -10.10 & -10.21 & -10.35 & -10.12 & -10.50 & -10.63 &   0.50 &  -0.25 &  -0.17 &  -0.05 &   0.08 &  -0.15 &   0.23 &   0.36 \\
     30 & He-c+H-sh & cool LBV & -10.76 & -10.01 & -10.09 & -10.20 & -10.33 & -10.10 & -10.48 & -10.62 &   0.50 &  -0.26 &  -0.18 &  -0.06 &   0.07 &  -0.16 &   0.22 &   0.35 \\
     31 & He-c+H-sh & cool LBV & -10.76 &  -9.94 &  -9.99 & -10.11 & -10.24 & -10.02 & -10.42 & -10.57 &   0.51 &  -0.31 &  -0.25 &  -0.14 &  -0.01 &  -0.23 &   0.17 &   0.32 \\
     32 & He-c+H-sh & cool LBV & -10.89 & -10.07 & -10.11 & -10.24 & -10.35 & -10.13 & -10.52 & -10.68 &   0.65 &  -0.17 &  -0.13 &   0.00 &   0.12 &  -0.10 &   0.29 &   0.45 \\
     33 & He-c+H-sh & cool LBV & -10.69 &  -9.92 &  -9.96 & -10.06 & -10.18 &  -9.98 & -10.34 & -10.49 &   0.47 &  -0.30 &  -0.26 &  -0.16 &  -0.04 &  -0.24 &   0.12 &   0.27 \\
     34 & He-c+H-sh & WN11(h) &  -9.33 &  -8.23 &  -8.07 &  -8.08 &  -7.93 &  -8.12 &  -7.85 &  -7.86 &  -0.88 &  -1.98 &  -2.14 &  -2.13 &  -2.28 &  -2.09 &  -2.36 &  -2.35 \\
     35 & He-c, H-sh & WN8(h) &  -8.70 &  -7.62 &  -7.39 &  -7.40 &  -7.30 &  -7.47 &  -7.20 &  -7.22 &  -1.49 &  -2.58 &  -2.80 &  -2.80 &  -2.90 &  -2.73 &  -2.99 &  -2.98 \\
     36 & He-c, H-sh & WN7(h) &  -8.06 &  -6.97 &  -6.71 &  -6.69 &  -6.58 &  -6.80 &  -6.48 &  -6.50 &  -2.12 &  -3.21 &  -3.47 &  -3.49 &  -3.60 &  -3.38 &  -3.70 &  -3.68 \\
     37 & He-c, H-sh & WN5(h) &  -7.57 &  -6.49 &  -6.18 &  -6.16 &  -6.05 &  -6.30 &  -6.00 &  -6.03 &  -2.60 &  -3.68 &  -3.98 &  -4.00 &  -4.12 &  -3.87 &  -4.17 &  -4.14 \\
     38 & He-c, H-sh & WN2(h) &  -6.79 &  -5.59 &  -5.29 &  -5.19 &  -5.04 &  -5.37 &  -4.79 &  -4.73 &  -3.35 &  -4.55 &  -4.86 &  -4.95 &  -5.10 &  -4.78 &  -5.36 &  -5.41 \\
     39 & He-c, H-sh & WN5(h) &  -6.72 &  -5.59 &  -5.26 &  -5.26 &  -5.13 &  -5.41 &  -5.11 &  -5.15 &  -3.41 &  -4.54 &  -4.86 &  -4.86 &  -4.99 &  -4.72 &  -5.01 &  -4.98 \\
     40 & He-c, H-sh & WN5(h) &  -7.46 &  -6.58 &  -6.26 &  -6.38 &  -6.32 &  -6.44 &  -6.47 &  -6.55 &  -2.64 &  -3.53 &  -3.84 &  -3.73 &  -3.79 &  -3.66 &  -3.64 &  -3.56 \\
     41 & He-c, H-sh & WN5(h) &  -7.31 &  -6.42 &  -6.12 &  -6.25 &  -6.19 &  -6.31 &  -6.36 &  -6.45 &  -2.75 &  -3.64 &  -3.95 &  -3.82 &  -3.88 &  -3.75 &  -3.70 &  -3.61 \\
     42 & He-c, H-sh & WN5(h) &  -7.16 &  -6.26 &  -5.96 &  -6.10 &  -6.04 &  -6.17 &  -6.23 &  -6.33 &  -2.85 &  -3.75 &  -4.05 &  -3.91 &  -3.97 &  -3.84 &  -3.77 &  -3.68 \\
     43 & He-c, H-sh & WN5(h) &  -6.93 &  -6.03 &  -5.73 &  -5.88 &  -5.81 &  -5.96 &  -6.02 &  -6.11 &  -2.97 &  -3.87 &  -4.16 &  -4.02 &  -4.09 &  -3.94 &  -3.88 &  -3.79 \\
     44 & He-c, H-sh & WN5 &  -6.96 &  -6.08 &  -5.81 &  -5.93 &  -5.88 &  -6.02 &  -6.08 &  -6.18 &  -2.88 &  -3.77 &  -4.04 &  -3.91 &  -3.97 &  -3.82 &  -3.76 &  -3.66 \\
     45 & He-c & WN5 &  -6.72 &  -5.83 &  -5.59 &  -5.71 &  -5.63 &  -5.80 &  -5.83 &  -5.93 &  -2.96 &  -3.84 &  -4.09 &  -3.97 &  -4.05 &  -3.88 &  -3.85 &  -3.75 \\
     46 & He-c & WO4 &  -5.71 &  -4.86 &  -4.76 &  -4.91 &  -4.73 &  -4.98 &  -4.94 &  -5.03 &  -3.89 &  -4.74 &  -4.84 &  -4.69 &  -4.87 &  -4.62 &  -4.66 &  -4.57 \\
     47 & He-c & WC4 &  -6.07 &  -5.39 &  -5.82 &  -6.14 &  -5.23 &  -6.01 &  -5.38 &  -5.46 &  -3.45 &  -4.13 &  -3.70 &  -3.39 &  -4.30 &  -3.51 &  -4.15 &  -4.06 \\
     48 & He-c & WC4 &  -6.34 &  -5.61 &  -5.39 &  -5.60 &  -5.18 &  -5.74 &  -5.18 &  -5.25 &  -3.01 &  -3.75 &  -3.97 &  -3.76 &  -4.17 &  -3.62 &  -4.18 &  -4.11 \\
     49 & He-c & WC4 &  -5.91 &  -4.88 &  -5.01 &  -5.23 &  -4.55 &  -5.25 &  -4.51 &  -4.60 &  -3.21 &  -4.24 &  -4.11 &  -3.89 &  -4.57 &  -3.87 &  -4.61 &  -4.53 \\
     50 & He-c & WO4 &  -5.41 &  -4.13 &  -4.78 &  -5.04 &  -3.95 &  -4.90 &  -4.05 &  -4.11 &  -3.65 &  -4.94 &  -4.28 &  -4.02 &  -5.11 &  -4.16 &  -5.02 &  -4.96 \\
     51 & He-c & WO1 &  -4.58 &  -3.14 &  -2.72 &  -2.73 &  -2.81 &  -2.88 &  -2.87 &  -2.96 &  -4.55 &  -6.00 &  -6.41 &  -6.40 &  -6.33 &  -6.26 &  -6.27 &  -6.18 \\
     52 & He-sh & WO1 &  -4.75 &  -3.26 &  -2.86 &  -2.83 &  -2.97 &  -2.98 &  -2.98 &  -3.09 &  -4.49 &  -5.98 &  -6.38 &  -6.41 &  -6.27 &  -6.26 &  -6.26 &  -6.15 \\
     53 & end C-c & WO1 &  -3.79 &  -2.69 &  -2.58 &  -2.61 &  -2.71 &  -2.66 &  -2.85 &  -3.07 &  -5.68 &  -6.78 &  -6.89 &  -6.86 &  -6.76 &  -6.81 &  -6.62 &  -6.40 \\
\hline
\end{tabular}
\end{minipage}
%\end{table*}
\end{sidewaystable*}

\begin{sidewaystable*}
%\begin{table*}
\begin{minipage}{\textwidth}
\scriptsize
%\tiny
%\footnotesize
\caption{Evolution of the absolute magnitudes and bolometric corrections of a non-rotating 60~\msun\ star in the {\it HST}/WFPC2 F170, F336W, F450W, F555W, F606W, and F814W filters.}
\label{absmagbc2}
\centering
\vspace{0.1cm}
\begin{tabular}{l c c c c c c c c c c c c c c}
%Non-rotating models\\
\hline\hline
Stage & Evol. phase &  Sp. Type &$M_{170}$ & $M_{336}$  & $M_{450}$ & $M_{555}$ & $M_{606}$ & $M_{814}$ & BC$_{170}$ & BC$_{336}$  & BC$_{450}$ & BC$_{555}$ & BC$_{606}$ & BC$_{814}$ \\ 
 & & & (mag) & (mag)& (mag)& (mag)& (mag)& (mag) & (mag)& (mag)& (mag)& (mag)& (mag) & (mag)\\
 \hline
 \hline
% WFPC2 filters
      1 & H-c & O3If$^*$c &  -8.93 &  -7.27 &  -5.52 &  -5.29 &  -5.21 &  -4.95&  -0.59 &  -2.26 &  -4.00 &  -4.23 &  -4.31 &  -4.58 \\
      2 & H-c & O3If$^*$c &  -9.01 &  -7.37 &  -5.63 &  -5.41 &  -5.33 &  -5.06&  -0.55 &  -2.20 &  -3.93 &  -4.16 &  -4.24 &  -4.50 \\
      3 & H-c & O3If$^*$c &  -9.13 &  -7.51 &  -5.78 &  -5.56 &  -5.49 &  -5.22&  -0.49 &  -2.11 &  -3.84 &  -4.06 &  -4.14 &  -4.40 \\
      4 & H-c & O4If$^*$c &  -9.32 &  -7.75 &  -6.04 &  -5.82 &  -5.75 &  -5.49&  -0.38 &  -1.95 &  -3.66 &  -3.88 &  -3.95 &  -4.21 \\
      5 & H-c & O5Ifc &  -9.57 &  -8.08 &  -6.40 &  -6.18 &  -6.12 &  -5.87&  -0.22 &  -1.70 &  -3.39 &  -3.60 &  -3.67 &  -3.92 \\
      6 & H-c & O6Iafc &  -9.79 &  -8.39 &  -6.73 &  -6.53 &  -6.46 &  -6.23&  -0.05 &  -1.45 &  -3.11 &  -3.31 &  -3.38 &  -3.61 \\
      7 & H-c & O7.5Iafc & -10.07 &  -8.80 &  -7.17 &  -6.99 &  -6.93 &  -6.72&   0.16 &  -1.11 &  -2.73 &  -2.92 &  -2.97 &  -3.18 \\  
      9 & H-c & B0.2 Ia & -10.38 &  -9.31 &  -7.74 &  -7.57 &  -7.53 &  -7.33&   0.42 &  -0.65 &  -2.22 &  -2.38 &  -2.43 &  -2.63 \\
     10 & H-c & B0.5 Ia$^+$ & -10.26 &  -9.39 &  -7.87 &  -7.72 &  -7.71 &  -7.52&   0.30 &  -0.57 &  -2.09 &  -2.24 &  -2.25 &  -2.44 \\
     11 & H-c & hot LBV & -10.27 &  -9.69 &  -8.20 &  -8.08 &  -8.08 &  -7.92&   0.30 &  -0.28 &  -1.77 &  -1.89 &  -1.89 &  -2.05 \\
     12 & H-c & hot LBV & -10.26 &  -9.94 &  -8.49 &  -8.39 &  -8.41 &  -8.33&   0.28 &  -0.04 &  -1.49 &  -1.59 &  -1.57 &  -1.65 \\
     13 & H-c & cool LBV &  -8.75 & -10.61 &  -9.59 &  -9.67 &  -9.79 &  -9.96&  -1.21 &   0.66 &  -0.36 &  -0.29 &  -0.17 &   0.01 \\
     14 & H-c & hot LBV & -10.25 &  -9.88 &  -8.42 &  -8.32 &  -8.36 &  -8.29&   0.32 &  -0.06 &  -1.52 &  -1.61 &  -1.58 &  -1.64 \\
     15 & H-c & hot LBV & -10.24 &  -9.61 &  -8.12 &  -8.05 &  -8.11 &  -8.05&   0.20 &  -0.43 &  -1.92 &  -1.99 &  -1.93 &  -1.98 \\
     16 & H-sh & hot LBV & -10.37 &  -9.95 &  -8.47 &  -8.40 &  -8.45 &  -8.41&   0.33 &  -0.09 &  -1.57 &  -1.64 &  -1.59 &  -1.64 \\
     17 & H-sh & cool LBV &  -9.21 & -10.79 &  -9.68 &  -9.71 &  -9.84 &  -9.97&  -0.84 &   0.74 &  -0.38 &  -0.35 &  -0.22 &  -0.08 \\
     18 & H-sh & cool LBV &  -8.76 & -10.68 &  -9.78 &  -9.87 &  -9.98 & -10.17&  -1.30 &   0.61 &  -0.29 &  -0.19 &  -0.08 &   0.11 \\
     19 & H-sh & cool LBV &  -8.65 & -10.67 &  -9.76 &  -9.85 &  -9.96 & -10.13&  -1.41 &   0.61 &  -0.30 &  -0.22 &  -0.11 &   0.06 \\
     20 & H-sh & cool LBV &  -8.61 & -10.65 &  -9.77 &  -9.86 &  -9.96 & -10.14&  -1.45 &   0.59 &  -0.29 &  -0.21 &  -0.10 &   0.07 \\
     21 & H-sh & cool LBV &  -8.59 & -10.65 &  -9.78 &  -9.86 &  -9.97 & -10.14&  -1.47 &   0.58 &  -0.29 &  -0.20 &  -0.10 &   0.07 \\
     22 & H-sh & cool LBV &  -8.55 & -10.63 &  -9.80 &  -9.88 &  -9.98 & -10.15&  -1.52 &   0.56 &  -0.27 &  -0.19 &  -0.09 &   0.08 \\
     23 & H-sh & cool LBV &  -8.44 & -10.54 &  -9.86 &  -9.94 & -10.03 & -10.19&  -1.63 &   0.47 &  -0.21 &  -0.13 &  -0.04 &   0.12 \\
     24 & H-sh & cool LBV &  -8.38 & -10.50 &  -9.89 &  -9.97 & -10.05 & -10.21&  -1.69 &   0.43 &  -0.18 &  -0.10 &  -0.02 &   0.14 \\
     25 & H-sh & cool LBV &  -8.32 & -10.44 &  -9.91 &  -9.99 & -10.07 & -10.22&  -1.75 &   0.37 &  -0.16 &  -0.08 &   0.00 &   0.15 \\
     26 & H-sh & cool LBV &  -7.92 & -10.18 &  -9.99 & -10.08 & -10.16 & -10.30&  -2.15 &   0.10 &  -0.08 &   0.01 &   0.08 &   0.23 \\
     27 & H-sh & cool LBV &  -7.77 &  -9.98 & -10.04 & -10.14 & -10.21 & -10.34&  -2.30 &  -0.10 &  -0.03 &   0.06 &   0.13 &   0.27 \\
     28 & He-c+H-sh & cool LBV &  -8.66 & -10.98 & -10.10 & -10.16 & -10.23 & -10.38&  -1.63 &   0.69 &  -0.19 &  -0.13 &  -0.06 &   0.09 \\
     29 & He-c+H-sh & cool LBV &  -8.70 & -10.98 & -10.06 & -10.12 & -10.20 & -10.35&  -1.57 &   0.72 &  -0.20 &  -0.15 &  -0.07 &   0.09 \\
     30 & He-c+H-sh & cool LBV &  -8.71 & -10.98 & -10.05 & -10.10 & -10.18 & -10.34&  -1.55 &   0.71 &  -0.21 &  -0.16 &  -0.08 &   0.08 \\
     31 & He-c+H-sh & cool LBV &  -9.11 & -11.04 &  -9.98 & -10.02 & -10.09 & -10.24&  -1.14 &   0.80 &  -0.27 &  -0.23 &  -0.16 &  -0.00 \\
     32 & He-c+H-sh & cool LBV &  -8.33 & -11.15 & -10.11 & -10.13 & -10.22 & -10.35&  -1.90 &   0.92 &  -0.13 &  -0.10 &  -0.01 &   0.12 \\
     33 & He-c+H-sh & cool LBV &  -9.00 & -10.99 &  -9.96 &  -9.98 & -10.06 & -10.19&  -1.22 &   0.77 &  -0.26 &  -0.24 &  -0.16 &  -0.03 \\
     34 & He-c+H-sh & WN11(h) & -10.40 &  -9.81 &  -8.25 &  -8.12 &  -8.13 &  -7.96&   0.19 &  -0.40 &  -1.96 &  -2.09 &  -2.08 &  -2.25 \\
     35 & He-c, H-sh & WN8(h) & -10.14 &  -9.22 &  -7.63 &  -7.47 &  -7.44 &  -7.32&  -0.06 &  -0.98 &  -2.57 &  -2.73 &  -2.75 &  -2.88 \\
     36 & He-c, H-sh & WN7(h) &  -9.90 &  -8.58 &  -6.98 &  -6.80 &  -6.75 &  -6.61&  -0.28 &  -1.60 &  -3.20 &  -3.38 &  -3.43 &  -3.57 \\
     37 & He-c, H-sh & WN5(h) &  -9.58 &  -8.12 &  -6.50 &  -6.30 &  -6.24 &  -6.11&  -0.59 &  -2.04 &  -3.66 &  -3.87 &  -3.93 &  -4.06 \\
     38 & He-c, H-sh & WN2(h) &  -8.99 &  -7.34 &  -5.59 &  -5.37 &  -5.29 &  -5.07&  -1.16 &  -2.80 &  -4.55 &  -4.78 &  -4.86 &  -5.07 \\
     39 & He-c, H-sh & WN5(h) &  -8.92 &  -7.30 &  -5.62 &  -5.41 &  -5.34 &  -5.22&  -1.21 &  -2.83 &  -4.50 &  -4.72 &  -4.79 &  -4.91 \\
     40 & He-c, H-sh & WN5(h) &  -9.42 &  -8.04 &  -6.61 &  -6.44 &  -6.41 &  -6.41&  -0.69 &  -2.07 &  -3.49 &  -3.66 &  -3.70 &  -3.70 \\
     41 & He-c, H-sh & WN5(h) &  -9.29 &  -7.90 &  -6.47 &  -6.31 &  -6.27 &  -6.29&  -0.78 &  -2.16 &  -3.59 &  -3.75 &  -3.79 &  -3.77 \\
     42 & He-c, H-sh & WN5(h) &  -9.14 &  -7.76 &  -6.32 &  -6.17 &  -6.13 &  -6.16&  -0.87 &  -2.25 &  -3.68 &  -3.84 &  -3.88 &  -3.85 \\
     43 & He-c, H-sh & WN5(h) &  -8.92 &  -7.55 &  -6.10 &  -5.96 &  -5.91 &  -5.94&  -0.98 &  -2.35 &  -3.80 &  -3.94 &  -3.99 &  -3.96 \\
     44 & He-c, H-sh & WN5 &  -8.96 &  -7.58 &  -6.15 &  -6.02 &  -5.96 &  -6.00&  -0.88 &  -2.26 &  -3.69 &  -3.82 &  -3.88 &  -3.85 \\
     45 & He-c & WN5 &  -8.75 &  -7.35 &  -5.91 &  -5.80 &  -5.73 &  -5.75&  -0.93 &  -2.33 &  -3.77 &  -3.88 &  -3.95 &  -3.93 \\
     46 & He-c & WO4 &  -9.18 &  -6.29 &  -4.98 &  -4.98 &  -4.85 &  -4.74&  -0.43 &  -3.32 &  -4.62 &  -4.62 &  -4.75 &  -4.86 \\
     47 & He-c & WC4 &  -8.51 &  -6.63 &  -5.53 &  -6.01 &  -5.84 &  -5.22&  -1.02 &  -2.89 &  -4.00 &  -3.51 &  -3.69 &  -4.31 \\
     48 & He-c & WC4 &  -8.53 &  -6.96 &  -5.73 &  -5.74 &  -5.43 &  -5.17&  -0.83 &  -2.40 &  -3.63 &  -3.62 &  -3.93 &  -4.18 \\
     49 & He-c & WC4 &  -7.88 &  -6.56 &  -5.00 &  -5.25 &  -5.00 &  -4.53&  -1.24 &  -2.56 &  -4.12 &  -3.87 &  -4.12 &  -4.59 \\
     50 & He-c & WO4 &  -7.24 &  -6.00 &  -4.24 &  -4.90 &  -4.74 &  -3.94&  -1.82 &  -3.07 &  -4.82 &  -4.16 &  -4.33 &  -5.12 \\
     51 & He-c & WO1 &  -6.27 &  -4.43 &  -3.23 &  -2.88 &  -2.76 &  -2.80&  -2.86 &  -4.70 &  -5.91 &  -6.26 &  -6.38 &  -6.34 \\
     52 & He-sh & WO1 &  -6.32 &  -4.61 &  -3.33 &  -2.98 &  -2.87 &  -2.95&  -2.92 &  -4.63 &  -5.91 &  -6.26 &  -6.37 &  -6.29 \\
     53 & end C-c & WO1 &  -5.84 &  -4.22 &  -2.74 &  -2.66 &  -2.62 &  -2.71&  -3.63 &  -5.25 &  -6.73 &  -6.81 &  -6.85 &  -6.76 \\
\hline
\end{tabular}
\end{minipage}
%\end{table*}
\end{sidewaystable*}

\section{Caveats}
\label{caveat}
One should have in mind that the results discussed in the previous sections are dependent on a number of physical ingredients of the models, concerning both the interior and atmosphere/wind of massive stars. Changes in any physical ingredient that has an impact on the position of the star on the HR diagram will affect the output spectrum, and possibly some of the conclusions reached here. For instance, the overshooting parameter is well known to affect the position of the star on the HR diagram \citep{maeder75}. 

Likewise, the evolution of a massive star is strongly dependent on mass loss, which affects not only the tracks in the HR diagram but also the spectroscopic appearance. The effects of mass loss on the interior evolution are well documented in the literature (see \citealt{chiosi86} and references therein), and exploring its effects on the spectroscopic appearance is computationally expensive and beyond the scope of this paper. We note though that a reduction in $\mdot$ during the MS would favor the appearance of O stars with luminosity class V and III and, instead of a long LBV phase,  BSGs/BHGs and A-type supergiants/hypergiants would be favored as well.

As noted before, wind clumping can also have a significant effect on the spectra of massive stars \citep{hillier91,schmutz97}. In the next subsection, we discuss how the choice of the clumping parameters ($f_\infty$ and $v_c$) affect our results.

\subsection{Effects of clumping}
\label{mdotclump}

We focus on the optical spectra of O stars and four spectral lines (H$\alpha$, \ion{He}{ii} \lam4687, \ion{He}{i} $\lambda4473$, and  \ion{He}{ii} $\lambda4543$) that are key for spectral type and $\mdot$ determinations. The effects of clumping on these and other spectral lines are well known and have been extensively discussed in the literature \citep{crowther02,bouret03,bouret05,bouret13,hillier03,fullerton06,martins07,martins09,martins12,oskinova07,sundqvist10,sundqvist11,surlan12,zsargo08}.

For O stars, our models use the mass-loss rate recipe from \citet{vink01}. Assuming that this prescription is not affected by clumping, \citet{repolust04} and \citet{mokiem07b} find that the observed H$\alpha$ strengths are consistent with the \citet{vink01} mass-loss rates if $f_\infty=0.2$ and clumps are formed immediately above the photosphere. For this reason, we adopt $f=0.2$ and $v_c=30~\kms$ for our O star models. Note, however, that clumping may affect the $\mdot$ prescription depending on the clumping scale length \citep{muijres11}.

\begin{figure}
\center
\resizebox{0.980\hsize}{!}{\includegraphics{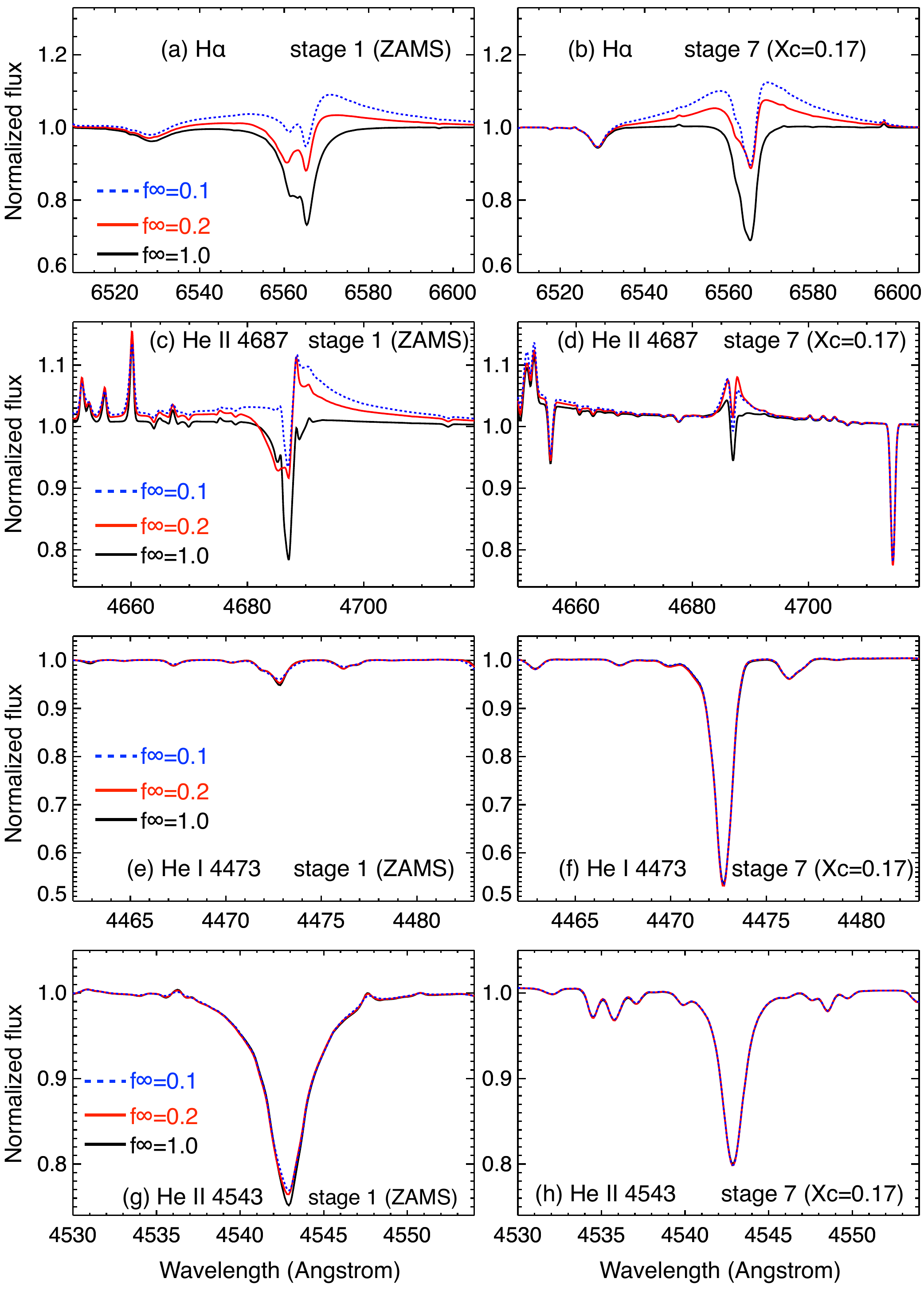}}\\
\caption{\label{lineclump} Normalized line profiles of H$\alpha$ (panels $a$ and $b$), \ion{He}{ii} \lam4687 ($c,d$) \ion{He}{i} $\lambda4473$ ($e,f$) and  \ion{He}{ii} $\lambda4543$ ($g,h$) for the evolutionary stages 1 (ZAMS; left panels) and 7 ($X_c=0.17$; right panels) using the \citet{vink01} mass-loss rates, but $f_\infty=0.1$ (dashed blue) , 0.2 (solid red), and 1.0 (solid black)). All models assume  $v_c=30~\kms$.
}
\end{figure}

To investigate the effects of clumping at the O-type stage, we computed models for stages 1 through 10 with the \citet{vink01} mass-loss rates, $v_c=30~\kms$,  but using different values of $f_\infty$ (0.1, 0.2, and 1.0). We show results for stages 1 (ZAMS) and 7 ($X_c=0.17$) in Fig. \ref{lineclump}, but models computed for the intermediate stages show similar behavior. Since $f$ is radially dependent, spectral lines are affected differently by changes in $f$ according to their formation region (e.g., \citealt{puls06}). Therefore, lines formed close to the photosphere, at low velocities, are less affected by clumping than lines formed in the wind. The strength of the H$\alpha$ emission, which is a recombination process, increases as $f_\infty$ decreases, as illustrated in Fig. \ref{lineclump}a,b. One can see that the H$\alpha$ line predicted by our evolutionary model with a fixed $\mdot$ is significantly affected by clumping.

In early O stars, the main criterium for determining the luminosity class is the strength of \ion{He}{ii} \lam4687 \citep{contialschuler71,walborn71}. Our models show that the \ion{He}{ii} \lam4687 profile is affected by clumping (Fig. \ref{lineclump}c,d), with stronger  \ion{He}{ii} \lam4687 emission seen as $f_\infty$ decreases. As a consequence, models with $f_\infty=0.1$ and $0.2$ both yield a luminosity class I at the ZAMS, while the unclumped model yield a luminosity class III.  At later stages in the MS (stage 7), the choice of $f_\infty$ does not change the luminosity class (both clumped and unclumped models yield a class I), although the shape of the \ion{He}{ii} \lam4687 line varies depending on $f_\infty$ .

This result suggests that once a certain amount of clumping is present, the luminosity class determination is not severely affected by the choice of $f_\infty$. Likewise, while the choice of $v_c$ can affect the amount of emission in H$\alpha$ and \ion{He}{ii} \lam4687, the luminosity class determination should be weakly affected by the precise choice of $v_c$ within reasonable values ($\sim20-300~\kms$), since significant wind contamination of \ion{He}{ii} \lam4687 is already present in unclumped models.

Let us look at how the diagnostic lines for spectral types of O stars (\ion{He}{i} $\lambda4473$ and  \ion{He}{ii} $\lambda4543$) are affected by clumping. These are lines formed mostly close to the photosphere, at low velocities, where clumps have not yet started to form.  Although a weak filling of the core of the line may occur, clumping does not affect these lines to an amount that would cause changes in the spectral type (Fig. \ref{lineclump}e,fg,h). For instance, both clumped and unclumped models  yield a spectral type O3 at the ZAMS and O7.5 at stage 7. 

For stars with dense winds, such as LBV and WRs, most or all lines are formed in the clumped stellar wind. As such, changes in $f$ significantly affect the output spectrum \citep{hillier91}. For a fixed $\mdot$, the strength of the emission lines in general increase with decreasing $f$, as long as no significant changes in the ionization structure occur. Thus, for small changes in $f_\infty$,  the spectrum, magnitudes, and colors are affected, but the spectroscopic classification (BHG, LBV, or WR) remains weakly affected. Large reductions in $f_\infty$ may cause a BHG to appear as an LBV or a WR to appear as an LBV. Therefore, the timescales for the different spectroscopic phases derived in this paper (Table \ref{duration}) should be taken with caution, as they depend on the choice of the clumping parameter. 

Finally, we stress that the real physical conditions in stellar winds are certainly more complicated than our simple exponential clumping law models. Wind clumping in O stars likely originates from line-driven instabilities \citep{owocki88}. Hydrodynamical simulations of line-driven instability predict non-monotonic, complex $f(r)$ \citep{runacres02, sundqvist12}, which were shown to have a significant effect in line profiles of O stars \citep{puls06, najarro11}. However, little is known as to the dependence of the clumping (structure) on stellar properties (but see \citealt{cantiello09}). These uncertainties warrant the simple clumping prescription adopted in our models.

\section{\label{conc} Concluding remarks}

In this paper we investigated the evolution of the interior, spectra, and photometry of a non-rotating 60~\msun\ star. For the first time, coupled stellar evolution and atmospheric modelings covering the full evolution of a massive star were developed. We employed the Geneva stellar evolution code and the CMFGEN atmospheric/wind code to produce observables out of stellar evolutionary calculations. Our main conclusions are summarized below.

\begin{enumerate}

\item We find that a non-rotating 60~\msun\ star has the following evolutionary sequence of spectral types: O3 I (ZAMS) $\rightarrow$ O4 I (mid H-core burning) $\rightarrow$ hot LBV (end H-core burning) $\rightarrow$ cool LBV (start He-core burning) $\rightarrow$ WNL $\rightarrow$ WNE $\rightarrow$ WC (mid He-core burning) $\rightarrow$ WO (end He-core burning until core collapse). 

\item During the MS evolutionary phase, when H-core burning occurs, the star has the following spectral types (durations in parenthesis): early O I (2.89 Myr), late O I (0.36 Myr), BSG (0.04 Myr),  BHG (0.08 Myr), and LBV (0.23 Myr). Thus, our models indicate that some LBVs can be H-core burning objects. During the MS, the star loses $8.5~\msun$ when it has an O-type spectrum, $1.3~\msun$ as a BHG, and $14.0~\msun$ as an LBV. The huge mass loss as an LBV is a consequence of the star crossing the bistability limit of line-drive winds, which occurs around 21000 K \citep{vink01}.

\item The post-MS evolution of a non-rotating 60~\msun\ star is comprised by a short H-shell burning, long He-core burning, and short He-shell burning, C-core burning, and multiple shell-burning stages. 

\item During the H-shell burning (0.007 Myr), the star is an LBV and loses 4~\msun. During the He-core burning, the surface temperature and chemical composition change dramatically, causing the star to show a variety of spectral types. It begins the He-core burning phase as an LBV and, in 0.10~Myr, 5.94~\msun\ is lost. It goes rapidly through spectroscopic phases of WNL with small amounts of H, WNE with small amounts of H, and WNE without H (WN 5). Most of the He-core burning (0.244 Myr) is spent as an early WC (WC 4), when 7.02~\msun\ of material is lost.

\item At the end of He-core burning, the surface temperature increases as the star contracts. Combined with the increase in O abundance at the surface, this implies a WO spectral type, with strong \ion{O}{vi} $\lambda3811$ emission. The star remains as  a WO, with extremely high $\teff$, until core-collapse.

\item With our approach of computing spectra out of evolutionary models, we investigated the duration of the different spectroscopic phases in a direct way. Compared to the results from \citet{georgy12a}, which employed the same evolutionary model but chemical abundance and $\teff$ criteria, we find a similar duration for the O and WR phases. However, we find that the star has a spectral type similar to LBVs for 0.235 Myr, which is relatively longer than commonly assumed for LBVs (a few 0.010 Myr; \citealt{bohannan97}). We stress that this result is valid only for a 60~\msun\ star without rotation and we anticipate that rotating models will have a much shorter duration for the LBV phase. Also, the long duration is extremely dependent on our spectroscopic definition of LBV and the \citealt{vink01} mass-loss recipe assumed.

\item Although we find similar duration as previous studies for the WR phase, the lifetimes of the different subtypes of WR is significantly different. We find that the duration of the WNL phase is about 20 times shorter than when using chemical abundance criteria. The duration of the WNE phase is increased by a factor of 2. The duration of the WC phase is also reduced by 16\%. Finally, we find that the endstage is characterized by a short WO phase that lasts for 0.038 Myr.

\item We present the photometric evolution of a non-rotating 60~\msun\ star from the ZAMS to the pre-SN stage. We find that the star becomes progressively brighter in all optical and near-IR filters during the MS, and reaches the peak of its brightness as an LBV during H-shell burning. It becomes rapidly faint in these filters as the He-core burning phase begins and $\teff$ increases. At the end of He-core burning and subsequent advanced phases, the star becomes even fainter in the optical and IR as $\teff$ surpasses 100000~\K. This has serious consequences for detecting progenitors of SN Ic in pre-explosion images, predicting that these progenitors  should be undetectable with the current magnitude limits \citep{gmg13}.

\item We computed the evolution of  number of  \ion{H}{} ($Q_0$), \ion{He}{i} ($Q_1$), and \ion{He}{ii} ($Q_2$) ionizing photons emitted per unit time. We found that $Q_0$ is roughly constant until the end of the MS ($Q_0=10^{49.5}$~photon/s), when the star becomes an LBV with low $\teff$. Our models indicate that $Q_0$ increases again at the beginning of He-core burning, remaining roughly constant until the pre-SN stage. $Q_1$ follows a similar qualitative trend, while $Q_2$ rapidly decreases during the MS and only becomes significant at the transition between the WNE and WCE phase, and before core-collapse, when the star is a WO.

\end{enumerate}

\begin{acknowledgements}
We thank the referee, Alex de Koter, for a careful review and a constructive report that improved the quality of the original manuscript. We are also thankful to John Hillier and Paco Najarro for discussions, sharing models, and making CMFGEN available, and Paul Crowther for comments, advices, and discussions. We thank Jes\'us Ma\'{i}z-Appel\'aniz for making CHORIZOS available. JHG is supported by an Ambizione Fellowship of the Swiss National Science Foundation. CG acknowledges support from the European Research Council under the EU Seventh Framework Program (FP/2007-2013) / ERC Grant Agreement n. 306901.

\end{acknowledgements}

\bibliography{../../refs}

\end{document}